\documentclass{ws-ijmpd}
\usepackage[super,compress]{cite}
\usepackage[colorlinks=green,linkcolor=blue]{hyperref}
\usepackage{cleveref}
\usepackage{color}
\usepackage{multirow}
\Crefrangeformat{equation}{Eqs. (#3#1#4)--(#5#2#6)}
\Crefname{equation}{Eq.}{Eqs.}
\Crefname{figure}{Fig.}{Figs.}
\Crefname{section}{Sec.}{Secs.}

\def\MLine#1{\par\hspace*{-\leftmargin}\parbox{\textwidth}{\[#1\]}}

\begin{document}

\markboth{S. Capozziello, R. D'Agostino, O. Luongo}
{ EXTENDED GRAVITY COSMOGRAPHY}

\catchline{}{}{}{}{}
%
%%%%%%%%%%%%%%%%%%%%%%%%%%%%%%%%%%%%%%%%%%%%%%%%%%%%%%%%%%%%%%%%%%%%

%\title{INSTRUCTIONS FOR TYPESETTING
%MANUSCRIPTS\footnote{For the title, try not to use more than 3 lines.
%Typeset the title in 10~pt Times roman, uppercase and boldface.}  }
%

\title{EXTENDED  GRAVITY COSMOGRAPHY}

%\author{FIRST AUTHOR\footnote{Typeset names in
%8~pt roman, uppercase. Use the footnote to indicate the
%present or permanent address of the author.}}
%
%\address{University Department, University Name, Address\\
%City, State ZIP/Zone,
%Country\footnote{State completely without abbreviations, the
%affiliation and mailing address, including country. Typeset in
%8~pt Times italic.}\\
%first\_author@university.edu}
%

\author{SALVATORE CAPOZZIELLO}

\address{Dipartimento di Fisica, Universit\`a di Napoli  ``Federico II'',  and
Istituto Nazionale di Fisica Nucleare, Sez. di Napoli, Via Cinthia 9, I-80126 Napoli, Italy, \\
Gran Sasso Science Institute, Via F. Crispi 7, I-67100, L'Aquila, Italy, \\
Laboratory for Theoretical Cosmology,\\
Tomsk State University of Control Systems and Radioelectronics (TUSUR),\\
 634050 Tomsk, Russia,\\
Tomsk State Pedagogical University, ul. Kievskaya, 60, 634061
Tomsk, Russia.\\
capozziello@na.infn.it}

\author{ROCCO D'AGOSTINO}

\address{Istituto Nazionale di Fisica Nucleare, Sez. di Roma ``Tor Vergata'', Via della Ricerca Scientifica 1, I-00133, Roma, Italy. \\
rocco.dagostino@roma2.infn.it}

\author{ORLANDO LUONGO}

\address{Istituto Nazionale di Fisica Nucleare, Laboratori Nazionali di Frascati, 00044 Frascati, Italy. \\
Scuola di Scienze e Tecnologie, Universit\`a di Camerino, 62032 Camerino, Italy. \\
NNLOT, Al-Farabi Kazakh National University, Al-Farabi av. 71, 050040 Almaty, Kazakhstan. \\
orlando.luongo@lnf.infn.it}

\maketitle

\begin{history}
\received{Day Month Year}
\revised{Day Month Year}
\end{history}

\begin{abstract}
%The abstract should summarize the context, content
%and conclusions of the paper in less than 200 words. It should
%not contain any references or displayed equations. Typeset the
%abstract in 8 pt Times roman with baselineskip of 10~pt, making
%an indentation of 1.5 pica on the left and right margins.

Cosmography  can be considered as a sort of  a model-independent approach to tackle the dark energy/modified gravity  problem. In this review, the success and the shortcomings of the $\Lambda$CDM model, based on General Relativity and standard model of particles, are discussed in view of the most recent observational constraints. The motivations for considering extensions and modifications of General Relativity are taken into account, with particular attention to $f(R)$ and $f(T)$  theories of gravity where dynamics is represented by curvature or torsion field respectively. The features of  $f(R)$ models are explored in  metric and Palatini formalisms. We discuss the connection between $f(R)$ gravity  and scalar-tensor theories highlighting the role of conformal transformations in the Einstein and Jordan frames. Cosmological dynamics of  $f(R)$ models is investigated through  the corresponding viability criteria. Afterwards, the equivalent formulation of General Relativity (Teleparallel Equivalent General Relativity) in terms of torsion and its extension to $f(T)$ gravity is considered. Finally, the cosmographic method is adopted to break the degeneracy among dark energy models. A novel approach, built upon rational  Pad\'e and Chebyshev polynomials, is proposed to overcome  limits of standard cosmography based on Taylor expansion.  The approach   provides   accurate model-independent approximations of the Hubble flow. Numerical analyses, based on Monte Carlo Markov Chain integration of  cosmic data, are presented to bound  coefficients of the cosmographic series. These techniques are thus applied to reconstruct $f(R)$ and $f(T)$ functions  and to frame the late-time expansion history of the universe with no \emph{a priori} assumptions on its equation of state. A comparison  between the $\Lambda$CDM cosmological model with $f(R)$ and $f(T)$ models is  reported.

\end{abstract}

\keywords{Extended gravity; cosmography; dark energy; cosmological observations.}

%\ccode{PACS numbers:}

\tableofcontents

\section{Introduction}
\label{sec:intro}

The present picture of the universe is based on the homogeneous and isotropic Friedmann-Lema\^{i}tre-Robertson-Walker (FLRW) metric, which represents a solution of the Einstein field equations of General Relativity (GR). The success of the Big Bang model \cite{Weinberg72} comes from its remarkable match with the available cosmological observations.
However, some shortcomings of this picture, emerged in the last thirty years, have made scientists doubt on the appropriateness of achieve a comprehensive picture of the universe simply based on GR and standard perfect fluid matter.
A crucial role in this respect is played by the relation between cosmology and quantum field theory.
The Big Bang singularity along with  issues such as the monopole, horizon, and flatness problems \cite{Guth81} undermine the standard model of particle physics and the standard model of cosmology as an adequate description of the  universe at high-energy regimes.
On the other hand, a fundamental theory to describe space-time in its full quantum aspects cannot be represented by a classical theory like GR. Therefore, the lack of a definitive quantum theory of gravity is the reason to consider alternative gravitational theories where GR can be reproduced in the semiclassical limit. The so-called Extended Theories of Gravity (ETG), based on corrections and extensions of the Einstein’s theory,  are the most fruitful paradigms following the aforementioned receipt. The idea behind this approach is essentially to consider some effective quantum gravity action adding higher order curvature invariants and scalar fields minimally or nonminimally coupled to the gravity sector recovering GR at local scales and in the weak field limit \cite{Capozziello-Faraoni10}. In this perspective, it is more correct to deal with \textit{Extended Gravity}\footnote{ A typical example of extended theory  is  $f(R)$ gravity. Assuming $f(R)=R$ means that GR is a particular theory in a wide family of model. On the other hand, considering $f(R)=R+\alpha R^2$ means that, if $R^2$ term is negligible, GR is recovered. Regarding $f(T)$ gravity, the situation is similar because $f(T)=T$ means recovering TEGR (Teleparallel Equivalent General Relativity). In this case, dynamics is given by the torsion scalar $T$, instead of the curvature scalar $R$. However the description is equivalent. We intend with  "modified gravity" or "alternative gravity", theories which do not reproduce GR in a given energy regime or choice of models.  This can be the case of some gauge theories of gravity.} instead of modified gravity\cite{Capozziello08}.
The presence of non-minimal couplings and high-order terms appears necessary in any scheme trying to unify fundamental interactions (\emph{e.g.} supergravity and superstring theories) \cite{Buchbinder92}.
These contributions come from first or higher-order loop corrections in the high curvature regime \cite{Buchbinder92}.
Quantization of matter fields in curved space-time   leads to corrections of the Hilbert-Einstein Lagrangian due to
the interactions between the background geometry and quantum scalar fields  \cite{Birrell82}.

A revision of standard cosmological scenarios is necessary also at late-time epochs. In fact, observations of Supernovae Ia (SNeIa) \cite{Perlmutter99,Riess-Schmidt98} suggested that the expansion of the universe has recently entered an accelerated phase that cannot be explained only by the dynamics of  ordinary matter and radiation  as constituents of the cosmic fluid \cite{roccoacc}.
On the other hand,  Cosmic Microwave Background (CMB) anisotropies \cite{WMAP9,Planck15}  strongly suggest a universe with flat spatial curvature.

Within the framework of GR, the simplest explanation for the cosmic speed up would be the well known cosmological constant \cite{Sahni00}, which defines the concordance $\Lambda$ Cold Dark Matter ($\Lambda$CDM) model.
Although very effective in fitting most of the cosmological data, the $\Lambda$CDM model  is plagued by some fundamental issues  related to its nature \cite{Weinberg89,Carroll00}.
One possible attempt to fix these problem  is to replace the cosmological constant with a slowly rolling scalar field, known as \textit{quintessence}  \cite{Padmanabhan03,Copeland06}. However, even the quintessence approach presents some issues related to the \textit{coincidence problem}.

Furthermore, there exists a different way to approach the cosmic acceleration problem. In fact, the observed behaviour of the late-time expansion might not be due to  new species in the cosmic fluid, but rather the signal of  a breakdown of standard gravity at infrared regimes. In this respect, modifications of the Friedmann equations give rise to alternative paradigms where effective models with generalizations of the gravity action  (e.g.  high-order curvature terms) can yield to quintessence behaviour \cite{Capozziello02}.
Moreover, the cosmological constant behaviour may be the consequence of including torsion fields starting from the so called Teleparallel Equivalent General Relativity (TEGR) \cite{Cai15}.

In these alternative approaches, the philosophy is that conceptual shortcomings in  cosmic evolution are overcome deriving  negative pressure scenarios naturally originated from the further geometric degrees of freedom that these models contain with respect to standard GR.

In this review paper,  we want to discuss how a cosmographic approach, besides observations coming from {\it Precision Cosmology},  can contribute to select self-consistent cosmological models based on extensions of GR and TEGR.

The structure of the paper is as follows.
In \Cref{sec:standard}, we review the concordance cosmological model and the issues related to the nature and origin of dark energy.
In \Cref{sec:ETG}, we discuss  $f(R)$ gravity  as a straightforward extension of GR introduced to approach shortcomings of the standard cosmological scenario.
In particular, we discuss  dynamics and observational viability of such theories in both metric and Palatini   formulations.
 Gravity with torsion  is considered in \Cref{sec:torsion}. Specifically, we present the teleparallel equivalent of Einstein's theory (TEGR) and extend the discussion to  generic functions of  torsion scalar in presence, eventually,  of scalar fields coupled to gravity.
In \Cref{sec:cosmography}, we present the cosmographic method as a model-independent tool to discriminate among  dark energy models. The limits of the standard cosmographic approach are discussed and  a new method, based on rational polynomials, is presened. The approach is aimed  to alleviate the convergence issues at high-redshift epochs.
Finally, in  \Cref{sec:f(R) cosmography,sec:f(T) cosmography},  the cosmographic method is applied to reconstruct  gravitational  action in a model-independent way starting from the cosmological constraints of the late-time universe.

Throughout the text, we use the metric signature $(+,-,-,-)$ and units such that $c = \hbar = 1$, unless differently specified. We also use the notation $\kappa\equiv 8\pi G=M_\text{P}^{-2}$, where $G$ is the Newton constant and $M_\text{P}$ is the reduced Planck mass.

\section{The  cosmological puzzle}
\label{sec:standard}

The standard cosmological model is  based on the \textit{cosmological principle}, which consists of two principles of spatial invariance. The first invariance is the isomorphism under translations. This means assuming the universe to be \textit{homogeneous} on large scales, with no special points and galaxies evenly distributed in space. The second invariance is the isomorphism under rotations. This implies an \textit{isotropic} universe with no special spatial directions, where the galaxies are evenly distributed in different angular directions at large scales.

The cosmological principle provides us with the simplest cosmological models, the homogeneous and isotropic universe described by the FLRW metric \cite{Friedmann,Lemaitre31,Robertson35,Walker37}:
\begin{equation}
ds^2\equiv g_{\mu\nu}dx^\mu dx^\nu=dt^2-a(t)^2\left[\dfrac{dr^2}{1-kr^2}+r^2(d\theta^2+\sin^2\theta\ d\phi^2)\right] ,
\label{FLRW metric}
\end{equation}
where $t$ is the cosmic time, $a(t)$ is the dimensionless scale factor normalized to unity at the present time ($a(t_0)=1$), and $k$ defines the spatial curvature:
\begin{equation}
k =\left\{
\begin{aligned}
&-1   \hspace{0.5cm} \text{open universe}, \\
&\hspace{0.4cm} 0 \hspace{0.5cm} \text{flat universe},  \\
&+1   \hspace{0.5cm} \text{closed universe}.
\end{aligned}
\right.
\end{equation}
To determine the dynamics of the gravitational field for a homogeneous and isotropic universe, we write the Einstein field equations:
\begin{equation}
R_{\mu\nu}-\dfrac{1}{2}Rg_{\mu\nu}=\kappa T_{\mu\nu}\ ,
\end{equation}
where $R_{\mu\nu}$ is the Ricci tensor, $R=g_{\mu\nu}R^{\mu\nu}$ is the Ricci (scalar) curvature. $T_{\mu\nu}$ is the energy-momentum tensor which, for a perfect fluid\footnote{A `perfect' fluid is an ideal fluid characterized by zero viscosity, no shear stresses and vanishing vorticity.}, takes the form
\begin{equation}
T_{\mu\nu}=(\rho+P)u_\mu u_\nu-Pg_{\mu\nu}\ ,
\label{energy-momentum tensor}
\end{equation}
where $\rho$ and $P(\rho)$ are the density and pressure of the \textit{barotropic} fluid, respectively, which depend on the cosmic time only in agreement with the symmetry properties of the FLRW metric.
The four-velocity field $u^\mu$ refer to an observer moving inside the light cone and, hence, it is normalized according to
\begin{equation}
g_{\mu\nu}u^\mu u^\nu=1 \ .
\end{equation}
In a reference frame which is at rest with  respect to the fluid $(u^i=0)$, the relation $u_0u^0=1$ holds  and one then has
\begin{equation}
T_{\mu\nu}=\text{diag}(\rho,-P,-P,-P)\ .
\label{eq:diag T}
\end{equation}

 SNeIa observations  at the end of the 90's indicated that the universe is currently undergoing a phase of accelerated expansion \cite{Perlmutter99,Riess-Schmidt98}. This implies that
\begin{equation}
\rho+3P<0 \ .
\end{equation}
Clearly, this condition cannot be satisfied if the cosmic fluid were made only of radiation and pressureless non-relativistic matter. Therefore, cosmological sources  have to include a further component with negative pressure $(P<-\rho/3)$, which is today dominant over the other species. This component is dubbed \textit{dark energy} \cite{Huterer01,Padmanabhan03,Caldwell03,Copeland06,Li11}. The simplest model that can describe the dark energy behaviour is a model with the cosmological constant $\Lambda$, characterized by the equation of state
\begin{equation}
w_\Lambda\equiv\dfrac{P_{\Lambda}}{\rho_{\Lambda}} =-1\ .
\label{eq:Lambda EoS}
\end{equation}
The gravitational contribution of the cosmological constant can be added into the Einstein-Hilbert action as
\begin{equation}
\mathcal S=\int d^4x\ \sqrt{-g} \left[\dfrac{1}{\kappa}\left(\frac{R}{2}-\Lambda\right)+\mathcal{L}_m\right] ,
\end{equation}
where $\mathcal{L}_m$ is the matter Lagrangian density.
The field equations are obtained by varying the above action with respect to the metric:
\begin{equation}
G_{\mu\nu}-\Lambda g_{\mu\nu}=\kappa T_{\mu\nu}\ ,
\label{eq:field eq with Lambda}
\end{equation}
where $G_{\mu\nu}\equiv R_{\mu\nu}-\frac{1}{2}g_{\mu\nu}R$ is the \textit{Einstein tensor}.
Writing \Cref{eq:field eq with Lambda} for the FLRW metric, one obtains the Friedmann equations as
\begin{align}
&H^2= \dfrac{\kappa}{3}\rho+\dfrac{\Lambda}{3}-\dfrac{k}{a^2}\ ,\label{Hubble1}\\
&\dfrac{\ddot a}{a}=-\dfrac{\kappa}{6}(\rho+3P)+\dfrac{\Lambda}{3}\ .
\end{align}
We thus define the density parameters associated to curvature, matter and cosmological constant as, respectively,
\begin{equation}
\Omega_k\equiv -\dfrac{k}{a^2H^2}, \  \Omega_{m}\equiv \dfrac{\kappa \rho_{m}}{3H^2}, \    \Omega_{\Lambda}\equiv-\dfrac{\Lambda}{3H^2}\,
\end{equation}
obeying the cosmic rule
\begin{equation}
1=\Omega_m+\Omega_k+\Omega_\Lambda\,,
\end{equation}
derived from \eqref{Hubble1}.
We can finally write the Hubble expansion rate in the form\footnote{We here include the contribution of radiation $\Omega_{r}$, which is usually neglected in the late-time epochs.}
\begin{equation}
H(a)=H_0\left[\dfrac{\Omega_{r0}}{a^4}+\dfrac{\Omega_{m0}}{a^3}+\dfrac{\Omega_{k0}}{a^2}+\Omega_{\Lambda 0}\right]^{1/2} ,
\end{equation}
where the subscript `0' denotes the corresponding present values of the density parameters.
The combinations of low-redshift data and CMB anisotropy measurements portray a universe with the following features:
\begin{itemize}
\item  vanishing spatial curvature: $\Omega_{k0}\approx 0$ ;
\item very small amount of residual radiation: $\Omega_{r0}\approx 5\times 10^{-5}$ ;
\item about 30\% of matter-energy density, mainly constituted of cold dark matter and a small contribution of baryonic matter: $\Omega_{m0}=\Omega_{cdm,0}+\Omega_{b0}\approx 0.3$, with $\Omega_{cdm,0}\approx 0.25$ and $\Omega_{b0}\approx 0.05$ ;
\item about $70\%$ of dark energy in the form of cosmological constant: $\Omega_{\Lambda0}\approx0.7$.
\end{itemize}
Such a "paradigm" is named $\Lambda$CDM model and represents the so-called \textit{concordance model of cosmology}.

\subsection{Issues with the $\Lambda$CDM model}

The simplest explanation for the accelerating universe provided by the cosmological constant, although very effective in fitting all the major cosmological observables, does not give a satisfactory physical interpretation of dark energy for a number of issues \cite{Carroll00}.
Particle physicists considered the possibility to identify the cosmological constant with the energy of the vacuum.
Assuming that the vacuum is a Lorentz-invariant state, its energy-momentum tensor takes the form
\begin{equation}
T_{\mu\nu}^\text{vac}=-\rho_\text{vac}g_{\mu\nu}\ ,
\end{equation}
where the vacuum energy density $\rho_\text{vac}$ is related to an isotropic pressure by
\begin{equation}
P_\text{vac}=-\rho_\text{vac}\ .
\label{eq:vacuum EoS}
\end{equation}
Comparing \Cref{eq:vacuum EoS} with \Cref{eq:Lambda EoS}, we find that they are formally equivalent:
\begin{equation}
\rho_\text{vac}=\rho_\Lambda\equiv \dfrac{\Lambda}{\kappa}\ .
\end{equation}
From the classical point of view, $\Lambda$ is simply a constant whose value should be determined through experiments.
These considerations, however, change once quantum mechanics enters the picture. In fact, from the Planck constant one can define a gravitational length scale named reduced Planck length:
\begin{equation}
L_\text{P}={M_\text{P}}^{-1}=\sqrt{\kappa}\ .
\end{equation}
We can thus think about quantum fluctuations in the vacuum.
For a non-interacting quantum field, each mode contributes to the vacuum energy and the net result is obtained by integrating over all the modes. This integral is in principle divergent, which implies that the vacuum energy is infinite. To avoid the ultraviolet divergence, one can introduce a cut-off and ignore any contribution above that. Then, one naturally would expect that this cut-off is related to the Planck scale by
\begin{equation}
\Lambda\sim {L_\text{P}}^{-2}\ ,
\end{equation}
so as to obtain
\begin{equation}
\rho_\Lambda\sim M_\text{P}^4 \sim (10^{18} \text{ GeV})^4\ .
\label{eq:theor vacuum}
\end{equation}
On the other hand, measurements of the cosmological constant over the last decades from observations of SNeIa and CMB anisotropies \cite{Planck15} indicate the following value for the vacuum energy:
\begin{equation}
\rho_\Lambda^{(\text{obs})}\sim(10^{-3}\ \text{eV})^4\ .
\label{eq:obs vacuum}
\end{equation}
Then, comparing (\ref{eq:theor vacuum}) with (\ref{eq:obs vacuum}), one gets
\begin{equation}
\rho_\Lambda^{(\text{obs})}\sim 10^{-120} \rho_\text{vac}.
\end{equation}
This embarrassing discrepancy of 120 orders of magnitude is known as \textit{the cosmological constant problem} \cite{Weinberg89,Peebles03}.

The second issue is called {\it coincidence problem} \cite{Zlatev99,Sahni02}. The concordance cosmological model provides values for the vacuum energy density and the matter density of the same order of magnitude. However, the two components have very different evolution histories:
\begin{equation}
\dfrac{\Omega_\Lambda}{\Omega_m}=\dfrac{\rho_\Lambda}{\rho_m}\propto a^3\ ,
\end{equation}
which implies that the current acceleration of the cosmic expansion started relatively recently.
It becomes immediately clear that the transition between a matter-dominated universe and a universe dominated by dark energy is quite fast. This means that the probability for an observer to live during a period when the two species have the same order of magnitude is very small. Therefore, there is no physical reason for us to be on the verge of such a special moment when these components have a similar order of magnitude.

Another further problem that compromises our understanding of the cosmic speed up concerns the discrepancy between the direct and indirect (model-dependent) measurements of the present expansion rate of the universe \cite{roccoH0}. Since the first determination by Hubble in 1929 \cite{Hubble29}, for decades astronomers derived values for $H_0$ in the range 50$\div$100 km/s/Mpc. Improved accuracy in the measurements of $H_0$ were made over the years thanks to a better control of systematics and the use of different calibration techniques. Using the period-luminosity relation for Cepheids to calibrate a number of secondary distance indicators such as SN Ia and the Tully-Fisher relation, the Hubble Space Telescope Key Project \cite{Freedman01} estimated $H_0=(73 \pm 8$) km/s/Mpc. The most recent direct estimate of $H_0$ has been provided in \refcite{Riess16}: $H_0=(73.24 \pm 1.74)$ km/s/Mpc. This value is in tension with the most recent result of the Planck collaboration \cite{Planck15} for the $\Lambda$CDM model,  $H_0=(67.51 \pm 0.64)$ km/s/Mpc, which represents so far the strongest constraint on $H_0$.
An alternative method to measure the Hubble constant, independent of the local distance ladder, is provided by strong gravitational lenses with time delays between the multiple images. Using this approach, the H0LiCOW collaboration estimated  $H_0=71.9^{+2.4}_{-3.0}$ km/s/Mpc for $\Lambda$CDM \cite{Bonvin16}. This value is in in agreement with the direct measurement of \cite{Riess16} but in tension with Planck.

During the past years, many attempts have been done to solve the dark energy problem. From particle physics point of view, the lack of observed supersymmetric partners of known particles in accelerators leads to assume that the scale at which supersymmetry was broken is of the order of $10^{3}$ GeV. This then implies the following estimate for the vacuum energy density:
\begin{equation}
\rho_\Lambda \sim M_\text{SUSY}^4 \sim (10^{12}\ \text{GeV})^4\ .
\end{equation}
This results is, however, still 60 orders of magnitude larger than the observed value (\ref{eq:obs vacuum}). Other approaches based on string theory or loop quantum gravity \cite{Weinberg89,Carroll00} require some fine-tuning and, in any case, fail to address the coincidence problem. In 2018 a mechanism for cancelling $\Lambda$ out has been proposed through the use of a symmetry breaking potential in a Lagrangian formalism in which matter shows a non-vanishing pressure \cite{LMmodel}. The model assumes that standard matter provides a pressure which counterbalances the action due to the cosmological constant. It has been shown that this mechanism permits to take vacuum energy as quantum field theory predicts, but removing the huge magnitude through a counterbalance term due to baryons and cold dark matter only. The approach is equivalent to have a dark fluid which degenerates with the standard cosmological model \cite{dopolm1,dopolm2,dopolm3,dopolm4} and enters the class of unified dark energy models \cite{dopolm5,dopolm6,dopolm7}.

\subsection{Dark energy}

Another approach that seeks for solving the cosmological constant problem is to consider dynamical properties of dark energy. A dynamical dark energy, however, should be able to mimic the cosmological constant at the present time, as required by cosmological observations. In this sense, similarly to the inflationary mechanism \cite{Linde82,Albrecht82}, but at different energies, the simplest candidate is a canonical scalar field, often dubbed  \textit{quintessence} \cite{Peebles-Ratra88,Caldwell98,Armendariz00}. For a homogeneous scalar field minimally coupled to gravity, the Klein-Gordon equation in FLRW space-time reads
\begin{equation}
\ddot{\phi}+3H\dot{\phi}+V'(\phi)=0 ,
\label{eq:Klein-Gordon}
\end{equation}
where $V(\phi)$ is the potential of the scalar field and the `prime' denotes derivative with respect to $\phi$. Thus, the energy density and pressure are given by, respectively,
\begin{align}
&\rho_\phi=\dfrac{1}{2}\dot{\phi}^2+V(\phi)\ , \label{eq:density scalar field}\\
&P_\phi= \dfrac{1}{2}\dot{\phi}^2-V(\phi)\label{eq:pressure scalar field}\ .
\end{align}
It is clear from \Cref{eq:density scalar field,eq:pressure scalar field} that $w_\phi=P_\phi/\rho_\phi$ approaches $-1$ if the \textit{slow-roll} condition $\dot{\phi}^2\ll V(\phi)$ is satisfied. Imposing this condition, from \Cref{eq:Klein-Gordon}, we  must  have $H\sim \sqrt{V''(\phi)}$. Thus, considering that $\sqrt{V''(\phi)}$ represents the effective mass of the scalar field $m_\phi$ and that the current value of $V(\phi)$ should be of the order of the observed $\Lambda$, one gets
\begin{equation}
m_\phi \sim 10^{-33}\ \text{eV}\ .
\label{eq:mass scalar field}
\end{equation}
Since the masses of scalar fields in quantum field theory are several orders of magnitude larger than the value of (\ref{eq:mass scalar field}), many doubts remain on whether quintessence could be an actual solution of the cosmological constant problem.

There are also attempts to solve the coincidence problem by adopting specific models of quintessence called \textit{tracker models} \cite{Ferreira98,Copeland98,Steinhardt99}. In these models, the coincidence problem is solved as the energy density of the scalar field has the same behaviour of the radiation and matter energy densities for a significant part of the cosmic evolution. These solutions do not suffer from fine-tuning problems related to initial conditions, even though they are dependent on the parameters of the potential.

\section{Extended theories of gravity: the case of $f(R)$ gravity}
\label{sec:ETG}
An alternative approach to address the dark energy issues is to consider modifications or extensions of the l.h.s of Einstein's field equations.
We here discuss such a  possibility  to cure the shortcomings of the concordance cosmological model. We start presenting some  historical reasons that brought first to consider extensions of GR at ultraviolet (UV) scales, and to infrared (IR) scales.

\begin{itemize}
\item \textbf{ UV scales}\\
Due to their empirical success in describing the physical phenomena, GR and Quantum Field Theory (QFT) represent the two main pillars which modern physics is built on. While GR is the theory of gravitating systems and non-inertial frames on large scales, QFT provides a description of the world on small scales and at high energy regimes. As a classical theory, GR does not take into account the quantum nature of matter; on the other hand, QFT assumes that the space-time contains quantum fields. The key point is, thus, to figure out how quantum fields behave in  presence of gravity or, in other words, whether these two theories are compatible.
Non-classical effects are expected to be relevant for gravity at Planck's scale, which is unfortunately unaccessible by current experiments. Nevertheless, investigating the fundamental nature of space-time on very small scales is unescapable  to shed light on the physics of the universe from the Big Bang to Planck's era.

At the end of the 1950s, the necessity to   build up some unified theory capable of describing all the fundamental interactions under the standard of QFT made recognize the need for a quantum theory of gravity.
So far, any unification scheme trying to include gravity has revealed unsuccessful or not completely satisfactory.
The difficulties are mainly due the fact that the gravitational field describes the background space-time where the same gravitational degrees of freedom, that is the space-time itself,   have to be considered as dynamical variables.
The assumed mutual interaction between geometry and quantum matter fields necessarily leads to modifications of the standard Einstein-Hilbert action, that is, to consider \textit{Extended Theories of Gravity} (ETG) \cite{review1,review2}. Such theories represent a semi-classical approach where GR is recovered in the low-energy limit. As GR, these models are gauge invariant and consist of adding higher-order curvature invariants (such as $R^2$, $R_{\mu\nu}R^{\mu\nu}$, $R_{\alpha\beta\mu\nu}R^{\alpha\beta\mu\nu}$, $R\square R$, $R\square^k R$) and minimally or non-minimally  coupled terms between scalar fields and geometry (such as $\phi^2 R$), which come out from the effective action of Quantum Gravity \cite{Nojiri07,Capozziello08}.

\item \textbf{ IR scales}\\
Einstein's theory has proven successful over many years of experimental tests. GR is in remarkable agreement with precision tests of gravity done in the solar system and consistent with gravitational waves detection \cite{GW}. However, GR has
not been tested independently on cosmological scales. The observational evidences that the main
amount of the present matter content of our universe is in the form of unknown particles that
are not included in the standard model of particles and interactions, and the discovery of the
present accelerated expansion of the universe, have led cosmologists to consider the possibility that
 GR might not be, in fact, the correct theory of gravity to describe the universe at  larger
scales. In order to address this issue, two different kinds of phenomena have been proposed:
the so-called `dark energy models' and modifications to GR. While the former introduce a new
fluid or field from which the apparent cosmological constant could originates, the latter refers to
modifying the l.h.s. of  Einstein's equations, i.e., GR itself,  by modifying or improving
the Einstein-Hilbert action.

The ETG theories have thus attracted great interest in cosmology. The related cosmological models, in fact, provide inflationary scenarios able to overcome the shortcomings of standard model based on GR, and the theoretical predictions match with the CMB observations \cite{Starobinsky80,Duruisseau83,La89}. Moreover, conformal transformations allow to reformulate the higher-order and non-minimally coupled terms  into  GR term plus one or multiple minimally coupled scalar fields \cite{Maeda89,Wands94,Capozziello98}.
However, modifications of standard gravitational theory are characterized by mathematical difficulties since the corrections to the standard Lagrangian increase the non-linearity of the field equations, which often produce differential equations higher than the second order.
\end{itemize}

\noindent The possibility to include higher-order curvature invariants in the gravitational action was firstly considered in the 1960s as an attempt to quantize gravity. It was shown that renormalization at the one-loop level  requires adding higher-order curvature terms to the Einstein-Hilbert action \cite{Utiyama62}. It was initially expected that such terms were suppressed by small couplings and their relevance was confined only to the strong gravity regimes. More recently, however, the dark energy problem related to the late-time acceleration of the universe has revived interest in considering these modifications as possible extensions of GR.
We can account for higher-order curvature invariants by generalizing the standard gravitational action to any function of the Ricci scalar:ì, that is:
\begin{equation}
\mathcal{S}=\dfrac{1}{2\kappa}\int d^4x\ \sqrt{-g}\ f(R) + \mathcal{S}_m(g_{\mu\nu},\psi)\ ,
\label{f(R):action metrico}
\end{equation}
where $\mathcal{S}_m$ is the action of the matter fields $\psi$. This example of ETG  is called $f(R)$ gravity \cite{Capozziello,Carroll04,Sotiriou10,DeFelice10,Nojiri11} and can be considered a straightforward example of extension or modification of GR.

There exists two variational approaches to derive the field equations of $f(R)$ gravity.
The standard procedure is to derived the field equations  by varying the gravitational  action with respect to the metric $g_{\mu\nu}$, which represents the only dynamical variable of the theory. In this standard approach, called \textit{metric formalism}, one assumes that the connection is symmetric ($\Gamma_{\mu\nu}^\alpha=\Gamma_{\nu\mu}^\alpha$) and metric compatible ($\nabla_\mu g_{\nu\alpha}=0$). This leads to the torsion-less Levi-Civita connection, which is completely determined by the metric components.

In principle, the metric and the connection are two independent quantities: the former governs the causal structure of space-time, while the latter defines the geodesic structure. This is the idea behind the \textit{Palatini formalism} \cite{Palatini19}, in which the action is varied with respect to both metric and connection. In the case of GR, the two formalisms are equivalent: the field equations for the connection gives exactly the Levi-Civita connections of the metric in the Einstein-Hilbert case.
The situation is, however, different for more general action including non-linear terms in $R$ or scalar fields non-minimally coupled to gravity.  In these cases, the two formalisms provide different field equations and different physics \cite{Ferraris94,Magnano94,Capozziello08,Borunda08}.

Finally, there is actually a third variational approach in which the matter action is assumed to be $\mathcal{S}_m(g_{\mu\nu},\Gamma^\lambda_{\mu\nu},\psi)$. This is a full \textit{metric-affine formalism} \cite{Hehl78,Capozziello07,Sotiriou07} and represents the most general case that reduces to metric or Palatini formalisms under certain assumptions.

In the following sections, we will derive the field equations for  $f(R)$ gravity  in the metric and Palatini formalisms. It is possible to show that the two versions of $f(R)$ gravity can be recast as   scalar-tensor theories with specific values of the Brans-Dicke parameter.

\subsection{The metric formalism and its viability conditions in cosmology}

\label{fRmetrico}

Ley us  now derive the field equations of $f(R)$ gravity in the metric formalism.  Varying the action \ref{f(R):action metrico} with respect to the metric $g_{\mu\nu}$, one obtains
\begin{equation}
 f'(R)R_{\mu\nu}-\frac{1}{2}f(R)g_{\mu\nu}- \left(\nabla_\mu\nabla_\nu -g_{\mu\nu}\square\right) f'(R)= \kappa T_{\mu\nu}\ ,
 \label{f(R):FE metrico}
\end{equation}
where
\begin{equation}
T_{\mu\nu}=\dfrac{-2}{\sqrt{-g}} \dfrac{\delta \mathcal S_m }{\delta g^{\mu\nu} }\  .
\end{equation}
Here, the `prime' denotes derivative with respect to $R$. The field equations (\ref{f(R):FE metrico}) are clearly fourth-order partial differential equations in the metric. When $f(R)$ is a linear function of $R$, the last two terms on left-hand side vanish and we recover GR.
Taking the trace of \Cref{f(R):FE metrico} yields
\begin{equation}
 f'(R)R-2f(R) + 3\square f'(R)  =\kappa T,
 \label{f(R):trace metrico}
\end{equation}
where $T=g^{\mu\nu}T_{\mu\nu}$. We note that $R$ and $T$ are related to each other through a differential equation, contrary to the algebraic relation $R=-\kappa T$ of GR. This means that solutions in $f(R)$ constitute a  larger set compared to Einstein's theory. It is useful to rewrite \Cref{f(R):FE metrico} in form of Einstein's equations with a total effective energy-momentum tensor accounting for matter and curvature terms:
\begin{equation}
G_{\mu\nu}=\kappa\left(T_{\mu\nu}^{(m)}+T_{\mu\nu}^{(curv)}\right) ,
\end{equation}
where we have identified
\begin{align}
T_{\mu\nu}^{(m)}&=\dfrac{T_{\mu\nu}}{f'(R)}\ ,\label{f(R):T_munu matter metrico} \\
T_{\mu\nu}^{(curv)}&=\frac{1}{\kappa f'(R)} \left[\frac{f(R)-Rf'(R)}{2} \,  g_{\mu\nu} + \nabla_{\mu}\nabla_{\nu}  f'(R)  -g_{\mu\nu} \square f'(R) \right] . \label{f(R):T_munu curvature metrico}
\end{align}
From \Cref{f(R):T_munu matter metrico}, we immediately find that the effective gravitational constant in $f(R)$ gravity is given as
\begin{equation}
G_{eff}=\dfrac{G}{f'(R)} \ ,
\label{f(R):G_eff}
\end{equation}
which imposes the condition $f'(R)>0$.

The $f(R)$ theories of gravity have been largely invoked in cosmology to explain the current acceleration of the universe without the need of dark energy. To study the cosmological evolution at the background level, we assume the FLRW metric restricting our attention to the flat case ($k=0$), which is favoured by the data \cite{Planck15}. Using the FLRW metric implies the following relation between the Ricci scalar and the Hubble parameter:
\begin{equation}
R=-6\left[\dfrac{\ddot a}{a}+\left(\dfrac{\dot a}{a}\right)^2\right]=-6(\dot{H}+2H^2)\ .
\label{eq:R-H}
\end{equation}
Furthermore, assuming that $T_{\mu\nu}^{(m)}$ is given by \Cref{eq:diag T}, the modified Friedmann equations read
\begin{align}
H^2&=\dfrac{\kappa}{3}\left[\dfrac{\rho_m}{f'(R)}+\rho_{curv}\right] , \label{f(R):first Friedmann metrico} \\
2\dot{H}+3H^2&=-\kappa\left[\dfrac{p_m}{f'(R)} + p_{curv}\right]\ ,\label{f(R):second Friedmann metrico}
\end{align}
where
\begin{align}
\rho_{curv}&=\dfrac{1}{f'(R)}\left[\dfrac{1}{2}\left(f(R)-Rf'(R)\right)-3H\dot{R}f''(R)\right] , \label{f(R):density curv metrico} \\
p_{curv}&=\dfrac{1}{f'(R)}\left[2H\dot{R}f''(R)+\ddot{R}f''(R)+(\dot{R})^2f'''(R)-\dfrac{1}{2}\left(f(R)-Rf'(R)\right)\right] \label{f(R):pressure curv metrico}
\end{align}
are the energy density and pressure of the effective curvature fluid, respectively. Thus, from \Cref{f(R):density curv metrico,f(R):pressure curv metrico} one obtains the effective equation of state
\begin{equation}
w_{DE}\equiv \dfrac{p_{curv}}{\rho_{curv}} =-1+\dfrac{\ddot{R}f''(R)+(\dot{R})^2f'''(R)-H\dot{R}f''(R)}{\left(f(R)-Rf'(R)\right)/2-3H\dot{R}f''(R)}\ ,
\label{f(R):w_DE metrico}
\end{equation}
which is supposed to fuel the effective dark energy fluid associated to the curvature. For $f(R)\propto R-2\Lambda$, $w_{DE}=-1$ as in the cosmological constant scenario.

Modelling matter as dust $(p_m=0)$, the conservation equation for the total energy density can be written as
\begin{equation}
\dot{\rho}_{tot}+3H(\rho_{tot}+p_{curv})=0\ ,
\label{f(R):cons tot energy metrico}
\end{equation}
where $\rho_{tot}=\rho_m/f'(R)+\rho_{curv}$.
Then, assuming no interaction between matter and curvature fluid, the conservation equation for the matter energy density is
\begin{equation}
\dot{\rho}_m+3H\rho_m=0\ ,
\label{f(R):cons matter metrico}
\end{equation}
whose solution gives the standard behaviour
\begin{equation}
\rho_m=\rho_{m0}a^{-3}=3H_0^2\Omega_{m0}(1+z)^3\ .
\label{f(R):sol matter metrico}
\end{equation}
Inserting \Cref{f(R):cons matter metrico} into \Cref{f(R):cons tot energy metrico} and using \Cref{f(R):sol matter metrico}, we obtain the continuity equation for the effective curvature fluid:
\begin{equation}
\dot \rho_{curv}+3H(1+w_{DE})\rho_{curv}=3H_0^2\Omega_{m0}(1+z)^3 \dfrac{\dot R f''(R)}{(f'(R))^2}\ .
\end{equation}
It is finally convenient to combine \Cref{f(R):first Friedmann metrico,f(R):second Friedmann metrico} into a single equation:
\begin{equation}
\dot{H}+\dfrac{1}{2f'(R)}\left[3H_0^2\Omega_{m0}(1+z)^3-H\dot{R}f''(R)+\ddot{R}f''(R)+(\dot{R})^2f'''(R)\right]=0\ ,
\end{equation}
where we have used the definitions (\ref{f(R):density curv metrico}) and (\ref{f(R):pressure curv metrico}).

In the last years, many attempts with the aim to construct quintessence-like $f(R)$ models were proved to produce both early and late-time acceleration \cite{review2}.

It has been shown that the model  $f(R)=R+\alpha R^2$ ($\alpha>0$) is consistent with the temperature anisotropies observed in CMB and it can be a viable alternative to the scalar field models of inflation \cite{Starobinsky80,Wang90}. The quadratic term $\alpha R^2$, in fact, gives rise to an asymptotically exact de Sitter solution, and inflation ends when it becomes subdominant with respect to the linear term $R$. However, this model is not suitable to explain the present cosmic acceleration because the quadratic term is much smaller than $R$ today.

Models of type $f(R)=R-\alpha R^{-n}\ (\alpha>0,n>0)$ were proposed to explain the late-time cosmic acceleration \cite{Capozziello02,Nojiri03,Carroll04}, but it has been shown that they do not satisfy local gravity constraints because
of the instability arising from negative value of $f''(R)$ \cite{Chiba03,Dolgov03,Soussa04}. Moreover, these models do not possess a standard matter-dominated epoch because of a large coupling between matter and dark energy \cite{Amendola07}. However, cosmological viability of $f(R)$ gravity as an ideal fluid and its compatibility with a matter dominated phase has been demonstrated for a large class of models \cite{Troisi}.

We can thus summarize the conditions that $f(R)$ models have to satisfy to  be  viable for dark energy. It has to be:

\begin{enumerate}
\item \MLine{f'(R)>0 \ , \hspace{0.5cm} R\geq R_0>0\ ,}
where $R_0$ is the value of the Ricci scalar at the present time. This condition is required in order to avoid negative values of the effective gravitational constant (cf. \Cref{f(R):G_eff}).
\item  \MLine{f''(R)>0 \ , \hspace{0.5cm} R\geq R_0>0\ .}
This arises from the constraints of gravity in the solar system \cite{Olmo05, Faraoni06}, and the consistency with the presence of a standard matter-dominated epoch \cite{Amendola07c}. Moreover, this condition guaranties the stability of cosmological perturbations.
\item \MLine{f'(R)\longrightarrow 1\ , \hspace{0.5cm} R\gg 1\ .}
This condition is expected to be fulfilled to ensure that viable $f(R)$ model tends to $\Lambda$CDM at large curvatures, as required by CMB observations \cite{Chiba07,Hu07,Appleby07}.
\end{enumerate}
 However, one of the main issues is to reconstruct early and late cosmology through the same approach. In the framework of  $f(R)$ gravity, as firstly reported in  \refcite{Cognola}, it is possible to select a class of realistic models describing inflation and the onset of late accelerated expansion.  Specifically,   power-law $f(R)$ gravity models, describing inflation, can be related to $\Lambda$CDM in a quite natural way \cite{Noj1}. In \refcite{Emilio}, exponential non-singular $f(R)$ models are discussed in order to connect early- and late-time accelerated expansions.

\subsection{The Palatini formalism and viability conditions}

As we have already discussed earlier in this section, the field equations can be derived by applying the variational principle to the metric and the connection, treated as independent variables. In the Palatini formalism, in fact, the curvature tensor is built up from independent connections.  To avoid confusion with the metric formalism, we denote the Ricci tensor constructed by  independent connections as $\mathcal{R}_{\mu\nu}$, and the corresponding  Ricci scalar as $\mathcal{R}=g^{\mu\nu}\mathcal{R}_{\mu\nu}$. The action thus takes the form
\begin{equation}
\mathcal S = \dfrac{1}{2\kappa}\int d^4x\ \sqrt{-g}\ f(\mathcal R) +\mathcal S_m(g_{\mu\nu}, \psi)\ ,
\label{f(R):action Palatini}
\end{equation}
which reduces to GR when $f(\mathcal R)=\mathcal R$. The variation of \Cref{f(R):action Palatini} with respect to the metric  provides
\begin{equation}
F({\mathcal R}) {\mathcal R}_{(\mu\nu)}-\frac{1}{2}f({\mathcal R})g_{\mu\nu}=\kappa T_{\mu\nu}\ ,
\label{f(R):variation metric Palatini}
\end{equation}
where $F(\mathcal{R})\equiv df/d\mathcal R$ and $(\mu\nu)$ denotes symmetrization over the indices $\mu$ and $\nu$.
Taking the trace of \Cref{f(R):variation metric Palatini} yields the following useful relation:
\begin{equation}
F(\mathcal{R})\mathcal{R}-2f(\mathcal{R})=\kappa T\ .
\label{f(R):trace FE Palatini}
\end{equation}
On the other hand, the variation with respect to the connection gives
\begin{equation}
\delta \mathcal{R}_{\mu\nu}=\bar{\nabla}_\lambda\delta \Gamma^\lambda_{\mu\nu}-\bar{\nabla}_\nu\delta \Gamma^\lambda_{\mu\lambda}\ ,
\end{equation}
where $\bar{\nabla}_\mu$ indicates the covariant derivative defined with respect to the independent connection $\Gamma^\lambda_{\mu\nu}$.
Therefore, variation of (\ref{f(R):action Palatini}) with respect to the connection yields
\begin{equation}
\bar \nabla_\lambda\left(\sqrt{-g}\ F({\cal R} )g^{\mu\nu}\right)-\bar \nabla_\sigma\left(\sqrt{-g}\ F({\cal R}) g^{\sigma(\mu}\right)\delta^{\nu)}_\lambda=0\ .
\end{equation}
Contracting the above equation over $\lambda$ and $\mu$ results in \cite{Vollick03}
\begin{equation}
\bar \nabla_\sigma\left(\sqrt{-g}\ F({\cal R})g^{\sigma\mu}\right)=0\ .
\label{f(R):conn new metric Palatini}
\end{equation}
This naturally leads to introduce a new metric conformally related to $g_{\mu\nu}$ being
\begin{equation}
\sqrt{-g}\ F(\mathcal{R})g^{\mu\nu}=\sqrt{-h}\ h^{\mu\nu}\ ,
\end{equation}
which implies
\begin{equation}
h_{\mu\nu}= F(\mathcal{R})g_{\mu\nu}\ .
\end{equation}
Thus, \Cref{f(R):conn new metric Palatini} becomes the definition of the Levi-Civita connection of the metric $h_{\mu\nu}$:
\begin{equation}
\Gamma^\lambda_{\mu\nu}= \frac{1}{2}
h^{\lambda\sigma}\left(\partial_\mu h_{\nu\sigma} +\partial_\nu
h_{\mu\sigma}-\partial_\sigma h_{\mu\nu}\right)\ .
\label{f(R):connection Palatini}
\end{equation}
The independent connection (\ref{f(R):connection Palatini}) can be written is terms of the metric $g_{\mu\nu}$ as
\begin{equation}
\Gamma_{\mu\nu}^\lambda=\{^\lambda_{\mu\nu}\}+\dfrac{1}{2F}\left[2\delta^\lambda_{(\mu}\partial_{\nu)}F-g_{\mu\nu}g^{\lambda\sigma}\partial_\sigma F\right] ,
\end{equation}
where $\{^\lambda_{\mu\nu}\}$ are the Christoffel symbols of the metric $g_{\mu\nu}$.
Considering how the  Ricci tensor transforms under conformal transformations, we can write
\begin{equation}
\mathcal R_{\mu\nu}=R_{\mu\nu}+\dfrac{3}{2}\left[\dfrac{(\nabla_\mu F)(\nabla_\nu F)}{F^2}\right]-\dfrac{1}{F}\left(\nabla_\mu\nabla_\nu+\dfrac{1}{2}g_{\mu\nu}\square\right)F \ ,
\label{f(R):Ricci tensor Palatini}
\end{equation}
and contracting with $g_{\mu\nu}$, one obtains
\begin{equation}
\mathcal{R}=R+\dfrac{3}{2}\left[\dfrac{\left(\nabla_\mu F\right)(\nabla^\mu F)}{F^2}\right]-\dfrac{3}{F}\square F\ .
\label{f(R):Ricci scalar Palatini}
\end{equation}
Note that, when $f(\mathcal R)=\mathcal{R}$, $F$ is constant and the theory reduces to GR as $\mathcal R_{\mu\nu}=R_{\mu\nu}$ and $\mathcal{R}=R$.
Finally, substituting \Cref{f(R):Ricci tensor Palatini,f(R):Ricci scalar Palatini} into \Cref{f(R):variation metric Palatini} leads to
\begin{align}
G_{\mu\nu}&=\dfrac{\kappa}{F}T_{\mu\nu}-\dfrac{1}{2}g_{\mu\nu}\left(\mathcal{R}-\dfrac{f}{F}\right)-\dfrac{3}{2F^2}\left[(\nabla_\mu F)(\nabla_\nu F)-\dfrac{1}{2}g_{\mu\nu}(\nabla_\mu F)(\nabla^\mu F)\right] \nonumber \\
&+\dfrac{1}{F}(\nabla_\mu\nabla_\nu-g_{\mu\nu}\square)F\ .
\label{f(R):FE Palatini}
\end{align}
Cosmic dynamics can be studied  assuming that the universe is described by the flat FLRW metric and it is filled with a perfect fluid with an energy-momentum tensor given by \Cref{eq:diag T}. Thus,
combining the modified Friedmann equations calculated as the $(0,0)$ and $(i,j)$ components of \Cref{f(R):FE Palatini}, one obtains \cite{Meng03}
\begin{equation}
\left(H+\dfrac{1}{2}\dfrac{\dot{F}}{F}\right)^2=\dfrac{1}{6}\left[\dfrac{\kappa(\rho+3p)}{F}+\dfrac{f}{F}\right] .
\label{f(R):gen Friedmann Palatini}
\end{equation}
Assuming that matter is dust and neglecting the contribution of radiation, we have $p=0$ and $\rho=\rho_m$. Then, the time derivative of \Cref{f(R):trace FE Palatini} reads
\begin{equation}
\dot{\mathcal R}(\mathcal RF_\mathcal{R}-2f)=\kappa\dot{\rho}_m\ ,
\label{f(R):dot trace FE Palatini}
\end{equation}
where we have used that $\dot{F}=F_\mathcal{R}\dot{\mathcal{R}}$, being $F_\mathcal{R}\equiv dF/d\mathcal{R}=d^2f/d\mathcal R^2$. Making use of the continuity equation $\dot{\rho}_m+3H\rho_m=0$ and again of \Cref{f(R):trace FE Palatini}, from \Cref{f(R):dot trace FE Palatini} one gets
\begin{equation}
\dot{\mathcal{R}}=-\dfrac{3H(\mathcal{R}F-2f)}{\mathcal{R}F_\mathcal{R}-2F}\ .
\end{equation}
Thus, substituting the above expression for $\dot{\mathcal{R}}$ into \Cref{f(R):gen Friedmann Palatini}, we finally obtain
\begin{equation}
H^2=\dfrac{1}{6F}\dfrac{2\kappa\rho_m+\mathcal{R}F-f}{\left[1-\dfrac{3}{2}\dfrac{F_\mathcal{R}(\mathcal{R}F-2f)}{F(\mathcal{R}F_\mathcal{R}-F)}\right]^2}\ ,
\label{f(R):gen Friedmann Palatini 2}
\end{equation}
where $\rho_m=3H_0^2\Omega_{m0}(1+z)^3$.

In the Palatini formalism, the field equations are second-order and are then free from the instabilities due to negative values of $f''(R)$ \cite{Meng1,Meng2}.
Several works addressing the dynamics of Palatini $f(R)$ gravity at  background level showed that the correct sequence of cosmological eras is realized even for the model $f(R)=R-\alpha R^{-n}$ with $n>0$ \cite{Fay07,AmarzguiouiPalatini}. Dark energy models from Palatini $f(R)$ gravity are not compatible with large-scale structure observations for substantial deviations from the $\Lambda$CDM model, because of a large coupling between non-relativistic matter and dark energy \cite{Koivisto06,LiPalatini,TsujikawaPalatini}. Also, the non-perturbative corrections to the matter action introduced by such a large coupling  appear in conflict with the Standard Model of particle physics \cite{Flanagan04}.

Moreover, while in metric $f(R)$ gravity the Cauchy problem is well-posed both in vacuo and with matter, in Palatini $f(R)$ gravity the Cauchy problem is unlikely to be well-formulated, unless for null derivatives of the trace of the energy-momentum tensor. This is due to the presence of higher derivatives of matter fields in the field equations \cite{Sotiriou10}. In any case, the well-position and the well-formulation of the Cauchy problem in Palatini $f(R)$ gravity can be correctly addressed considering specific forms of sourcing fluids \cite{vignolo-cauchy}.

\subsection{Equivalence between $f(R)$ gravity and scalar-tensor theories}

Similarly to classical mechanics where one can redefine variables in order to make equations easier to handle, in field theory, it is also possible to redefine fields and rewrite  action and field equations in a different form.
Theories that, under a suitable transformation of fields, preserve  action and equations of motion are said dynamically equivalent.
Such theories give  the same results and can be seen as different representations of the same theory. In this section, we show the equivalence between  $f(R)$ gravity and scalar-tensor theories of gravity.

A general scalar-tensor theory of gravity is described by the action
\begin{equation}
\mathcal{S}_{ST}=\dfrac{1}{2\kappa}\int d^4 x \sqrt{-g} \left[\phi R-\dfrac{\omega(\phi)}{\phi}\ \partial_\mu \phi\ \partial^\mu \phi-V(\phi)\right] +\mathcal S_m(g_{\mu\nu},\psi)\ ,
\label{f(R):action ST}
\end{equation}
where $V$ is the potential of the scalar field $\phi$ and $\omega(\phi)$ is some arbitrary function of $\phi$.
Varying action \eqref{f(R):action ST} with respect to the metric
provides
\begin{align}
G_{\mu\nu}=\dfrac{\kappa}{\phi}T_{\mu\nu} -\dfrac{V(\phi)}{2\phi} g_{\mu\nu} +\dfrac{1}{\phi}\left(\nabla_{\mu}\!\nabla_{\nu}\phi -g_{\mu\nu} \square\phi \right)+\dfrac{\omega(\phi)}{\phi^2}\left(\nabla_\mu\phi \nabla_\nu\phi-\dfrac{1}{2}g_{\mu\nu}\nabla^\alpha\phi\nabla_\alpha\phi\right) ,
\label{f(R):var ST wrt metric}
\end{align}
while variation with respect to the scalar field yields
\begin{align}
\square\phi=\dfrac{\phi}{2\omega(\phi)}\left[V'(\phi)-R\right]+\dfrac{1}{2}\left[\dfrac{1}{\phi}-\dfrac{\omega'(\phi)}{\omega(\phi)}\right]\nabla^\mu\phi\ \nabla_\mu\phi
\label{f(R):var ST wrt scalar}
\end{align}
The trace of \Cref{f(R):var ST wrt metric} can be used to replace $R$ in \Cref{f(R):var ST wrt scalar} obtaining thus
\begin{equation}
\left[2\omega(\phi)+3\right]\square\phi=\kappa\ T+\phi\ V'(\phi)-\omega'(\phi)\ \nabla^\alpha\phi\ \nabla_\alpha\phi-2V(\phi)\ .
\label{f(R):trace ST}
\end{equation}
From the general action \Cref{f(R):action ST} one can retrieve a Brans-Dicke-like  theory \cite{Brans61} with a scalar-field potential by setting $\omega(\phi)=\omega_{BD}$:
\begin{equation}
\mathcal{S}_{BD}=\dfrac{1}{2\kappa}\int d^4 x\ \sqrt{-g} \left[\phi R-\dfrac{\omega_{BD}}{\phi}\ \partial_\mu\phi\ \partial^\mu\phi-V(\phi)\right] +\mathcal S_m(g_{\mu\nu},\psi)\ ,
\label{Brans-Dicke action Jordan}
\end{equation}
where the Brans-Dicke parameter $\omega_{BD}$ plays the role of a coupling constant.

The equivalence between $f(R)$ gravity, in the metric formalism, and scalar-tensor theories can be achieved as follows.
 We can introduce a new scalar field $\chi$ and consider the following action \cite{Chiba03}:
\begin{equation}
\mathcal{S}_{met}=\frac{1}{ 2\kappa }\int\ d^4 x \sqrt{-g}
\left[ f(\chi)+f'(\chi)(R-\chi)\right]+ \mathcal{S}_m(g_{\mu\nu},\psi)\ .
\label{f(R):equiv action metrico}
\end{equation}
Varying with respect to $\chi$ yields
\begin{equation}
f''( \chi )(R-\chi)=0\ ,
\end{equation}
which implies that $\chi=R$ if $f''(\chi)\neq 0$. This reproduces action (\ref{f(R):action metrico}) and proves that the theory is dynamically equivalent to the original. Then, one can redefine the field $\chi$ by setting
\begin{equation}
\begin{aligned}
\phi&=f'(\chi)\ , \\
V(\phi)&=\phi\chi(\phi)-f(\chi(\phi))\ .
\end{aligned}
\label{f(R):redef field}
\end{equation}
Hence, (\ref{f(R):equiv action metrico}) takes the form
\begin{equation}
\mathcal{S}_{met}=\dfrac{1}{2\kappa}\int d^4 x \sqrt{-g} \left[\phi R-V(\phi)\right] +\mathcal S_m(g_{\mu\nu},\psi)\ ,
\label{f(R):action omega_BD=0}
\end{equation}
which is equivalent to a  Brans-Dicke-like theory with $\omega_{BD} = 0$ \cite{Wands94}.
In such a case, field equations (\ref{f(R):var ST wrt metric}) read
\begin{equation}
G_{\mu\nu}=\dfrac{\kappa}{\phi}T_{\mu\nu} -\dfrac{1}{2\phi} g_{\mu\nu} V(\phi)+\dfrac{1}{\phi}\left(\nabla_{\mu}\!\nabla_{\nu}\phi -g_{\mu\nu} \square\phi \right) ,
\end{equation}
and \Cref{f(R):trace ST} becomes
\begin{equation}
3\square\phi +2V(\phi) -\phi\ V'(\phi)=\kappa\ T\ .
\end{equation}
Furthermore, as usual in scalar-tensor theories, one can perform a conformal transformation and move from the \textit{Jordan frame} to the \textit{Einstein frame}. In fact, through the conformal transformation
\begin{equation}
\tilde g_{\mu\nu}=f'(R)\ g_{\mu\nu}\equiv \phi\ g_{\mu\nu}\ ,
\label{f(R):conf transf}
\end{equation}
and the redefinition of field
\begin{equation}
\tilde{\phi}=\sqrt{\dfrac{2\omega_{BD}+3}{2\kappa}}\ln\left(\dfrac{\phi}{\phi_0}\right) ,
\end{equation}
we obtain the Einstein frame, in which the new field $\tilde{\phi}$ has a kinetic energy and it is minimally coupled to gravity:
\begin{equation}
\mathcal{S}_{BD}^{(Ein)}=\int d^4x \, \sqrt{-\tilde{g}} \, \left[\frac{\tilde{R}}{2\kappa}-\frac{1}{2}\ \partial^{\mu}\tilde{\phi}\ \partial_{\mu}\tilde{\phi} -U(\tilde{\phi} )\right] .
\end{equation}
For $\omega_{BD}=0$, corresponding to $f(R)$ gravity in the metric formalism, one has
\begin{align}
&\phi \equiv f'(R)=e^{\sqrt{ \frac{2\kappa}{3}} \tilde{\phi}} \ , \\
&U( \tilde{\phi} )=\frac{R f'(R)-f(R)}{2\kappa \left( f'(R) \right)^2}\ ,
\end{align}
and the action reads
\begin{equation}
\mathcal{S}^{(Ein)}_{met}=\int  d^4x \, \sqrt{-\tilde{g}} \,
\left[\frac{\tilde{R}}{2\kappa }-\frac{1}{2} \,
\partial^\mu\tilde{\phi}\ \partial_\mu\tilde{\phi} -U( \tilde{\phi} )
\right]+S_m(e^{-\sqrt{\frac{2\kappa}{3}}\,\tilde{\phi}}\
\tilde{g}_{\mu\nu},\psi)\ .
\label{action metrico Einstein}
\end{equation}
We want to stress that actions (\ref{f(R):action metrico}), (\ref{f(R):action omega_BD=0}) and (\ref{action metrico Einstein}) are equivalent representations of the same theory.
However, the issue on which conformal frame (Jordan or Einstein) is the `physical' one has been the subject of much debate and the answers are still controversial. A detailed discussion on this can be found in \refcite{Faraoni07} and the references therein.

Let us now examine the equivalence between the Palatini formulation of $f(R)$ gravity and a scalar-tensor theory. Adopting a similar procedure to the one presented above, we consider the action for the field $\chi$:
\begin{equation}
\mathcal{S}_{Pal}=\frac{1}{2\kappa}\int d^4 x \sqrt{-g} \left[f(\chi)+f'(\chi)(\mathcal{R}-\chi)\right] +\mathcal{S}_m(g_{\mu\nu}, \psi)\ .
\label{f(R): equiv action Palatini}
\end{equation}
A variation with respect to $\chi$ yields $\chi=\mathcal{R}$. Then, redefining the field as in \Cref{f(R):redef field}, the action takes the form
\begin{equation}
\mathcal{S}_{Pal}=\dfrac{1}{2\kappa}\int d^4 x \sqrt{-g} \left[\phi \mathcal R-V(\phi)\right] +\mathcal S_m(g_{\mu\nu},\psi)\ .
\label{f(R):action omega_BD=-3/2}
\end{equation}
and \Cref{f(R):Ricci scalar Palatini} in terms of the new field $\phi$ reads
\begin{equation}
\mathcal{R}=R+\dfrac{3}{2\phi^2}(\nabla_\mu\phi)(\nabla^\mu\phi)+\dfrac{3}{\phi}\square\phi\ .
\end{equation}
Therefore, plugging the above relation into \Cref{f(R):action omega_BD=-3/2} and neglecting a total divergence, one obtains
\begin{equation}
\mathcal{S}_{Pal}=\frac{1}{2\kappa}\int d^4 x \sqrt{-g} \left(\phi
 R+\frac{3}{2\phi}\ \partial_\mu \phi\ \partial^\mu \phi-V(\phi)\right)+S_m(g_{\mu\nu}, \psi).
\label{action BD=-3/2}
\end{equation}
Comparing this with (\ref{Brans-Dicke action Jordan}) we deduce that $f(R)$ gravity in the Palatini formalism is equivalent to a Brans-Dicke theory with $\omega_{BD}=-3/2$. Thus, the field equations that one obtains from varying the action with respect to the metric and the scalar field are, respectively,
\begin{align}
G_{\mu\nu}&=\dfrac{\kappa}{\phi} T_{\mu\nu} -\dfrac{3}{2\phi^2}
\left(\nabla_\mu \phi \nabla_\nu \phi -\dfrac{1}{2}g_{\mu\nu} \nabla^\lambda \phi \nabla_\lambda \phi \right)+\dfrac{1}{\phi}   (\nabla_\mu\nabla_\nu \phi-g_{\mu\nu} \square \phi)-\dfrac{V(\phi)}{2\phi}g_{\mu\nu}\ , \label{f(R):var BD=-3/2 metric}\\
\square\phi& = \dfrac{\phi}{3}(R-V'(\phi))+\dfrac{1}{2\phi}\ \nabla^\mu\phi\
\nabla_\mu\phi\ .
\label{f(R):var BD=-3/2 scalar}
\end{align}
Moreover, \Cref{f(R):trace ST} takes the simpler form
\begin{equation}
2V(\phi)-\phi\ V'(\phi)=\kappa\ T\ .
\end{equation}
Finally, performing conformal transformation (\ref{f(R):conf transf}) one can rewrite action \ref{action BD=-3/2} in the Eistein frame:
\begin{equation}
\mathcal{S}^{(Ein)}_{Pal}=\int d^4x \, \sqrt{-\tilde{g}}  \, \left[\frac{\tilde{R}}{2\kappa}-U(
\phi ) \right] +S_m(\phi^{-1}\tilde{g}_{\mu\nu},\psi)\ ,
\end{equation}
where
\begin{equation}
U(\phi)=\dfrac{1}{2\kappa}\dfrac{V(\phi)}{\phi^2}\ .
\end{equation}
An important issue has to be stressed at this point. The equivalence between $f(R)$ and scalar-tensor  gravity   can be lost, even mathematically,  in the presence of singularities. As discussed in \refcite{Briscese}, big rip singularities can emerge in these models related to phantom scalar fields.  Furthermore, it is possible to demonstrate that 
 $f(R)$ gravity singularities in Jordan and Einstein frames correspond \cite{Baha1}.
Finally, even if equivalence is fulfilled, the physical interpretation
may be different in the two frames \cite{Troisi,Baha2}.

Recently, it has been considered the possibility to combine metric and Palatini formalism considering a theory of gravity as
\begin{equation}
        {\cal F}=R+f({\cal R})\,,
\end{equation}
where $R$ is formulated in metric formalism and $f({\cal R})$, that is the extra terms with respect to GR, are formulated in Palatini formalism \cite{hybrid1,hybrid2,hybrid3}. This combined approach (the so called Hybrid Gravity) allows to bypass  some shortcomings of both the scenarios \cite{hybrid4}.

\section{Gravity with torsion}
\label{sec:torsion}

The issue about the symmetry of the space-time connection has led to consider the role of torsion in the description of the gravitational interaction.
Quantum effects are not taken into account in a classical theory as GR.
However, those effects cannot be neglected one deals with any theory involving gravity at a fundamental level.
A straightforward generalization including in GR matter spin fields is obtained when one considers a four-dimensional space-time manifold with torsion. In such a picture, mass-energy and spin are, respectively, the sources of curvature and torsion.
A relevant example towards this direction is represented by the Einstein-Cartan-Sciama-Kibble (ECSK) theory \cite{Hehl76}.
Also, higher dimensional paradigms such as Kaluza-Klein theories \cite{Green87,Hammond94,deSabbata91} take into account torsion in unification schemes with gravity and electromagnetism.
Moreover, torsion must be included in any gravity theory with the presence of twistors \cite{deSabbata96,Howe98}, and in supergravity where curvature is considered together with torsion and matter fields \cite{Hull93,Papadopoulos95}.

Besides, several authors take seriously into account the role played by torsion in the early universe with the observational consequences at the present time. The repulsive contributions of torsion to the energy-momentum tensor yield cosmological models which are free from singularities. \cite{Gonner84,Chatterjee93,Wolf95}.
Topological defects originated from torsion, in a universe characterized by phase transitions, \cite{Ross89,deAndrade97,Vignolo15} are reflected today into the angular momenta of cosmic structures.
Furthermore, the energy-momentum contribution of torsion influences the cosmological perturbations giving rise to characteristic lengths in the spectrum  \cite{Capozziello98b}.
The presence of torsion also modifies the evolution equations of shear, expansion and other kinematic quantities \cite{Hawking73}.

To describe the dynamics of space-time with torsion, it is possible to  introduce  tetrad fields $e_{A}^{\mu}$. Denoting by $A,B,C\dots$ the coordinates of the tangent space-time, the tetrad fields are dynamical variables which form an orthonormal basis at each point $x^\mu$ of the manifold \cite{Capozziello01}.
One thus defines the co-tetrad field $e_{\mu}^{A}$ with the following properties:
\begin{align}
\label{delta1}
  e_{A}^{\mu}e_{\nu}^{A} &= \delta^{\mu}_{\nu}\,, \\
\label{delta2}
 e_{A}^{\mu}e_{\mu}^{B} &= \delta^{B}_{A}\,.
\end{align}
The metric of tetrad fields is
\begin{equation}\label{metrad}
 \eta_{AB}=\eta^{AB}= \text{diag}(1,-1,-1,-1),
\end{equation}
from which one can construct the metric tensor as
\begin{equation}\label{metrica}
 g_{\mu\nu} =\eta_{AB}e_{\mu}^{A}e_{\nu}^{B}.
\end{equation}
We thus consider simple bivectors which are obtained by skew-symmetric tensor product of two vectors.
A bivector $B^{\mu\nu}$ is simple if it satisfies the condition
\begin{equation}
\label{simbiv}
  B^{[\mu\nu}B^{\rho]\sigma}=0.
\end{equation}
It is possible to construct the $N(N-1)/2$ simple bivectors through tetrad vectors in a $N$-dimensional manifold as
 \begin{equation}\label{simple}
 F_{AB}^{\mu\nu}= e_{A}^{[\mu}e_{B}^{\nu]},
\end{equation}
while any bivector $B^{\mu\nu}$ takes the form
\begin{equation}
\label{biv} B^{\mu\nu} = B^{AB}e_{A}^{\mu}e_{B}^{\nu},
\end{equation}
being $B^{AB}=-B^{BA}$.

From the antisymmetric part of the affine connection, we define the torsion tensor $T_{\mu\nu}^{\phantom{\mu\nu}\rho}$ as
\begin{eqnarray}
\label{t1}
 T_{\mu\nu}^{\phantom{\mu\nu}\rho} = \frac{1}{2} \left( \Gamma_{\mu\nu}^{\rho}
-\Gamma_{\nu\mu}^{\rho} \right)
\equiv\Gamma_{
[\mu\nu]}^{\rho}\,.
\end{eqnarray}
We note that $T_{\mu\nu}^{\phantom{\mu\nu}\rho}=0$ is postulated in GR.
It is often useful to define the {\it contorsion tensor} as
\begin{eqnarray}
\label{t3}
 K_{\mu\nu}^{\phantom{\mu\nu}\rho} = -T_{\mu\nu}^{\phantom{\mu\nu}\rho}
-T^{\rho}_{\phantom{\rho}\mu\nu}
+T^{\phantom{\mu}\rho}_{\nu\phantom{\rho}{\mu}} =
-K^{\phantom{\mu}\rho}_{\mu\phantom{\rho}\nu} \,,
\end{eqnarray}
and the {\it modified torsion tensor} as
\begin{eqnarray}
\label{t4}
 \hat{T}_{\mu\nu}^{\phantom{\mu\nu}\rho} = T_{\mu\nu}^{\phantom{\mu\nu}\rho}
+2\delta_{[\mu}^{\ \ \rho} T_{\nu]} ~,
\end{eqnarray}
where $T_\mu\equiv T_{\mu\nu}^{\phantom{\mu\nu}\nu}$.
From the above definition, we can write the affine connection as
\begin{eqnarray} \label{t5}
 \Gamma_{\mu\nu}^{\rho} = \left\{^{\rho}_{\mu\nu} \right\}
-K_{\mu\nu}^{\phantom{\mu\nu}\rho}\,,
\end{eqnarray}
where $\left\{^{\rho}_{\mu\nu}\right\}$ are the Christoffel symbols of the
symmetric Levi-Civita connection defined in GR.
Since torsion is present in the affine connection, the covariant derivatives of a scalar field $\phi$ do not commute:
\begin{equation}\label{scalarder}
 \tilde
\nabla_{[\mu}\tilde\nabla_{\nu]}\phi=-T_{\mu\nu}^{\phantom{\mu\nu}\rho}\tilde\nabla_{\rho}
\phi ~,
\end{equation}
while one has the following relations for a vector $v^\mu$ and a covector $w_\mu$:
\begin{align}
\label{doppiader1}
 (\tilde\nabla_{\mu}\tilde\nabla_{\nu} -\tilde\nabla_{\nu}\tilde\nabla_{\mu}) v^{\rho} &=
R_{\mu\nu\sigma}^{\phantom{
abd}\rho}v^{\sigma}
 -2T_{\mu\nu}^{\phantom{\mu\nu}\sigma}\tilde\nabla_{\sigma}v^\rho ~, \\
\label{doppiader2}
 (\tilde\nabla_{\mu}\tilde\nabla_{\nu} -\tilde\nabla_{\nu}\tilde\nabla_{\mu}) w_{\sigma}
&=
R_{\mu\nu\rho}^{\phantom{
\mu\nu\rho}\sigma}w_{\sigma}
 -2T_{\mu\nu}^{\phantom{\mu\nu}\sigma}\tilde\nabla_{\sigma}w_{\rho} ~,
\end{align}
where the Riemann tensor is given as
\begin{equation}
\label{Riemann}
 R_{\mu\nu\rho}^{\phantom{\mu\nu\rho}\sigma} =\partial_{\mu}\Gamma_{\nu\rho}^{\sigma}
-\partial_{\nu}\Gamma_{\mu\rho}^{\sigma}
+\Gamma_{\mu \lambda}^{\sigma}
\Gamma_{\nu\rho}^{\lambda} -\Gamma_{\nu\lambda}^{\sigma}\Gamma_{\mu\rho}^{\lambda} ~.
\end{equation}
Also, torsion contributes to the Riemann tensor as
\begin{equation}
\label{riexpanded}
 R_{\mu\nu\rho}^{\phantom{\mu\nu\sigma}\sigma} =
R_{\mu\nu\rho}^{\phantom{\mu\nu\sigma}\sigma}(\{\})
-\nabla_{\mu}K_{\nu\rho}^{\phantom{\nu]\rho}\sigma}
+\nabla_{\nu}K_{\mu\rho}^{\phantom{\mu\rho}\sigma}
+ K_{\mu \lambda}^{\phantom{\mu\lambda}\sigma}K_{\nu\rho}^{\phantom{\nu\rho}\lambda}
-K_{\nu \lambda}^{\phantom{\nu \lambda}\sigma}K_{\mu\rho}^{\phantom{\mu\rho}\lambda} ~,
\end{equation}
where $R_{\mu\nu\rho}^{\phantom{\mu\nu\rho}\sigma}(\{\})$ is the tensor of the
symmetric connection. Here, $\tilde\nabla$ and $\nabla$ denote the
covariant derivative with and without torsion, respectively.
Thus, the Ricci tensor reads
\begin{equation}
\label{ricci}
 \!\!\!
 R_{\mu\nu} = R_{\mu\nu}(\{\}) -2\nabla_{\mu}T_{\rho}
+\nabla_{\nu}K_{\mu\rho}^{\phantom{\mu\rho}\nu}
+K_{\mu \lambda}^{\phantom{\mu \lambda}\nu}K_
{\nu\rho}^{\phantom{\nu\rho}\lambda} -2T_\lambda K_{\mu\rho}^{\phantom{\mu\rho}\lambda}
\end{equation}
and the Ricci scalar is given by
\begin{equation}
\label{curvscalar}
 R =R(\{\}) -4\nabla_{\mu}T^{\mu} +K_{\rho \lambda\nu}K^{\nu \rho\lambda} -4T_\mu T^\mu ~.
\end{equation}

\begin{figure}[!h]
\begin{center}
\includegraphics[scale=0.5]{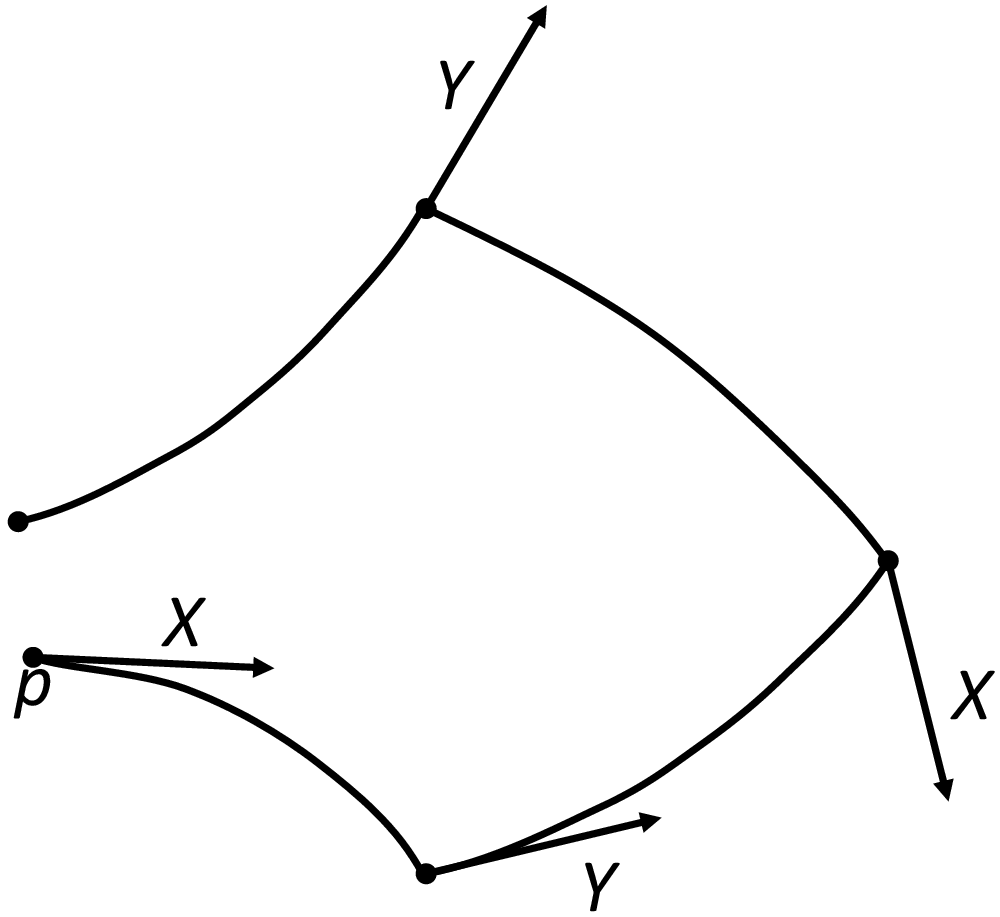}
\caption{Representation of parallelograms determined by torsion.}
\label{fig:parallelogramma}
\end{center}
\end{figure}

To understand the geometrical meaning of torsion, we can think in terms of parallelograms breaking \cite{Trautman73}.
In a curved space, an infinitesimal parallelogram is formed when two parts of geodesics are displaced one along the other. Thus, curvature determines the difference obtained by the parallel transport of a field across both paths.
However, the same procedure applied in a twisted space produces a gap between the extremities of the two geodesics, \emph{i.e.} breaking the infinitesimal parallelogram \Cref{fig:parallelogramma}.

\subsection{Teleparallel Equivalent of General Relativity}

Bearing in mind the considerations made above, let us here summarize the so called  Teleparallel Equivalent of General Relativity (TEGR)  as an alternative approach to describe the gravitational interaction (see \cite{Cai15} for a review).
This scenario was first studied by Einstein himself as an equivalent alternative to GR and it represents a gauge theory for the translation group \cite{Moller61,Maluf94}.
Within this approach, tetrad fields are used to define the free-curvature Weitzenb\"ock connection \cite{Weitzenbock23}.
It is worth mentioning that curvature and torsion are properties of the connection and, within the same space-time, it is possible to define several different connections \cite{Kobayashi63}.
While the Riemmanian structure is related to the Levi-Civita connection, the teleparallel structure is related to Weitzenb\"ock
connection. These geometrical structures are linked to the gravitational interaction due to its universality.

Although gravity can be equivalently described in terms of curvature and torsion, conceptual differences occur.
In teleparallel theories, torsion accounts for the gravitational interaction acting like a {\it force}, rather than providing a geometric picture of space-time as in GR.
In the teleparallel version of GR, in fact, the geodesic equation can be seen as the Lorentz force law of electrodynamics.

To describe the teleparallel equivalent of GR, we adopt the notation in which the Greek indices $(\mu, \nu, \rho, \dots = 0,1,2,3)$ are
related to space-time, and the capital Latin indices $(A,B,C,
\dots = 0,1,2,3)$ denote the tangent space. We assume that the tangent space is Minkoskian with metric
\begin{equation}
\eta_{AB}=\mbox{diag}(+1,-1,-1,-1)\ .
\end{equation}
Introducing the translation generators $P_A = \partial /\partial x^A$, one can define a local translation on the
tangent-space as follows:
\begin{equation}
\delta x^{A} = \delta\alpha^BP_Bx^{A} \; ,
\end{equation}
Then, for a given matter field $\Psi$,  we define its gauge covariant derivative as
\begin{equation}
{\mathcal D}_\mu \Psi = e^B{}_{\mu} \; \partial_B \Psi \; ,
\end{equation}
where
\begin{equation}
e^B{}_{\mu} = \partial_{\mu}x^B + A^B{}_{\mu}\ ,
\label{2.17}
\end{equation}
with $A^B{}_{\mu}$ being the gauge potentials.
As in the standard Abelian gauge theories, the field strength reads
\begin{equation}
F^B{}_{\mu \nu} = \partial_{\mu} A^B{}_{\nu}
- \partial_{\nu}A^B{}_{\mu} \; ,
\label{core}
\end{equation}
satisfying the following relation:
\begin{equation}
[{\mathcal D}_{\mu}, {\mathcal D}_{\nu}] \Psi =
  F^B{}_{\mu \nu} P_B \Psi \; .
\end{equation}
The teleparallel structure on space-time is induced by nontrivial tetrad fields, which allow to define the \textit{Weitzenb\"ock connection}:
\begin{equation}
\hat{\Gamma}^{\rho}{}_{\mu \nu} = e_A{}^{\rho}\partial_{\nu}e^A{}_{\mu} \;  ,
\label{carco}
\end{equation}
characterized by torsion with no curvature.
As a consequence, the Weitzenb\"ock covariant derivative of the tetrad field is identically zero:
\begin{equation}
\nabla_{\nu}e^A{}_{\mu} \equiv \partial_{\nu}e^A{}_{\mu} -
\hat{\Gamma}^{\rho}{}_{\mu \nu} \, e^A{}_{\rho} = 0 \; .
\label{cacd}
\end{equation}
The above condition is known as absolute parallelism.
Moreover, one can write the expression for the torsion related  the Weitzenb\"ock connection as
\begin{equation}
T^{\rho}{}_{\mu \nu} = \hat{\Gamma}^{\rho}{}_{\nu \mu} -
\hat{\Gamma}^{\rho}{}_{\mu \nu} \; ,
\label{tor}
\end{equation}
and the corresponding gravitational "force" is
\begin{equation}
F^A{}_{\mu \nu} =  e^A{}_{\rho}T^{\rho}{}_{\mu \nu} \; .
\end{equation}
One can use a nontrivial tetrad field to define also the torsionless Levi-Civita connection of the space-time metric:
\begin{equation}
{\stackrel{\circ}{\Gamma}}{}^{\sigma}{}_{\mu \nu} = \frac{1}{2}
g^{\sigma \rho} \left[ \partial_{\mu} g_{\rho \nu} + \partial_{\nu}
g_{\rho \mu} - \partial_{\rho} g_{\mu \nu} \right] \; .
\label{lci}
\end{equation}
The relation between the Weitzenb\"ock and the Levi-Civita connections is
\begin{equation}
\hat{\Gamma}^{\rho}{}_{\mu \nu} =
{\stackrel{\circ}{\Gamma}}{}^{\rho} {}_{\mu \nu} +
K^{\rho}{}_{\mu \nu} \; ,
\label{rela}
\end{equation}
where $K^{\rho}{}_{\mu \nu}$
\begin{equation}
K^{\rho}{}_{\mu \nu} = {\textstyle \frac{1}{2}} \left( T_{\mu}{}^{\rho}{}_{\nu}
+ T_{\nu}{}^{\rho}{}_{\mu} - T^{\rho}{}_{\mu \nu} \right)
\label{contorsiontensor}
\end{equation}
is the contorsion tensor.
From the identity
\begin{equation}
{R}^{\rho}{}_{\lambda \mu \nu} = \partial_\mu
\hat{\Gamma}^{\rho}{}_{\lambda \nu} + \hat{\Gamma}^{\rho}{}_{\sigma
\mu} \; \hat{\Gamma}^{\sigma}{}_{\lambda \nu} - (\mu  \leftrightarrow \nu)
\equiv 0 \ ,
\label{r}
\end{equation}
substituting the expression for $\hat{\Gamma}^{\rho}{}_{\mu \nu}$ given in (\ref{rela}), we find
\begin{equation}
{R}^{\rho}{}_{\lambda \mu \nu} =
{\stackrel{\circ}{R}}{}^{\rho}{}_{\lambda \mu \nu} +
Q^{\rho}{}_{\lambda \mu \nu} \equiv 0 \; ,
\label{relar}
\end{equation}
where ${\stackrel{\circ}{R}}{}^{\rho}{}_{\lambda \mu \nu}$ is the Riemann tensor of
the Levi-Civita connection, and $Q^{\rho} {}_{\lambda \mu \nu}$ is expressed in terms of Weitzenb\"ock connection only:
\begin{equation}
Q^{\rho} {}_{\lambda \mu \nu} = {D}_{\mu}{}{K}^{\rho}{}_{\lambda \nu} -
{D}_{\nu}{}{K}^{\rho}{}_{\lambda \mu} + {K}^{\sigma}{}_{\lambda \nu}
\; {K}^{\rho}{}_{\sigma \mu} - {K}^{\sigma}{}_{\lambda \mu} \;
{K}^{\rho}{}_{\sigma \nu}
\label{qdk}
\end{equation}
Here, $D_\mu$ is the teleparallel covariant derivative, whose explicit form can be obtained by operating on a space-time vector
$V^\mu$:
\begin{equation}
D_\rho \, V^\mu \equiv \partial_\rho V^\mu +
\left( \hat{\Gamma}^\mu{}_{\lambda \rho} - K^\mu{}_{\lambda \rho} \right) V^\lambda \; .
\label{tcd}
\end{equation}
The equivalence between the teleparallel and the Riemann descriptions is clearly expressed in \Cref{relar}: the contribution
 from the
Levi-Civita connection (${\stackrel{\circ}{R}}{}^{\rho}{}_{\lambda \mu \nu}$) compensates  the one from the Weitzenb\"ock connection ($Q^{\rho}{}_{\lambda \mu \nu}$), so that ${R}^{\rho}{}_{\lambda \mu \nu}$ is identically zero.

Therefore, we can write the Lagrangian of the gravitational field as
 \begin{equation}
{\cal L}_G =
\frac{e}{2\kappa} \; S^{\rho \mu \nu} \; T_{\rho \mu \nu} \; ,
\label{gala}
\end{equation}
where $e = {\rm det}(e^{a}{}_{\mu})$ and
\begin{equation}
S^{\rho \mu \nu} = - S^{\rho \nu \mu} \equiv \dfrac{1}{2}
\left( K^{\mu \nu \rho} - g^{\rho \nu} \; T^{\lambda \mu}{}_{\lambda} + g^{\rho \mu}
\; T^{\lambda \nu}{}_{\lambda} \right)
\label{Ssuperpotdef}
\end{equation}
Using relation (\ref{rela}) and identifying $e = \sqrt{-g}$, the Lagrangian (\ref{gala}) results to be equivalent to the Einstein-Hilbert Lagrangian of GR modulo divergence:
\begin{equation}
{\cal L} = - \frac{1}{2\kappa} \;  \sqrt{-g} \, {\stackrel{\circ}{R}} \; .
\end{equation}
The  teleparallel version of the gravitational field is obtained by  varying  ${\cal L}_G$  with respect to the gauge
field $A_B{}^\rho$:
\begin{equation}
\partial_\sigma(e S_B{}^{\sigma \rho}) -
 4 \pi G  \, (e j_B{}^{\rho}) = 0 \; ,
\label{tfe1}
\end{equation}
where $S_B{}^{\sigma \rho} \equiv e_B{}^{\lambda}
S_{\lambda}{}^{\sigma \rho}$. The quantity
\begin{equation}
e j_B{}^{\rho} \equiv \frac{\partial {\cal L}_G}{\partial e^B{}_{\rho}} = -
\frac{1}{4 \pi G} \, e e_B{}^{\lambda} S_{\mu}{}^{\nu \rho}
T^\mu{}_{\nu \lambda} + e_B{}^{\rho} {\cal L}_G
\label{ptem1}
\end{equation}
represents the gauge current which coincides with the energy and momentum of the gravitational field \cite{Andrade00}.
The quantity $e S_B{}^{\sigma\rho}$ is called  {\it superpotential} as its derivative provides the gauge
current $e j_B{}^{\rho}$, which is conserved:
\begin{equation}
\partial_\rho (e j_B{}^\rho) = 0 \; .
\label{conser1}
\end{equation}
Using the following identity
\begin{equation}
\partial_\rho e \equiv e {\stackrel{\circ}{\Gamma}}{}^{\nu}{}_{\nu \rho} =
e \left( {\hat\Gamma}^{\nu}{}_{\rho \nu} - K^{\nu}{}_{\rho \nu} \right)  ,
\label{id1}
\end{equation}
\Cref{conser1} can be written as
\begin{equation}
D_\rho \, j_B{}^\rho \equiv \partial_\rho j_B{}^\rho +
\left( {\hat\Gamma}^\rho{}_{\lambda \rho} - K^\rho{}_{\lambda \rho} \right)
j_B{}^\lambda = 0 \; ,
\label{conser2}
\end{equation}

It is interesting to relate the above gauge approach with canonical GR.
To do that, we use \Cref{carco} to express $\partial_\rho e_A{}^\lambda$ and rewrite \Cref{tfe1} in the form
\begin{equation}
\partial_\sigma(e S_\lambda{}^{\sigma \rho}) -
 4 \pi G  \, (e t_{\lambda}{}^{\rho}) = 0 \; ,
\label{tfe2}
\end{equation}
where
\begin{equation}
e t_{\lambda}{}^{\rho} =
\frac{  e}{4 \pi G} \, \Gamma^{\mu}{}_{\nu \lambda} S_{\mu}{}^{\nu \rho}
+ \delta_\lambda{}^{\rho} {\cal L}_G
\label{ptem2}
\end{equation}
represents the standard energy-momentum pseudotensor of the gravitational field
\cite{Virbhadra90,Shirafuji96}.
Using \Cref{rela}, one can rewrite \Cref{tfe2} in terms of the Levi-Civita connection.
Thus, from the equivalence of the Lagrangians, we can reproduce the Einstein field equations:
\begin{equation}
\frac{e}{2} \left[{\stackrel{\circ}{R}}_{\mu \nu} -
\frac{1}{2} \, g_{\mu \nu}
{\stackrel{\circ}{R}} \right] = 0 \; .
\end{equation}
This result proves the equivalence of the two approaches and justifies the name ``Teleparallel Equivalent of General Relativity'' (TEGR) \cite{Unzicker05}.

\subsection{$f(T)$ gravity}
Among all the models suggested to describe the late-time accelerated expansion, the \emph{teleparallel} description of gravity has recently reached much attention \cite{rocco,Capozziello15,Aviles13,Abedi17}.
As for the $f(R)$ theories of gravity, interesting scenarios arise when one replaces the torsion scalar with a generic function $f(T)$ \cite{Ferraro07,Bengochea09,Linder10}.
In particular, one considers the action
\begin{equation}
\mathcal S=\int d^4 x\ e \left[\dfrac{f(T)}{2\kappa}+\mathcal{L}_m \right]\,,
\label{f(T):action}
\end{equation}
where $e=\sqrt{-g}=\det(e_\mu^A)$. The field equations are thus obtained by varying the action (\ref{f(T):action}) with respect to the vierbein fields:
\begin{equation}
e_A^\rho{S_\rho}^{\mu\nu}(\partial_\mu T)f''+\left[\dfrac{1}{e}\partial_\mu(e e_A^\rho {S_\rho}^{\mu\nu})-e_A^\lambda {T^\rho}_{\mu\lambda}{S_\rho}^{\nu\mu}\right]f' +\dfrac{1}{4}e_A^\nu f=\dfrac{\kappa}{2}e_A^\rho {{T^{(m)}}_\rho}^\nu\ ,
\end{equation}
where ${{T^{(m)}}_\rho}^\nu$ represents the energy-momentum tensor of matter, and the `primes' indicate derivatives with respect to $T$.

Assuming a spatially flat FLRW background manifold, the vierbein takes the form
\begin{equation}  \label{weproudlyuse}
 e_{\mu}^A=\mathrm{diag}(1,a,a,a) \ ,
\end{equation}
while the dual vierbein is given by $e^{\mu}_A=\mathrm{diag}(1,a^{-1},a^{-1},a^{-1})$. From this choice, one abtains the well-known metric
\begin{equation}
 ds^{2}=dt^{2}-a^{2}(t)\delta_{ij}dx^idx^j\,.
\label{metriccosmo}
\end{equation}
Under this assumption, the torsion scalar is related to the Hubble parameter through
\begin{equation}
T=-6H^2\ .
\label{eq:T-H}
\end{equation}
Moreover, considering a perfect fluid for matter and neglecting radiation, the modified Friedmann equations are thus given as
\begin{align}
H^2&=\dfrac{1}{3}(\rho_m+\rho_T)\ , \label{f(T):Fried1} \\
2\dot{H}+3H^2&=-\dfrac{1}{3}(p_m+p_T)\ , \label{f(T):Fried2}
\end{align}
where $\rho_T$ and $p_T$ are, respectively, the torsional energy density and pressure:
\begin{align}
\rho_T&=Tf'-\dfrac{f}{2}-\dfrac{T}{2}\,, 	\label{eq:rho_T}\\
p_T&=\dfrac{f-Tf'+2T^2f''}{2(f'+2Tf'')}\,.\label{eq:p_T}
\end{align}
The above quantities can be then used to define an effective dark energy fluid with equation of state parameter
\begin{equation}
w_{DE}\equiv \dfrac{p_T}{\rho_T}=-1+\dfrac{(f-2Tf')(f'+2Tf''-1)}{(f+T-2Tf')(f'+2Tf'')}\ .
\end{equation}
In particular, when $f'=1$, one gets $w_{DE}=-1$ which corresponds to the $\Lambda$CDM case.

The prescription described above can be even extended to consider \emph{teleparallel dark energy models}, in which a scalar field non-minimally coupled to gravity is responsible for the cosmic acceleration \cite{Geng12,Xu12,D'Agostino18}.
Although a single scalar field is commonly employed, multiple field models can be also considered to explain late-time acceleration and inflation \cite{Vardanyan15}.
In GR, through a suitable conformal transformation, the Lagrangian can be written in a particular frame, \emph{i.e.} Einstein frame, in which the coupling does not show up.
The situation is different in teleparallel gravity, where no Einstein frames exist even in the case of a single field model \cite{Yang11,Wright16}.

We thus can write the generic action for a scalar field $\phi $ and the kinetic term $X\equiv\nabla_\mu \phi \, \nabla^\mu \phi$ in the form \cite{Abedi18}
\begin{equation}
\mathcal S= \int {d}^4 x\, e \left[\frac{1}{2} f(T,\phi,X) +\mathcal{L}_{m} \right] , \label{action1}
\end{equation}
Variation of the above action with respect to the vierbein fields gives
\begin{align}
\Theta ^{\phantom{A} \mu}_{A}=\ &\frac{1}{2}f e_{A}^{\phantom{A} \mu}+ f_{T} \left[ e^{-1}\partial_{\nu} \left( ee^{\phantom{A}
	\rho}_{A}S^{\phantom{\rho} \mu \nu}_{\rho} \right)-e^{\phantom{A} \gamma}_{A}S^{\rho \beta
	\mu}T_{\rho \beta \gamma} \right]
+ e^{\phantom{A} \rho}_{A}S^{\phantom{\rho}
	\nu \mu}_{\rho} \partial_{\nu} f_T \nonumber \\	&
- \frac{1}{2} f_{X}
e_{A}^{\phantom{A} \nu} \, \partial^{\mu} \phi \,\partial_{\nu} \phi\ , \label{field1}
\end{align}
where $f_T\equiv\partial f/\partial T $, $ f_X\equiv\partial f/\partial X$ and
\begin{equation}
\Theta ^{\phantom{\nu}
\mu}_{\nu}=e^{A}_{\phantom{A} \nu} \Theta^{\phantom{A} \mu}_{A}=-e^{A}_{\phantom{A}
\nu}\, \delta \mathcal{L}_{m}/\delta e^{A}_{\phantom{A} \mu}
\end{equation}
is the matter energy-momentum tensor. On the other hand, varying the action (\ref{action1}) with respect to the scalar field yields
\begin{equation}
\square \phi +\frac{\partial_\alpha f_X}{f_X}\, \partial^\alpha \phi -\frac{f_\phi}{f_X}=0 \label{field2}\ ,
\end{equation}
where $\square \phi=\partial_{\mu}\big(\sqrt{-g}\partial^{\mu}\phi \big) / \sqrt{-g}$.
Then, we can write a covariant representation of \Cref{field1} as
\begin{equation}
\Theta^{\phantom{\alpha} \mu}_{\alpha}=f_T G^{\phantom{\alpha} \mu}_{\alpha}+ \frac{1}{2} \delta^{\mu}_{\alpha} (f-f_T
T) + S^{\phantom{\alpha} \nu \mu}_{\alpha}  \partial_{\nu} f_T -\frac{1}{2} f_{X}\; \partial^{\mu} \phi \;
\partial_{\alpha} \phi, \label{field8}
\end{equation}
being $G_\alpha^{\phantom{\alpha}\mu}$ the Einstein tensor.
Requiring the symmetry and the local invariance of the energy-momentum tensor under Lorentz transformation, one obtains that
\begin{equation}
\left( S_{\alpha}^{\phantom{\alpha} \lambda\nu}g^{\alpha\mu}-S_{\alpha}^{\phantom{\alpha} \lambda\mu
}g^{\alpha\nu} \right)  \partial_{\lambda} f_T =0 \label{cons1}.
\end{equation}
The form of the field equations under the assumptions (\ref{weproudlyuse}) and (\ref{metriccosmo}) is
\begin{align}
\rho_{m}& =3H^2 f_T-\frac{1}{2} (f-f_T T)+\frac{1}{2} f_{X} \dot{\phi}^{2}  , \label{edensity}
\\
p_{m}&=- \big(3H^2 + 2 \dot{H} \big) f_T+\frac{1}{2} (f-f_T T) -H \, \partial_{0} f_T, \label{pressure}
\\
0&=\ddot{\phi}+\left( 3H+\frac{\partial_0 f_X}{f_X}\right)\dot{\phi}-\frac{f_\phi}{f_X},
\end{align}
under the hypothesis that $\phi$ depends only on time at the background level.
By means of the definitions
\begin{align}
&\rho_{DE}= 3H^2 (1-f_T)+\frac{1}{2} (f-f_T T) - \frac{1}{2}f_X \dot{\phi}^2, \nonumber \\
&p_{DE}= -(3H^2+2\dot{H}) (1-f_T)-\frac{1}{2} (f-f_T T)+H\, \partial_0 f_T \ ,
\end{align}
Eqs.~\eqref{edensity} and \eqref{pressure} read
\begin{align}
\rho_{DE}+\rho_{m}& = 3H^2,
\nonumber \\
p_{DE}+p_{m}&= -\big( 3H^2+2\dot{H} \big),
\end{align}
Therefore, we can finally obtain the dark energy equation of state as $w_{DE} = p_{DE}/\rho_{DE}$:
\begin{equation}
w_{DE}=-\frac{1+\dfrac{2\dot{H}}{3H^2}+ \dfrac{w_m\rho_{m}}{3H^2}}{1-\dfrac{\rho_{m}}{3H^2}}\ ,
\end{equation}
where $p_m =w_m\rho_m$.

\section{Cosmography}

\label{sec:cosmography}
The above $f(R)$ and $f(T)$ gravity are some examples of the possibilities to address the problem of  accelerated speed of the observed universe under the standard of geometry. However,
the degeneracy among the cosmological models invoked to feature the dark energy behaviour has made clear the need of model-independent techniques to describe the expansion of the universe.
Among all reasonable approaches, \textit{cosmography} has recently attracted a lot of attention \cite{Visser,Dunsby16}.
This model-independent technique relies only on the observational assumptions of the cosmological principle  and permits the study of the dark energy evolution without the need of assuming a specific cosmological model \cite{Weinberg72,Harrison76,j1,j2}.
The standard cosmographic approach is based on Taylor expansions of observables which can be directly compared to data, and the outcomes of such a procedure are independent of any equation of state postulated to study the cosmic evolution.
For these reasons, cosmography turns out to be a powerful tool to break the degeneracy among cosmological models and it is currently widely adopted to understand universe's kinematics \cite{j3,j4,j5,j6,j7,j8,j9,j10,j11,j12,j13,j14,j15,j17,j18,j19,j20,j21,j22,j23,j24,j25,j26,j27,j28,j29,j30,j31,j32,j33,j34,j35,j36}.

The cosmological principle demands the scale factor as the only degree of freedom governing the universe.
One can thus expand $a(t)$ in Taylor series around the present time:
\begin{equation}
a(t)=1+\sum_{k=1}^{\infty}\dfrac{1}{k!}\dfrac{d^k a}{dt^k}\bigg | _{t=t_0}(t-t_0)^k\ .
\label{eq:scale factor}
\end{equation}
The above expansion defines the so-called cosmographic series \cite{Cattoen08}:
\begin{subequations}
\begin{align}
&H(t)\equiv \dfrac{1}{a}\dfrac{da}{dt} \ , \hspace{1cm} q(t)\equiv -\dfrac{1}{aH^2}\dfrac{d^2a}{dt^2}\ ,  \label{eq:H&q} \\
&j(t) \equiv \dfrac{1}{aH^3}\dfrac{d^3a}{dt^3} \ , \hspace{0.5cm}  s(t)\equiv\dfrac{1}{aH^4}\dfrac{d^4a}{dt^4}\ ,    \label{eq:j&s}
\end{align}
\end{subequations}
which are known as the \textit{Hubble}, \textit{deceleration}, \textit{jerk} and \textit{snap} parameters\footnote{One may, in principle, consider high-order terms in the series. However, we here limit our study up to the \emph{snap} parameter, as the current observations are not able to  properly constrain the next order terms \cite{Aviles12}.}.
These quantities are used to study the dynamics of the late-time universe.
The physical properties of the coefficients can be deduced by the shape of the Hubble expansion.
In particular, the sign of the parameter $q$ indicates whether the universe is accelerating or decelerating.
The sign of $j$ determines the change of the universe's dynamics, a positive value indicating the occurrence of a transition time during which the universe modifies its expansion.
Moreover, the value of $s$ is necessary to discriminate between an evolving dark energy term or a cosmological constant behaviour.

From the definition $z=a^{-1}-1$ and \Cref{eq:scale factor}, one obtains the Taylor expansion of the luminosity distance:
\begin{align}
d_L(z)=&\ \dfrac{z}{H_0}\bigg[1+\dfrac{z}{2}(1-q_0) -\dfrac{z^2}{6}\left(1-q_0-3q_0^2+j_0\right)+ \nonumber \\
&\hspace{0.8cm}+\dfrac{z^3}{24}\left(2-2q_0-15q_0^2-15q_0^3+5j_0+10q_0j_0+s_0\right)+\mathcal{O}(z^4)\bigg].
\label{eq:luminosity distance}
\end{align}
The above expression can be used to constrain the cosmographic series and frame the expansion of the universe without resorting to any \emph{a priori} cosmological model. In fact, one can insert \Cref{eq:luminosity distance} into
\begin{equation}
H(z)=\left[\dfrac{d}{dz}\left(\dfrac{d_L(z)}{1+z}\right)\right]^{-1}\,,
\label{eq:Hubble rate}
\end{equation}
and find
\begin{subequations}
\begin{align}
H(z)&\simeq H_0\left[1+H^{(1)}z+H^{(2)}\dfrac{z^2}{2}+H^{(3)}\dfrac{z^3}{6}\right]\,,\\
H^{(1)}&=1+q_0\,,\\
H^{(2)}&=j_0-q_0^2\,,\\
H^{(3)}&=3q_0^2+3q_0^3-j_0(3+4q_0)-s_0\,.
\label{eq:Taylor H(z)}
\end{align}
\end{subequations}
Useful relations are obtained by expressing the cosmographic series in terms of the time-derivatives of the Hubble rate:
\begin{align}
q&=-1-\dfrac{\dot{H}}{H}\ ,\\
j&=1+\dfrac{\dot{H}+\ddot{H}}{H^2}\ , \\
s&=1+\dfrac{2}{H^2}\left(3\dot{H}+2\ddot{H}\right)+\dfrac{1}{H^4}\left(3H^2+\dddot{H}\right) .
\end{align}
The above equations permit to calculate the cosmografic coefficients for a given cosmological model.

\subsection{Limits of standard cosmography}

Although powerful and simple to apply, the cosmographic method is plagued with several shortcomings which limit the possibility to use it in certain circumstances.
The main problem concerns the inability of the currently available cosmological data to put tight constraints on the cosmographic parameters and fix the kinematic expansion of the universe especially at early stages.
Also, the arbitrary order of truncation of the Taylor series might compromise the predictive power of cosmography.
Another important issue is the degeneracy between all of the cosmographic coefficients. The impossibility to measure them separately but only the sum leads to different results depending on the probability distribution associated with each coefficient.

Moreover, the role of spatial curvature is crucial in cosmographic constraints. In fact, $\Omega_{k}\neq0$ causes a dark energy equation of state evolving with time \cite{Clarkson07,Pavlov13}, which fixes  \emph{a priori} the dark energy term.
Due to the close relation between luminosity distance and spatial curvature, one is forced to fix the value of $\Omega_k$ in order to constrain the cosmographic parameters. In doing so, the resulting series is the expression of the universe with that assumed curvature.
Furthermore, $\Omega_{k}$ is strongly degenerate with the other cosmographic coefficients, which cannot be measured independently if the curvature is not fixed.
On the one hand, assuming a precise value of $\Omega_{k}$ may affect the dark energy reconstruction while, on the other hand, convergence issues could arise if $\Omega_{k}$ is not postulated \emph{a priori}.

\subsection{Improving standard cosmography}

The standard cosmographic approach suffers from severe issues when high-redshift data are used to study the dark energy behaviour \cite{j36}.
The restricted convergence of the Taylor series makes this method poorly predictive for cosmographic analysis at $z>1$.

\subsubsection{The method of Pad\'e polynomials}

A way to overcome these restrictions is offered by Pad\'e rational polynomials  \cite{Baker66}.
The method of Pad\'e approximations is built up from the standard Taylor series of a generic function $f(z)$:
\begin{equation}
f(z)=\sum_{i=0}^\infty c_iz^i\,,
\label{eq:power series}
\end{equation}
where $c_i=f'(0)/i!$.  We thus define the $(n,m)$  Pad\'e approximation of $f(z)$ as
\begin{equation}
P_{n,m}(z)=\dfrac{\displaystyle{\sum_{i=0}^{n}a_i z^i}}{1+\displaystyle{\sum_{j=1}^{m}b_j z^j}}\,.
\label{def Pade}
\end{equation}
whose Taylor expansion agrees with \Cref{eq:power series} to the highest possible order, i.e.
\begin{equation}
\left\{
\begin{aligned}
&P_{n,m}(0)=f(0)\,,\\
&P_{n,m}'(0)=f'(0)\,,\\
&\vdots	\\	
&P_{n,m}^{(n+m)}(0)=f^{(n+m)}(0)\,.
\end{aligned}
\right.
\end{equation}
The $n+1$ independent coefficients in the numerator and $m$ independent coefficients in the denominator of \Cref{def Pade} make $n+m+1$ the number of total unknown terms. These can be determined by imposing
\begin{equation}
f(z)-P_{n,m}(z)=\mathcal{O}(z^{n+m+1})\ ,
\end{equation}
from which one obtains
\begin{equation}
(1+b_1z+\hdots +b_mz^m)(c_0+c_1z+\hdots)=a_0+a_1z+\hdots+a_nz^n +\mathcal{O}(z^{n+m+1})\,.
\label{coeff}
\end{equation}
Then, equating the coefficients with the same power provides a set of $n+m+1$ equations for the $n+m+1$ unknown terms $a_i$ and $b_i$:
\begin{equation}
\left\{
\begin{aligned}
&a_i=\sum_{k=0}^i b_{i-k}\ c_{k} \ ,  \\
&\sum_{j=1}^m b_j\ c_{n+k+j}=-b_0\ c_{n+k}\ , \hspace{0.5cm} k=1,\hdots, m \ .
\end{aligned}
\right .
\end{equation}
Recent applications of Pad\'e approximations in the cosmological context have shown the good properties of this technique to alleviate the convergence problems at high redshifts \cite{Gruber14,Wei14,Dutta18}.  Also in this approach all physical information got from data, i.e. the cosmographic series, is based on assuming cosmic homogeneity and isotropy only.
We can summarize the advantages of Pad\'e rational approximations as follows:
\begin{itemize}
  \item the series can heal bad convergence issues in the data ranges;
  \item the series can decrease error propagations outside the interval $z<1$;
  \item the series can be calibrated by choosing appropriate orders depending on the specific situation.
\end{itemize}
Nevertheless, the Pad\'e polynomials suffer from some issues:
\begin{itemize}
  \item the convergence of the series is not known a priori, and directly comparing with data is necessary in order to the specify the appropriate order;
  \item possible poles characteristic of the series within the observational domain may limit the convergence;
  \item there could exist a degeneration among different series.
\end{itemize}

A detailed study of Pad\'e approximations in the cosmographic context has been performed in \refcite{Aviles14}, where the authors showed the advantages of this method by analyzing several Pad\'e expansions and comparing them with different cosmological observables. They obtained bounds on the cosmographic series and investigated how to reduce the errors systematics and to overcome degeneracy between the cosmological parameters.
We report in \ref{sec:Pade approx} some explicit expressions of the Pad\'e approximations of the luminosity distance.

\subsubsection{The method of Chebyshev polynomials}

We have seen that Pad\'e method
still leaves a degree of subjectivity in the choice of the highest orders
of expansion. In addition, the Pad\'e treatment works much better as
one has to approximate non-smooth functions in which other numerical
methods fail. This happens as one needs to approximate flexes
or discontinuities in domains. Unfortunately, this is not the case of
cosmic distances. So that, from the one hand it is possible to heal
the convergence problem, but from the other hand one conceptually
uses Pad\'e series to approximate well-defined cosmic distances,
albeit no poles are effectively involved.

To alleviate these caveats, we proposed in \refcite{rocco3} a new cosmographic method based on ratios of
Chebyshev polynomials. We showed that they reduce systematics on fitted coefficients, and candidate as a
serious alternative to Taylor and Pad\'e series in cosmology.
The Chebyshev polynomials $T_n(z)$ are defined as
\begin{equation}
T_n(z)=\cos(n\theta)\,,
\end{equation}
where $\theta=\arccos(z)$ and $n\in\mathbb{N}_0$. They are orthogonal polynomials with respect to the function  $w(z)=(1-z^2)^{-1/2}$ for $|z|\leq1$ \cite{Chebyshev} such that
\begin{equation}
\int_{-1}^{1}T_n(z)T_m(z) w(z)=
\begin{cases}
\pi\ , & n=m=0 \vspace{0.2cm}\\
\dfrac{\pi}{2} \delta_{nm}\ , & \text{otherwise}
\end{cases}
\end{equation}
It is possible to generate the Chebyshev polynomials from the recurrence relation:
\begin{equation}
T_{n+1}(z)=2zT_n(z)-T_{n-1}(z)\ .
\end{equation}
The explicit expressions up to the fifth order are the following:\footnote{We here truncate up to the fifth order, since additional contributions go beyond this treatment. In so doing, we arrive to analyse up to snap parameter $s_0$.}
\begin{equation}
\begin{aligned}
&T_0(z) = 1\ , \\
&T_1(z) = z \ ,\\
&T_2(z) = 2z^2 - 1 \ , \\
&T_3(z) = 4z^3 - 3z \ ,\\
&T_4(z) = 8z^4 - 8z^2 + 1 \ ;
\end{aligned}
\label{eq:T_k}
\end{equation}
which will be employed to build the new expression for $d_L(z)$.
The powers of $z$ can be expressed in terms of the Chebyshev polynomials as
\begin{equation}
z^n=2^{1-n}\sum_{k=0}^{[n/2]}a_k\binom{n}{k}T_{n-2k}(z)\ ,
\end{equation}
for $n>0$, being $[n/2]$ the integer part of $n/2$,  $a_k=1/2$ if $k=n/2$ and $a_k=1$ if $a_k\neq n/2$.

Suppose $f(z)\in L_w^2$, where $L_w^2$ is the Hilbert space of the square-integrable functions with respect to $w^{-1}(z)\ dz$.
If the truncated Taylor series of $f(z)$ around the point $z=0$, $g(z)$, is known, it is possible to obtain the polynomial of degree $n$, $\sum_{k=0}^n c_k T_k$ , giving the best approximation of $f(z)$ in the interval $[-1,1]$ in $L_w^2$. Then, the Chebyshev series expansion of $f(z)$ can be written as
\begin{equation}
f(z)=\sum_{k=0}^{\infty} c_k T_k(z)\ ,
\label{eq:f(z)}
\end{equation}
where
\begin{align}
\begin{cases}
c_0= \dfrac{1}{\pi}\displaystyle{\int_{-1}^{1}}g(z)\ T(z)\ w(z)\ dz \ , \vspace{0.2cm} \\
c_k= \dfrac{2}{\pi}\displaystyle\int_{-1}^{1}g(z)\ T(z)\ w(z)\ dz\ , \hspace{0.5cm} k>0\ . \label{eq:c_k}
\end{cases}
\end{align}
Therefore, the $(n,m)$ rational Chebyshev approximant is
\begin{equation}
R_{n,m}(z)=\dfrac{\displaystyle{\sum_{i=0}^n}\ a_i T_i(z)}{1+\displaystyle{\sum_{j=1}^m}\ b_j T_j(z)}\ .
\label{eq:rational Chebyshev}
\end{equation}
Equating \Cref{eq:f(z)} and \Cref{eq:rational Chebyshev} up to the $(n+m)$-th Chebyshev polynomial, one obtains the unknown coefficients $a_k$ and $b_k$:
\begin{equation}
f(z)=R_{n,m}(z)+\mathcal{O}(T_{n+m+1})\ .
\end{equation}
By doing so, one gets
\begin{equation}
(1+b_1T_1+\hdots+b_mT_m)(c_0+c_1T_1+\hdots)= a_0+a_1T_1+\hdots +a_nT_n+\mathcal{O}(T_{n+m+1})\ .
\label{eq:prod coeff}
\end{equation}
The products of The Chebyshev polynomials on the left hand side of \Cref{eq:prod coeff} can be obtained through the trigonometric identity
\begin{equation*}
\cos(n\theta)\cos(m\theta)=\dfrac{1}{2}\Big[\cos\big[(n+m)\theta\big]+\cos\big[(n-m)\theta\big]\Big],
\end{equation*}
leading to
\begin{equation}
T_n(z)T_m(z)=\dfrac{1}{2}\Big[T_{n+m}(z)+T_{|n-m|}(z)\Big] .
\label{eq:prod Chebyshev}
\end{equation}
Hence, equating the terms with the same degree of $T$'s one has
\begin{equation}
\left\{
\begin{aligned}
&a_i=\dfrac{1}{2}\sum_{j=0}^{m}{}^{\prime}\ b_j(c_{i+j}+c_{|i-j|})=0 \ , \hspace{0.5cm} i=0,\hdots,n   \\
&\sum_{j=0}^{m}{}^{\prime}\ b_j(c_{i+j}+c_{|i-j|})=0 \ , \hspace{0.5cm} i=n+1,\hdots,n+m\ .
\end{aligned}
\right .
\end{equation}
The above formalism can be easily generalized for $z$ in an arbitrary interval $[a,b]$.
To do that, one can define the generalized Chebyshev polynomials $T^{[a,b]}_n(z)=\cos(n\theta)$, where $z$ is the new variable
\begin{equation}
z=\dfrac{a(1-\cos\theta)+b(1+\cos\theta)}{2}\ .
\end{equation}
This is obtained by means of
\begin{equation}
\cos\theta=\dfrac{2z-(a+b)}{b-a}\ ,
\end{equation}
so that $\theta\in[-\pi,\pi]$ while $z\in[a,b]$.
From the ordinary Chebyshev polynomials it is possible to obtain the generalized polynomials through
\begin{equation}
T_n^{[a,b]}(z)=T_n\left(\dfrac{2z-(a+b)}{b-a}\right) .
\end{equation}
$T_n^{[a,b]}(z)$ form an orthogonal set with respect to the weighting function \cite{Obsieger13}
\begin{equation}
w_{[a,b]}(z)=[(z-a)(b-z)]^{-1/2}\ ,
\end{equation}
so that
\begin{equation}
\langle T_m^{[a,b]},T_n^{[a,b]}\rangle=\int_a^b dz\ w_{[a,b]}(z)\ T_n^{[a,b]}(z)\ T_m^{[a,b]}(z)\ .
\end{equation}
Then, the orthogonality condition reads
\begin{equation}
\langle T_m^{[a,b]},T_n^{[a,b]}\rangle=
\begin{cases}
\pi\ , & n=m=0 \vspace{0.2cm}\\
\dfrac{\pi}{2} \delta_{nm}\ , & \text{otherwise}
\end{cases}
\end{equation}
since $T_n^{[a,b]}(z)=\cos(n\theta)$ and $d\theta=-w_{[a,b]}dz$.

To approximate the luminosity distance with Chebyshev polynomials, we need to calculate the coefficients $c_k$ in \Cref{eq:c_k} where $g(z)$ is the Taylor expansion (\ref{eq:luminosity distance}). The fourth-order Chebyshev expansion of the luminosity distance reads
\begin{equation}
d_L(z)=\dfrac{1}{H_0}\sum_{n=0}^4 c_n T_n(z)\ ,
\label{eq:dL new}
\end{equation}
where the coefficients $c_n$ are:
\begin{equation*}
\begin{aligned}
&c_0=\dfrac{1}{64} \Big[18 + 5 j_0 (1 + 2 q_0) - 3 q_0 \big(6 + 5 q_0 (1 + q_0)\big) + s_0\Big] ,\\
&c_1=\dfrac{1}{8}\left(7 - j_0 + q_0 + 3 q_0^2\right) ,\\
&c_2=\dfrac{1}{48}\Big[14 + 5 j_0 (1 + 2 q_0) - q_0 \big(14 + 15 q_0 (1 + q_0)\big) + s_0\Big] , \\
&c_3=\dfrac{1}{24}\big(-1 - j_0 + q_0 (1 + 3 q_0)\big) , \\
&c_4=\dfrac{1}{192}\Big[2 + 5 j_0(1 + 2 q_0) - q_0\big(2 + 15 q_0(1 + q_0)\big) + s_0\big] .
\end{aligned}
\end{equation*}
We report some explicit expressions  of the rational Chebyshev approximations of $d_L(z)$ in \ref{sec:Cheb approx}. High-order polynomials  leading to more accurate approximations are characterized by more complicated forms. The most suitable choice of Chebyshev approximation lies on assuming the correct set of coefficients which avoids one to encounter poles in the numerical analyses. This strategy can be performed by simply requiring no poles in the investigated redshift domain. Moreover, the underlying request over coefficient priors also gives an indication on which are the most viable orders to use in Chebyshev expansions.

\begin{figure}
\begin{center}
\includegraphics[width=0.8\textwidth]{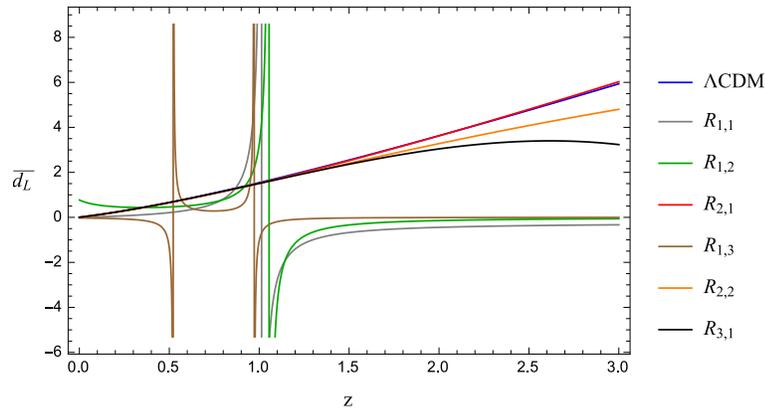}
\caption{Dimensionless luminosity distance in terms of the redshift in the case of rational Chebyshev approximations ($R_{1,1}$), ($R_{1,2}, R_{2,1}$) and ($R_{1,3}, R_{2,2}, R_{3,1}$) orders; it is possible to notice the comparison with the $\Lambda$CDM model. Choosing a set of values for the free parameters enables to get the correct expansion orders in Chebyshev analyses.}
\label{fig:dL_Rnm}
\end{center}
\end{figure}

We compare the various Chebyshev approximations with the $\Lambda$CDM luminosity distance to check their accuracy.
In the case of the standard model, the cosmographic series are calculated in terms of $\Omega_{m0}$:
\begin{equation}
\begin{aligned}
&q_{0,\Lambda\text{CDM}}= -1+\dfrac{3}{2}\Omega_{m0}\ ,\\
&j_{0,\Lambda\text{CDM}}=1\ ,\\
&s_{0,\Lambda\text{CDM}}=1-\dfrac{9}{2}\Omega_{m0}\ .
\end{aligned}
\label{eq:cosmog param LCDM}
\end{equation}
As an indicative example, we fix $\Omega_{m0}=0.3$. From \Cref{eq:cosmog param LCDM} one then get
\begin{equation}
\left\{
\begin{aligned}
&q_0=-0.55\ , \\
&j_0=1\ , \\
&s_0=-0.35\ .
\end{aligned}
\right.
\label{eq:indicat values}
\end{equation}
Adopting the values of \Cref{eq:indicat values}, in \Cref{fig:dL_Rnm} we show $\overline{d_L}(z)\equiv H_0\times d_L(z)$ for different degrees of Chebyshev approximations.

\subsubsection{The convergence radius of rational approximations}

To verify the effective improvement of the new cosmographic technique in approximating cosmic distances, it is necessary to test the stability of Chebyshev approximations at high-redshift domains. Therefore, one can study the convergence radius $\rho$ of the various cosmographic methods.

\begin{figure}[h]
\begin{center}
\includegraphics[width=0.7\textwidth]{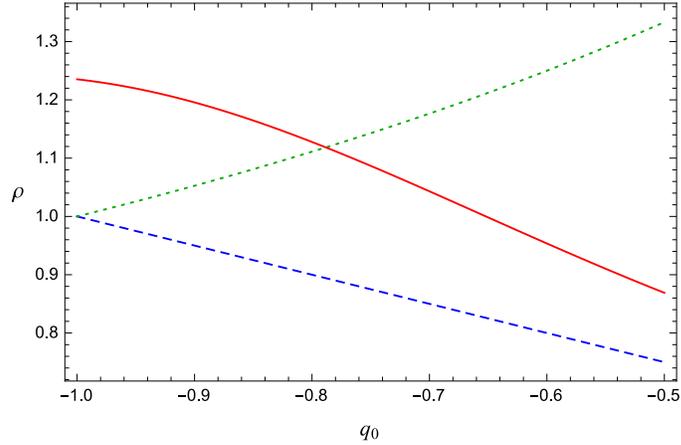}
\caption{Convergence radii for different orders. In particular, for second-order Taylor (dashed curve) and equivalent (1,1) Pad\'e (dotted curve) and (1,1) rational Chebyshev (solid curve). In the picture of Chebyshev approximation we took $j_0=2$, $s_0=-1$.}
\label{fig:radii}
\end{center}
\end{figure}

As an example, we compare the convergence radius of the (1,1) rational Chebyshev approximation of $d_L(z)$ with the second-order Taylor and the (1,1) Pad\'e approximations. From \Cref{eq:rational Chebyshev,eq:T_k}, it follows
\begin{equation}
R_{1,1}(z)=\dfrac{a_0 T_0(z)+a_1T_1(z)}{1+b_1T_1(z)}=\dfrac{a_0+a_1 z}{1+b_1 z}\ ,
\label{eq:R_11}
\end{equation}
where $\{a_0,a_1,b_1\}$ are expressed in terms of the series given in \Cref{eq:R11}. One can recast \Cref{eq:R_11} as
\begin{equation}
R_{1,1}=\dfrac{a_0}{1+b_1 z}+\dfrac{a_1}{b_1}\left(1-\dfrac{1}{1+b_1 z}\right) ,
\end{equation}
which leads to
\begin{equation}
R_{1,1}=\dfrac{a_1}{b_1}+\left(a_0-\dfrac{a_1}{b_1}\right)\sum_{n=0}^{\infty}(-b_1)^n z^n\ .
\label{eq:new R_11}
\end{equation}
The convergence radius of the geometric series in \Cref{eq:new R_11} is thus
\begin{equation}
\rho_{R_{1,1}}=\dfrac{1}{|b_1|}=\bigg|\dfrac{-3 (7 - j_0 + q_0 + 3 q_0^2)}{14 + 5 j_0 (1 + 2 q_0) - q_0 \big(14 + 15 q_0 (1 + q_0)\big) + s_0}\bigg|.
\end{equation}
For the (1,1) Pad\'e approximation of $d_L(z)$, similar calculations yield
\begin{equation}
\rho_{P_{1,1}}=\dfrac{2}{1-q_0}\ ,
\end{equation}
while, in the case of the second-order Taylor series, one has
\begin{equation}
\rho_{d_{L,2}}=\dfrac{1-q_0}{2}\ .
\end{equation}
The proper procedure should make use of fitting results over the cosmographic coefficients to compute $\rho_{R_{1,1}}, \rho_{P_{1,1}}$ and $\rho_{d_{L,2}}$. However, an immediate check can be done assuming the reference values given by \Cref{eq:cosmog param LCDM},  in which case one finds
\begin{equation}
\left\{
\begin{aligned}
\rho_{R_{1,1}}&=1.014\ , \\
\rho_{P_{1,1}}&= 1.290\ ,\\
\rho_{d_{L,2}}&=0.775\ .
\end{aligned}
\right.
\end{equation}
The above results demonstrate the improvements obtained in the case of rational polynomials.
In \Cref{fig:radii} we show the convergence radii for Taylor, Pad\'e and Chebyshev polynomials using a different calibration with respect to the concordance paradigm.

\subsection{Observational constraints}

\begin{figure}[h]
\begin{center}
\includegraphics[width=1\textwidth]{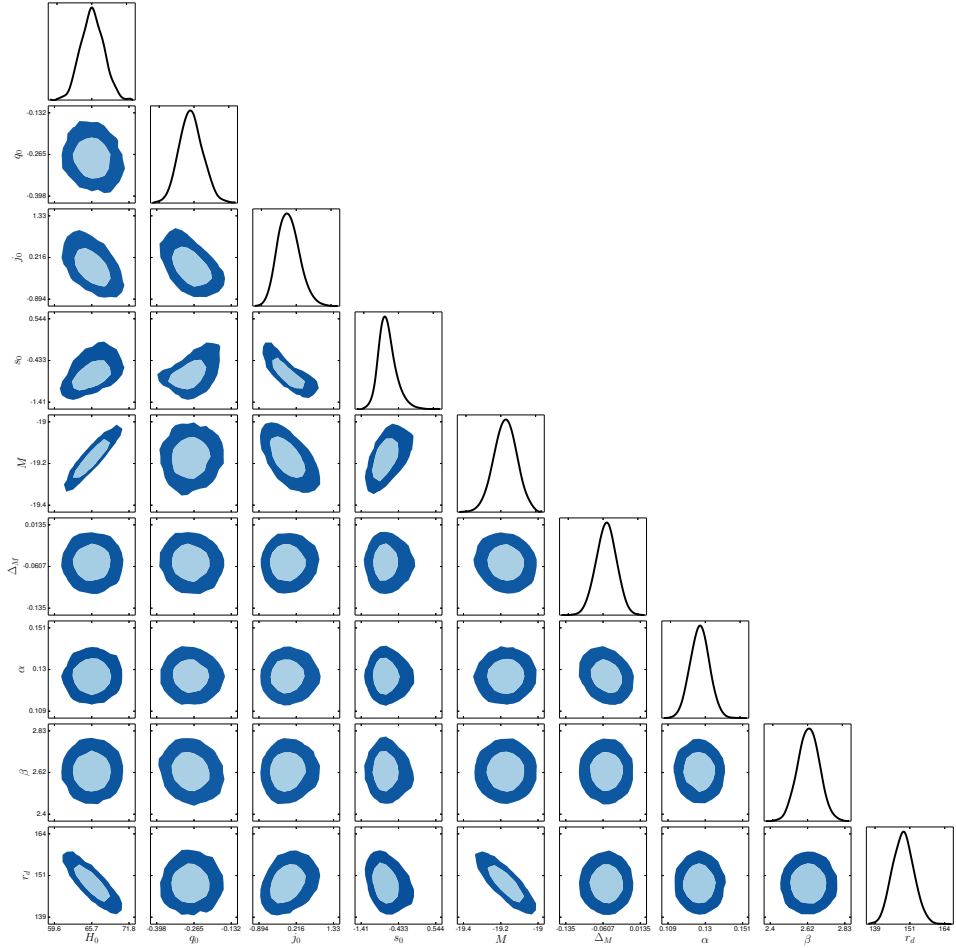}
\caption{1$\sigma$ and 2$\sigma$ confidence level contours and posterior distributions inferred from the MCMC analysis by combining SN+OHD+BAO data surveys. The results have been obtained for fourth-order Taylor approximation of $d_L$. The units of $H_0$ are Km/s/Mpc, whereas $r_d$ in Mpc.}
\label{fig:Taylor}
\end{center}
\end{figure}

\begin{figure}[h]
\begin{center}
\includegraphics[width=1\textwidth]{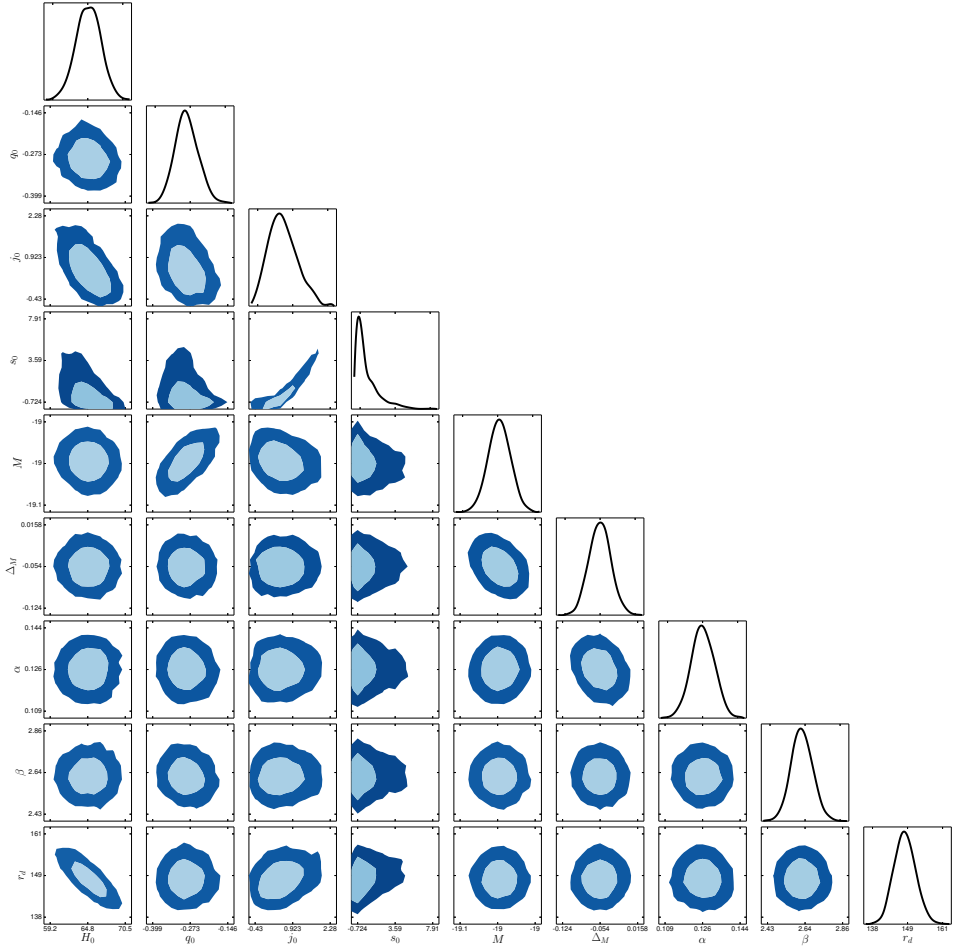}
\caption{68\% and 95\% confidence levels and corresponding contours with posterior distributions determined from the MCMC analysis. Here, we considered a combined SN+OHD+BAO survey for the (2,2) Pad\'e approximation of $d_L$. $H_0$ is written in Km/s/Mpc, while $r_d$ in Mpc.}
\label{fig:Pade}
\end{center}
\end{figure}

\begin{figure}[h]
\begin{center}
\includegraphics[width=1\textwidth]{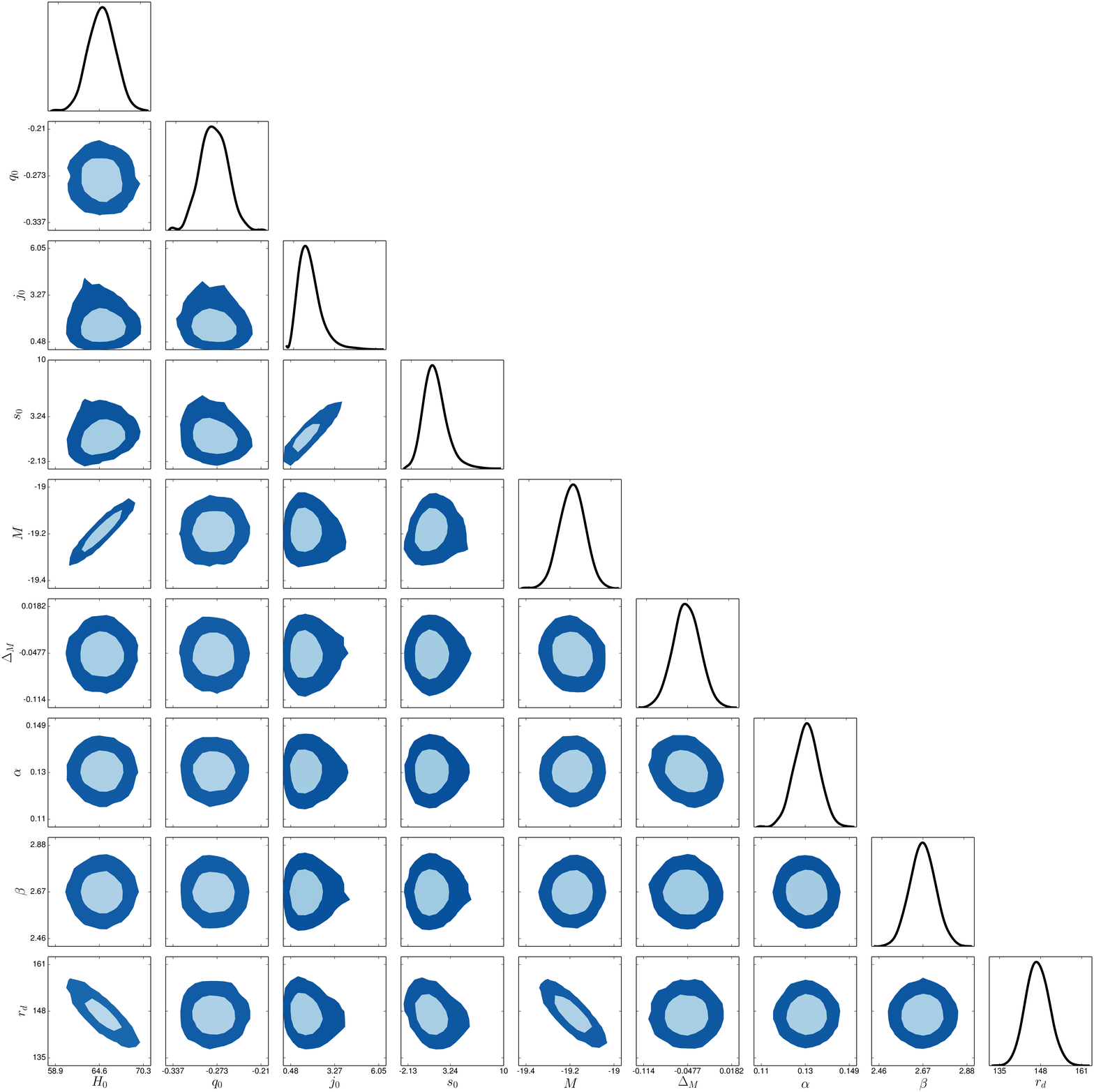}
\caption{Contours and posterior distributions for 68\% and 95\% confidence levels. We got these plots from the MCMC analysis of the whole SN+OHD+BAO data set in the case of (2,1) rational Chebyshev approximation of $d_L$, with  $H_0$ in Km/s/Mpc, and $r_d$ in Mpc.}
\label{fig:Chebyshev}
\end{center}
\end{figure}

In \refcite{rocco3} we tested the new method of rational Chebyshev polynomials against other cosmographic approaches by performing a Markov Chain Monte Carlo (MCMC) integration on the combined likelihood of the SN Ia JLA data \cite{Betoule14}, and other low-redshift measurements such as the Observational Hubble data \cite{Jimenez02} (OHD) and Baryon Acoustic Oscillations \cite{Eisenstein05} (BAO) (see \Cref{tab:OHD} and \Cref{tab:BAO}).
Assuming the uniform priors listed in \Cref{tab:priors}, we show in \Cref{tab:joint fit} the results of the joint analysis obtained through the Metropolis numerical algorithm implemented by the Monte Python code \cite{MontePython}.
We also show, in \Cref{fig:Taylor}, \Cref{fig:Pade}, and \Cref{fig:Chebyshev}, the marginalized contours and posterior distributions for the different cosmographic techniques. As shown in \Cref{tab:errors}, the relative uncertainties for the cosmographic parameters are clearly reduced in the case of rational Chebyshev polynomials compared to the other approximation methods.

An interesting fact to note is that, by construction, one uses Chebyshev polynomials with lower orders than Taylor series and Pad\'e approximants. This mostly reduces the computational difficulties in implementing cosmic data, although does not accurately fixes the highest-order parameter  in the approximation. This is the case of $s_0$ whose error bars are not significantly improved adopting Chebyshev polynomials. To overcome this issue, it would be enough to increase the Chebyshev order to better fix $s_0$ than Taylor and Pad\'e treatments.

\begin{table}
\small
\renewcommand{\arraystretch}{1.5}
\tbl{Parameter priors used for MCMC, with $H_0$ in units of Km/s/Mpc and $r_d$ in units of Mpc.}
{\begin{tabular}{c c  }
\hline
\hline
Parameter & Prior \\
\hline
$H_0$  & $(50,90)$ \\
$q_0$ & $(-10,10)$ \\
$j_0$ & $(-10,10)$\\
$s_0$ & $(-10,10)$\\
$M$ & $(-20,-18)$ \\
$\Delta_M$  & $(-1,1)$ \\
$\alpha$  & $(0,1)$ \\
$\beta$  & $(0,5)$ \\
$r_d$  & $(140,160)$ \\
\hline
\hline
\end{tabular}
 \label{tab:priors}
 }
\end{table}

\begin{table}
\small
\setlength{\tabcolsep}{0.5em}
\renewcommand{\arraystretch}{1.5}
\tbl{1 and 2 $\sigma$ confidence level got from the MCMC analysis using SN+OHD+BAO data surveys in the case of fourth-order Taylor, (2,2) Pad\'{e} and (2,1) rational Chebyshev polynomial approximations of $d_L$.}
{\begin{tabular}{c| c c c| c c c| c c c }
\hline
\hline
\multirow{2}{*}{Parameter} & \multicolumn{3}{c|}{Taylor} &  \multicolumn{3}{c|}{Pad\'e} &  \multicolumn{3}{c}{Rational Chebyshev} \vspace{-0.2cm} \\
& \footnotesize Mean & \footnotesize $1\sigma$ & \footnotesize$2\sigma$ & \footnotesize Mean & \footnotesize $1\sigma$ & \footnotesize $2\sigma$  & \footnotesize Mean & \footnotesize $1\sigma$ & \footnotesize $2\sigma$ \\
\hline
$H_0$ & $65.80$ & $^{+2.09}_{-2.11}$ & $^{+4.22}_{-4.00} $  & $64.94$ & $^{+2.11}_{-2.02}$ & $^{+4.12}_{-4.13}$ & $64.95$ & $^{+1.89}_{-1.94}$ & $^{+3.77}_{-3.77}$ \\
$q_0$ & $-0.276$ & $^{+0.043}_{-0.049}$ & $^{+0.093}_{-0.091}$ & $-0.285$ & $^{+0.040}_{-0.046}$ & $^{+0.087}_{-0.084}$  & $-0.278$ & $^{+0.021}_{-0.021}$ & $^{+0.041}_{-0.042}$ \\
$j_0$ & $-0.023$ & $^{+0.317}_{-0.397}$ & $^{+0.748}_{-0.685}$ & $0.545$ & $^{+0.463}_{-0.652}$ & $^{+1.135}_{-1.025}$ &  $1.585$ & $^{+0.497}_{-0.914}$ & $^{+1.594}_{-1.453}$ \\
$s_0$ & $-0.745$ & $^{+0.196}_{-0.284}$ & $^{+0.564}_{-0.487}$ & $0.118$ & $^{+0.451}_{-1.600}$ & $^{+3.422}_{-1.921}$ & $1.041$ & $^{+1.183}_{-1.784}$ & $^{+3.388}_{-3.087}$ \\
$M$ &  $-19.16$ & $^{+0.07}_{-0.07}$ & $^{+0.14}_{-0.14}$ & $-19.03$ & $^{+0.02}_{-0.02}$ & $^{+0.05}_{-0.05}$ & $-19.17$ & $^{+0.07}_{-0.07}$ & $^{+0.13}_{-0.13}$ \\
$\Delta_M$  & $-0.054$ & $^{+0.023}_{-0.022}$ & $^{+0.044}_{-0.045}$ & $-0.054$ & $^{+0.022}_{-0.023}$ & $^{+0.045}_{-0.045}$ & $-0.050$ & $^{+0.022}_{-0.022}$ & $^{+0.044}_{-0.045}$  \\
$\alpha$  & $0.127$ & $^{+0.006}_{-0.006}$ & $^{+0.012}_{-0.012}$ & $0.127$ & $^{+0.006}_{-0.006}$ & $^{+0.012}_{-0.012}$  & $0.130$ & $^{+0.006}_{-0.006}$ & $^{+0.012}_{-0.012}$  \\
$\beta$  & $2.624$ & $^{+0.071}_{-0.068}$ & $^{+0.136}_{-0.140}$ & $2.625$ & $^{+0.065}_{-0.069}$ & $^{+0.137}_{-0.135}$ &  $2.667$ & $^{+0.068}_{-0.069}$ & $^{+0.137}_{-0.135}$ \\
$r_d$ & $149.2$ & $^{+3.7}_{-4.1}$ & $^{+7.7}_{-7.5}$  & $148.6$ & $^{+3.5}_{-3.8}$ & $^{+7.5}_{-7.1}$ & $147.2$ & $^{+3.7}_{-4.0}$ & $^{+7.8}_{-7.5}$ \\
\hline
\hline
\end{tabular}
 \label{tab:joint fit}
 }
\end{table}

\begin{table}
\small
\setlength{\tabcolsep}{1.2em}
\renewcommand{\arraystretch}{1.5}
\tbl{68\% and 95\% errors on the cosmographic outputs got by MCMC analysis in which we used SN+OHD+BAO data in the case of fourth-order Taylor, (2,2) Pad\'{e} and (2,1) rational Chebyshev polynomial approximations of $d_L$.}
{\begin{tabular}{c| c c| c c| c c }
\hline
\hline
\multirow{2}{*}{Parameter} & \multicolumn{2}{c|}{Taylor}  & \multicolumn{2}{c|}{Pad\'e} & \multicolumn{2}{c}{Rational Chebyshev}  \vspace{-0.2cm} \\
&\footnotesize $1\sigma$ & \footnotesize $2\sigma$ &  \footnotesize $1\sigma$ &   \footnotesize $2\sigma$  &   \footnotesize $1\sigma$ &  \footnotesize $2\sigma$\\
\hline
$H_0$ & $3.19\%$ & $6.25\%$  & $3.17\%$ & $6.35\%$ & $2.95\%$ & $4.11\%$ \\
$q_0$ & $16.8\%$ & $33.5\%$ & $15.1\%$  & $ 30.1\%$ & $7.66\%$ & $14.8\%$ \\
$j_0$ & $1534\%$ & $3079\%$ & $102\%$ & $198\%$ & $44.5\%$ &	$96.1\%$ \\
$s_0$ & $32.2\%$ & $70.5\%$ & $866\%$ & $2258\%$ & $142\%$ & $311\%$	\\
\hline
\hline
\end{tabular}
 \label{tab:errors}
}
\end{table}

\subsection{The $ E_i$s method}

In this subsection it is relevant to cite a possible approach which consists in testing cosmography with the Hubble parameter, without making use of rational approximations. Formally, the Hubble function in Taylor series around $z=0$ is
\begin{equation} \label{Eofz}
 E(z) \equiv \frac{H(z)}{H_0} = \sum_i \frac{1}{i!}E_i z^i
\end{equation}
with $E_i = H^{(i)}(z)/H_0 |_{z=0}$. Hereafter we baptize with \emph{eis} coefficients the first four terms in the expansions, which read:
\begin{eqnarray} \label{eisOfsf}
 E_0 &=& 1, \nonumber\\ E_1 &=& 1 + q_0,  \nonumber\\ E_2 &=& -q_0^2 + j_0, \\
   E_3 &=& 3 q_0^2 + 3 q_0^3 -j_0 (4 q_0+3)-s_0. \nonumber
\end{eqnarray}
To reduce systematics, a possible trick is to use directly the Taylor expansion of $H(z)$ \cite{orlH} within the comoving distance $ \eta(z) =  \int_{0}^{z} \frac{dz'}{H(z')}$ and integrate numerically to obtain the luminosity distance. Details of numerical simulations and strategies have been reported in \cite{EIS}, in which the estimation of the \emph{eis} parameters through a hierarchical manner has been performed by:

\begin{equation} \label{EisModel}
   \tilde{d}_L(z;E_1,E_2,E_3) = \begin{cases}
               \tilde{d}^{(1)}_L(z)               & z< z_{low}\\
               \tilde{d}^{(2)}_L(z)               & z_{low}< z < z_{mid}\\
               \tilde{d}^{(3)}_L(z)               & z_{mid}< z < z_{high}.
           \end{cases}
\end{equation}
A good choice for redshift cut-offs is
\begin{equation}\label{redshiftscuts}
z_{low} = 0.05, \qquad z_{mid} = 0.4, \qquad z_{high} = 0.9.
\end{equation}
With simple considerations, adopting binning procedure for Ei's and standard cosmography, one can mix the two approach to reduce significantly the error propagations of every cosmographic analysis.
As a genuine example, a module for the code CosmoMC \cite{Lewis:2002ah} to draw the likelihood distributions for all the methods is available at
\href{https://github.com/alejandroaviles/EisCosmography}{https://github.com/alejandroaviles/EisCosmography}; also, all the
simulated catalogs and further statistics can be found there.

\begin{figure}
\begin{center}
\includegraphics[width=3.2in]{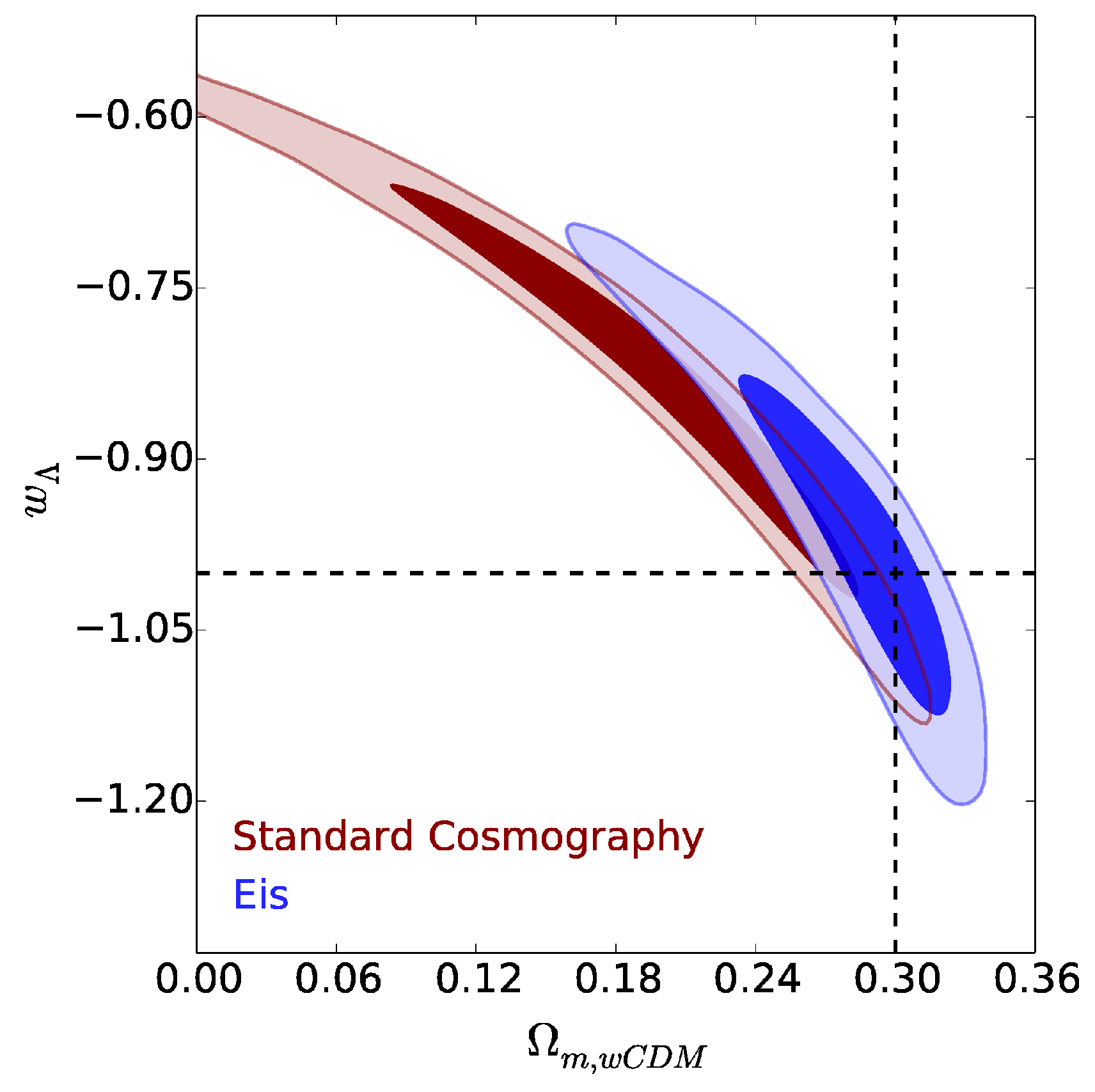}
\caption{2D confidence regions for $\Omega_m$ and $w$, as a pure example for comparing standard cosmography and Eis procedures. It is possible to notice that standard cosmography is unpredictive at $2\sigma$.}
\label{qaz}
\end{center}
\end{figure}

A simple representation of the improvements of the method can be found in Fig. \eqref{qaz}

The need of additional methods, different from standard approaches, is essential to overcome the broadening of coefficients due to systematics and error propagation. Often this problem has been imputed to the lacking of cosmic data. However, discussions over this issue are still open \cite{cruzless}. An intriguing open challenge would be \emph{unifying} the Ei's method with rational approximations, either with rational approximations or in the framework of extended and/or modified theories of gravity.

\section{Model-independent reconstruction of $f(R)$ gravity}
\label{sec:f(R) cosmography}

The above cosmographic analysis can be adopted to reconstruct  dark energy  models deriving the functional forms of Lagrangians from observational data. The method can be considered as a sort of \textit{Inverse Scattering} to approach the cosmological problem.
In this section, we  reconstruct, from cosmography,  the $f(R)$ gravity action both in the metric and in the Palatini formalisms, without postulating any specific functional form.
A standard procedure in the $f(R)$ studies consists of assuming the gravity action and then finding out the dynamics  by solving the modified Friedmann equations. The standard  approach relies on postulating the form of $f(R)$ \textit{a priori}, which determines the cosmological model. In what follows, instead, we present a model-independent method  to reconstruct the functional form of the action.
To this end, we use rational polynomials to obtain accurate cosmographic approximations of the luminosity distance up to high redshifts. We shall study the late-time expansion history of the universe and discuss possible departures from GR and then the  $\Lambda$CDM model.

\subsection{The metric formalism case}

\begin{figure}[h!]
\begin{center}
\includegraphics[width=0.7\textwidth]{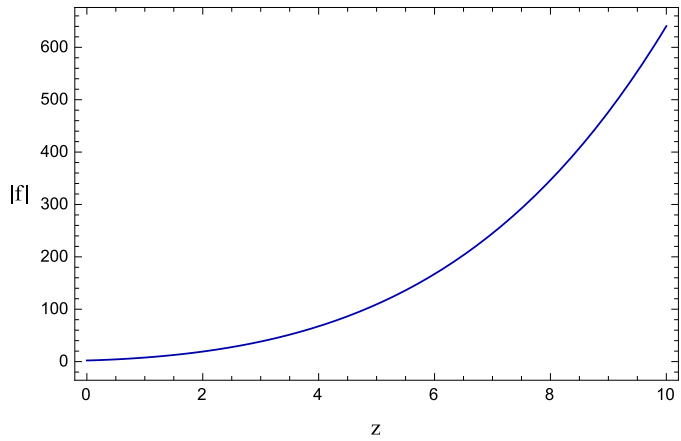}
\caption{Plot of the numerical shape of $|f(z)|$ approximated by using the (2,1) Pad{\'e} polynomial.}
\label{fig:f(z) Pade}
\end{center}
\end{figure}

Let us start  with the model-independent reconstruction of the $f(R)$ action in the metric formalism \cite{rocco4}.
The determination of $f(R)$ through the match with cosmic data has been subject of a wide discussion in the last years \cite{Hu07,Starobinski07,Cognola08}.
In particular, the method of Taylor-expanding $f(R)$ for $R$ approaching its late-time values is limited by the short range of redshift characteristic of observational data. Also, the truncation of the Taylor polynomial reproducing the  $f(R)$ function unavoidably introduces errors in the analysis.
In this respect, Pad\'e polynomials may offer a possible solution to the convergence problem.
Motivated by the results already obtained \cite{Aviles14}, let us consider the (2,1) Pad{\'e} approximation of the Hubble rate:
\begin{align}
H_{21}(z)=\ &\Big[2H_0(1 + z)^2 \big(3 + z + j_0 z - q_0 (3 + z + 3 q_0 z)\big)^2 \Big]\times \Big[18 (q_0 - 1)^2 + 6 (q_0 - 1)\nonumber \\
& \big(-5 - 2 j_0 + q_0 (8 + 3 q_0)\big) z+\Big(14 + 2 j_0^2 + j_0 \big(7 - q_0 (10 + 9 q_0)\big)+   q_0 \big(-40  \nonumber \\
& + q_0 (17 + 9 q_0 (2 + q_0))\big)\Big) z^2\Big]^{-1}\,.
\label{eq:H21}
\end{align}
We note that that $H_{21}(z)$ is expressed in terms of the cosmographic series up to the jerk parameter, whereas the third-order Taylor approximation contains also the snap parameter.
To apply our strategy, we first convert the time derivatives and the derivatives with respect to $R$ into derivatives with respect to $z$ according to the prescription
\begin{align}
\dfrac{d\mathfrak{F}}{dt}&=-(1+z)H\mathfrak{F}_z\,,	\label{f(R):dt-dz}	\\
\,\nonumber\\
\dfrac{\partial \mathfrak{F}}{\partial R}&=\dfrac{1}{6} \Big[(1 + z)H_z^2 +H\left(-3H_z +(1+z)H_{zz}\right)\Big]^{-1}\mathfrak{F}_z\ ,  \label{f(R):dR-dz}
\end{align}
where $\mathfrak{F}(z)$ is an arbitrary function and we denote derivatives with respect to the redshift by the subscripts `$z$'.
Then, after determining the values of the cosmographic parameters, one can combine \Cref{f(R):first Friedmann metrico} and \Cref{f(R):density curv metrico} and use \Cref{eq:R-H}.
This provides us with the following second-order differential equation for $f(z)$:
\begin{align}
H^2f_z&=\Big[-(1+z)H_z^2+H\big(3H_z-(1+z)H_{zz}\big)\Big]\Bigg[-6 H_0^2 (1 + z)^3 \Omega_{m0} - f \nonumber \\
& -\dfrac{Hf_z \left(2 H - (1 + z) H_z\right)}{(1 + z)H_z^2 +H\left(-3H_z+ (1 + z)H_{zz}^2\right)}-\dfrac{f_{zz}\big((1 +z)H_z^2 + H (-3H_z + (1 + z)H_{zz})\big)}{\big[(1+z)H_z^2+H\big(-3H_z+(1+z)H_{zz}\big)\big]^2} \nonumber \\
&\times (1 + z) H^2-\dfrac{(1 + z) H^2\Big(f_z\big(2H_z^2-3(1+z)H_zH_{zz}+H(2H_{zz}-(1+z)H_{zzz})\big)\Big) }{{\big[(1+z)H_z^2+H\big(-3H_z+(1+z)H_{zz}\big)\big]}^2} \Bigg]\,.
\label{f(R):f(z)}
\end{align}
The initial conditions needed to solve the above equation can be obtained by combining the condition $f'(R_0)=1$ together with evaluating  \Cref{f(R):density curv metrico,f(R):pressure curv metrico,f(R):first Friedmann metrico} at the present time:
\begin{align}
&f_0=R_0+6H_0^2(\Omega_{m0}-1)\ , \label{f(R):f0}\\
&f_z\big|_{z=0}=R_z\big|_{z=0}\ . \label{f(R):fp0}
\end{align}
In what follows, we fix $\Omega_{m0}=0.3$. Regarding the cosmographic parameters, for the (2,1) Pad{\'e} approximation, we use the results found in \refcite{Aviles14}:
\begin{equation}
\left\{
\begin{aligned}
&h=0.7064^{+0.0277}_{-0.0263}\,, \\
&q_0=-0.4712^{+0.1224}_{-0.1106}\,,\\
&j_0=0.593^{+0.216}_{-0.210}\,,\\
\end{aligned}
\right .
\label{eq:cosm param Pade}
\end{equation}
and for the third-order Taylor expansion:
\begin{equation}
\left\{
\begin{aligned}
&h=0.7253^{+0.0353}_{-0.0351}\,,\\
&q_0=-0.6642^{+0.2050}_{-0.1963}\,,\\
&j_0=1.223^{+0.644}_{-0.664}\,,\\
&s_0=0.394^{+1.335}_{-0.731}\,.
\end{aligned}
\right .
\label{eq:cosmogr param Taylor}
\end{equation}
Hence, we can reconstruct $f(z)$ numerically by inserting \Cref{eq:H21} into \Cref{eq:f(z)}.
The $f(R)$ function resulting from the reconstruction procedure will be negative due to the metric signature adopted in the analysis.
Consistently, $f(z)$ must be a negative function as for the case of upper bound results of (\ref{eq:cosm param Pade}).
In \Cref{fig:f(z) Pade} we show the numerical reconstruction of $f(z)$ for the (2,1) Pad{\'e} approximation.

\begin{figure}[h!]
\begin{center}
\includegraphics[width=0.7\textwidth]{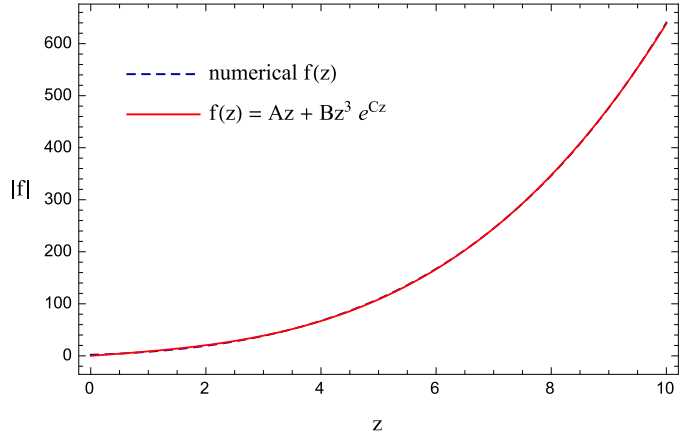}
\caption{Comparison between $|f(z)|$ and the functional form \Cref{f(R):approx f} provided by (2,1) Pad{\'e} approximation.}
\label{fig:approx f(z)}
\end{center}
\end{figure}

To find the analytical match for $f(z)$, we considered the following test-functions with three free constant coefficients ($\mathcal{A}$, $\mathcal{B}$, $\mathcal{C}$):
\begin{subequations}
\begin{align}
Exponential\nonumber\\
f_1(z)&=\mathcal A z+\mathcal B z^3 e^{\mathcal C z}	               \label{f(R):exp}\\
f_2(z)&=\mathcal A+\mathcal B z^2 \sinh(1+\mathcal C z)				\label{f(R):sinh}\\
f_3(z)&=\mathcal A z+\mathcal B z^3 \cosh(\mathcal C z)		        \label{f(R):cosh}\\
f_4(z)&=\mathcal Az^2+\mathcal B z^4\tanh(\mathcal C z)	            \label{f(R):tanh}\\
Trigonometric\nonumber\\
f_5(z)&=\mathcal Az^3+\mathcal B z^5 \sin(1+\mathcal C z)         \label{f(R):sin}\\
f_6(z)&=\mathcal Az^3+\mathcal B z^4 \cos(1+ \mathcal C z)                           \label{f(R):cos}\\
f_7(z)&=\mathcal Az+\mathcal B z^2 \tan(\mathcal C z)	                \label{f(R):tan}\\
Logarithmic\nonumber\\
f_8(z)&=\mathcal Az+\mathcal B z^3 \ln(1+\mathcal C z) 		       \label{f(R):ln}
\end{align}
\end{subequations}
Then, we perform  the $\mathcal{F}$-\textit{statistics} \cite{James13}:
\begin{equation}
\mathcal F=\dfrac{(\text{TSS}-\text{RSS})/p}{\text{RSS}/(n-p-1)}\,,
\label{eq:F-statistics}
\end{equation}
where
\begin{align}
&\text{TSS}=\sum_{i=1}^{n}(y_i-\bar{y})^2\,, \\
&\text{RSS}=\sum_{i=1}^n(y_i-\hat{y}_i)^2\,,
\end{align}
and
\begin{equation}
\bar{y}=\dfrac{1}{n}\sum_{i=1}^n y_i\,.
\end{equation}
Here, $y_i$ are the observed value while $\hat{y}_i$ are the values predicted by the model;  $n$ is the number of observations and $p$ the number of predictors.
The goodness of the model is tested by comparing the null hypothesis (the explanatory power of the model is null as all the regression coefficients are zero) with the case in which there exists at least one non-zero regression coefficient.
While in other tests such as $t$-\textit{statistics} and $p$-\textit{value} the goodness of the model is measured by looking for any association between the individual variables and the response, in the $\mathcal{F}$-\textit{statistics} the model is tested through the joint explanatory power of its predictors. For large $p$ it may happen, in fact, that the $p$-\textit{values} are small even when there is no real association between the predictors and the response.
Furthermore, the advantage of the $\mathcal{F}$-\textit{statistics} with respect to $\mathcal{R}^2$-\textit{test}\footnote{It is worth noticing that, a part the abuse of notation,  $\mathcal{R}^2$ is not the $\mathcal{R}$ scalar curvature of Palatini formalism.} relies on the presence of the number of predictors. Adding more predictors to the model makes $\mathcal{R}^2$ always increase, even if the association between those variables and the response is weak.
It is actually possible to express the $\mathcal{F}$-\textit{statistics} in terms of $\mathcal{R}^2$ as
\begin{equation}
\mathcal{F}=\dfrac{\mathcal{R}^2/p}{(1-\mathcal{R}^2)/(n-p-1)}\,.
\end{equation}
The higher the values of $\mathcal{F}$, the higher the evidence against the null hypothesis, for which we expect very small  $\mathcal{R}^2$ and $\mathcal{F}$.
\begin{table}[h!]
\small
\setlength{\tabcolsep}{1.5em}
\renewcommand{\arraystretch}{1.5}
\tbl{$\mathcal{F}$-\textit{statistics} on \Crefrange{f(R):exp}{f(R):ln} for the (2,1) Pad{\'e} approximation.}
{\begin{tabular}{c|c|c}
\hline
\hline
Test-function & $(\mathcal A, \mathcal B, \mathcal C)$ & $\mathcal F (\times 10^6)$  \\
\hline
$f_1(z)$ & $(-8.078, -0.530, 0.005)$  &  $31.7 $ \\
$f_2(z)$ & $( -6.147, -2.148, 0.080)$ & $13.5$  \\
$f_3(z)$ & $(-8.046, -0.541, 0.025)$ & $3.637$  \\
$f_4(z)$ & $(-3.699,  0.027,  -562.2)$& $4.535$  \\
$f_5(z)$ & $(-0.708,  -0.001, 1.095)$ & $0.118$   \\
$f_6(z)$ & $(-0.717, -0.008,  0.)$ & $0.142$  \\
$f_7(z)$ & $(-41.30, 0.002, 1.000)$ & $0.026$   \\
$f_8(z)$ & $(-11.69, -0.208, 1.182)$ & $1.484$\\
\hline
\hline
\end{tabular}
\label{tab:F-Pade}
}
\end{table}
In this study, $p=3$ is the number of free parameters and $n=1000$ are the points generated from the numerical solution of $f(z)$.
The results shown in \Cref{tab:F-Pade} indicate that the best analytical match for $f(z)$ is
\begin{equation}
f(z)=\mathcal A z+\mathcal B z^3 e^{\mathcal C z}\,,
\label{f(R):approx f}
\end{equation}
where
\begin{equation}
(\mathcal A, \mathcal B, \mathcal C)=(-8.078, -0.530, 0.005)\,.
\label{f(R):coeff 1}
\end{equation}
In \Cref{fig:approx f(z)} we show the comparison between the numerical and the analytical solutions of $f(z)$ in the domain $z\in[0,10]$.

\begin{figure}[h!]
\begin{center}
\includegraphics[width=0.68\textwidth]{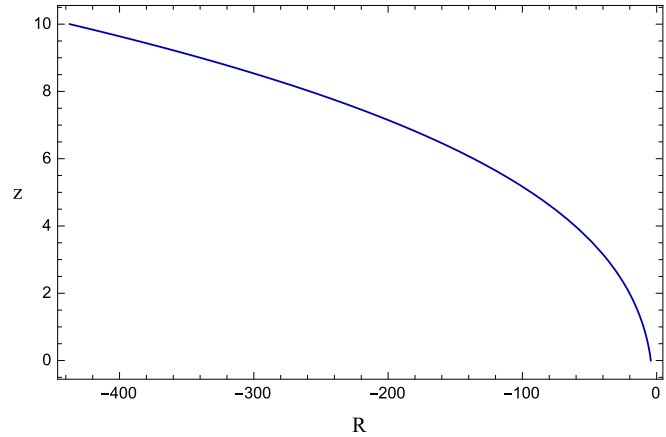}
\caption{Framing  $z(R)$ out for (2,1) Pad{\'e} polynomial.}
\label{fig:z(R)}
\end{center}
\end{figure}

\begin{figure}[h!]
\begin{center}
\includegraphics[width=0.7\textwidth]{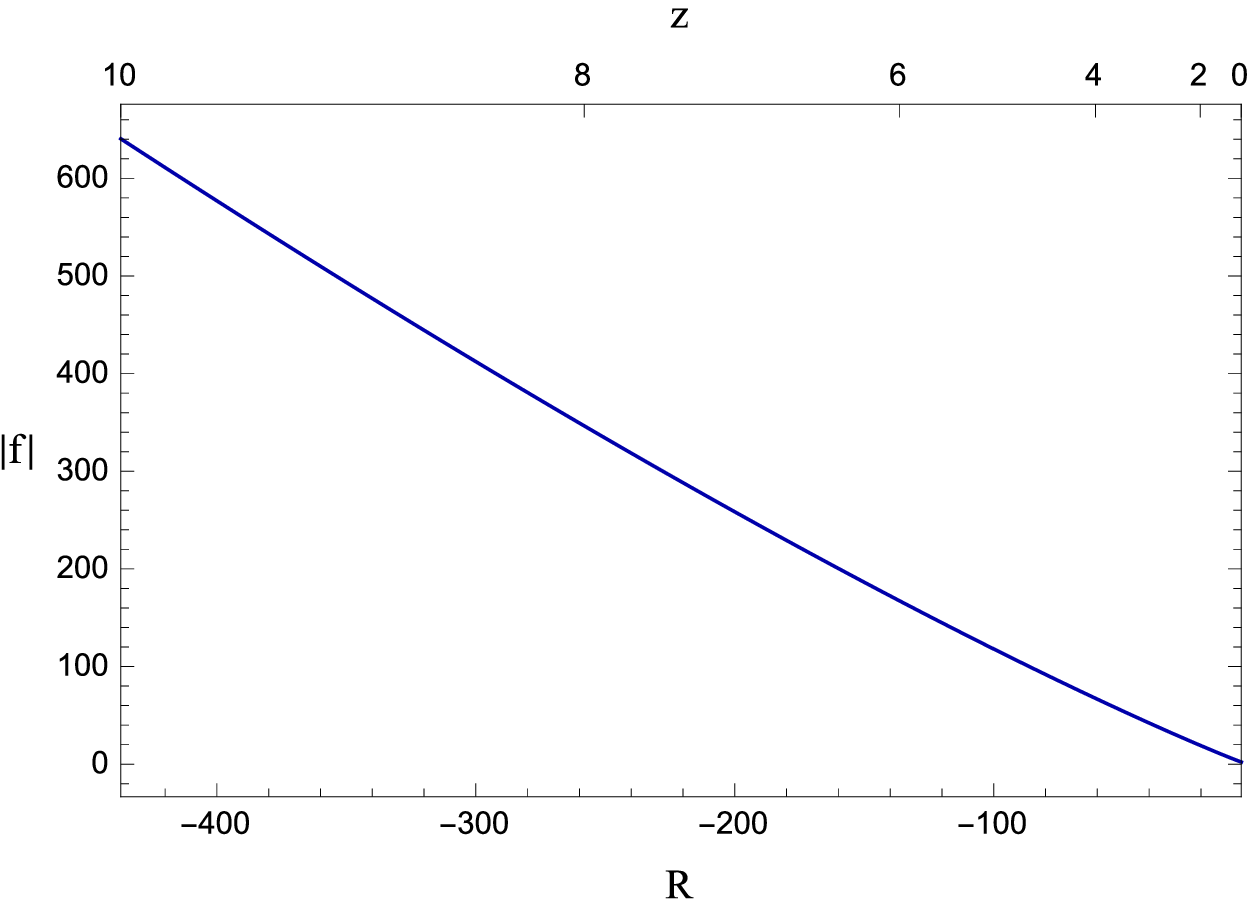}
\caption{Reconstructed $|f(R)|$ function developed in the case of (2,1) Pad{\'e} polynomial inside $z\in[0,10]$}
\label{fig:f(R)}
\end{center}
\end{figure}

To determine $f(R)$, we need to invert the function $R(z)$ and insert back into \Cref{f(R):approx f}.
This procedure can be only done numerically due to the impossibility for an analytical inversion of \Cref{eq:H21}.
Therefore, we used \Cref{eq:R-H} to find $z(R)$ (see \Cref{fig:z(R)}), which we plugged into \Cref{f(R):approx f} to finally obtain $f(R)$ (see \Cref{fig:f(R)}).

\begin{figure}[h!]
\begin{center}
\includegraphics[width=0.7\textwidth]{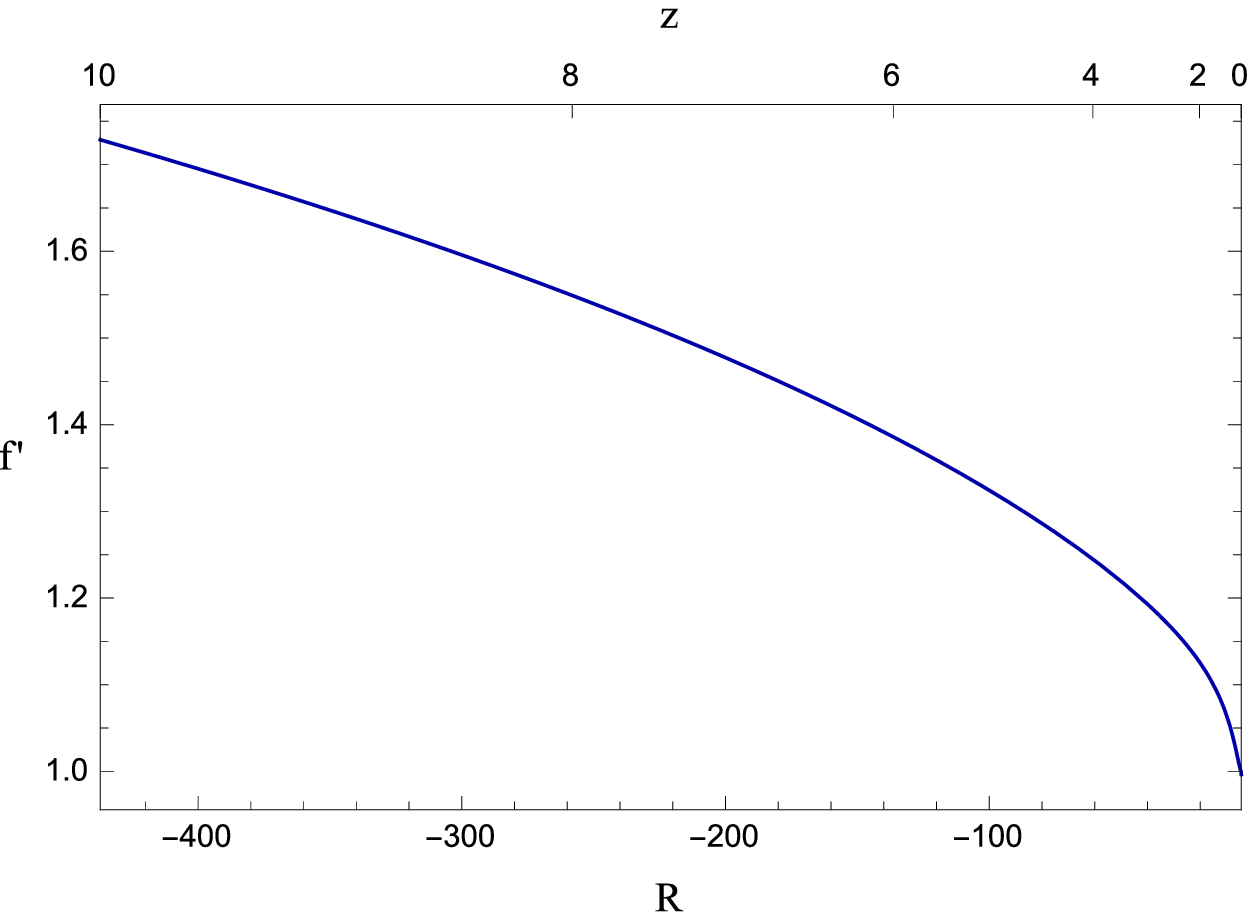}
\caption{Behaviour of $df/dR$ as byproduct of (2,1) Pad{\'e} polynomial inside $z\in[0,10]$.}
\label{fig:f'(R)}
\end{center}
\end{figure}

\begin{figure}[h!]
\begin{center}
\includegraphics[width=0.7\textwidth]{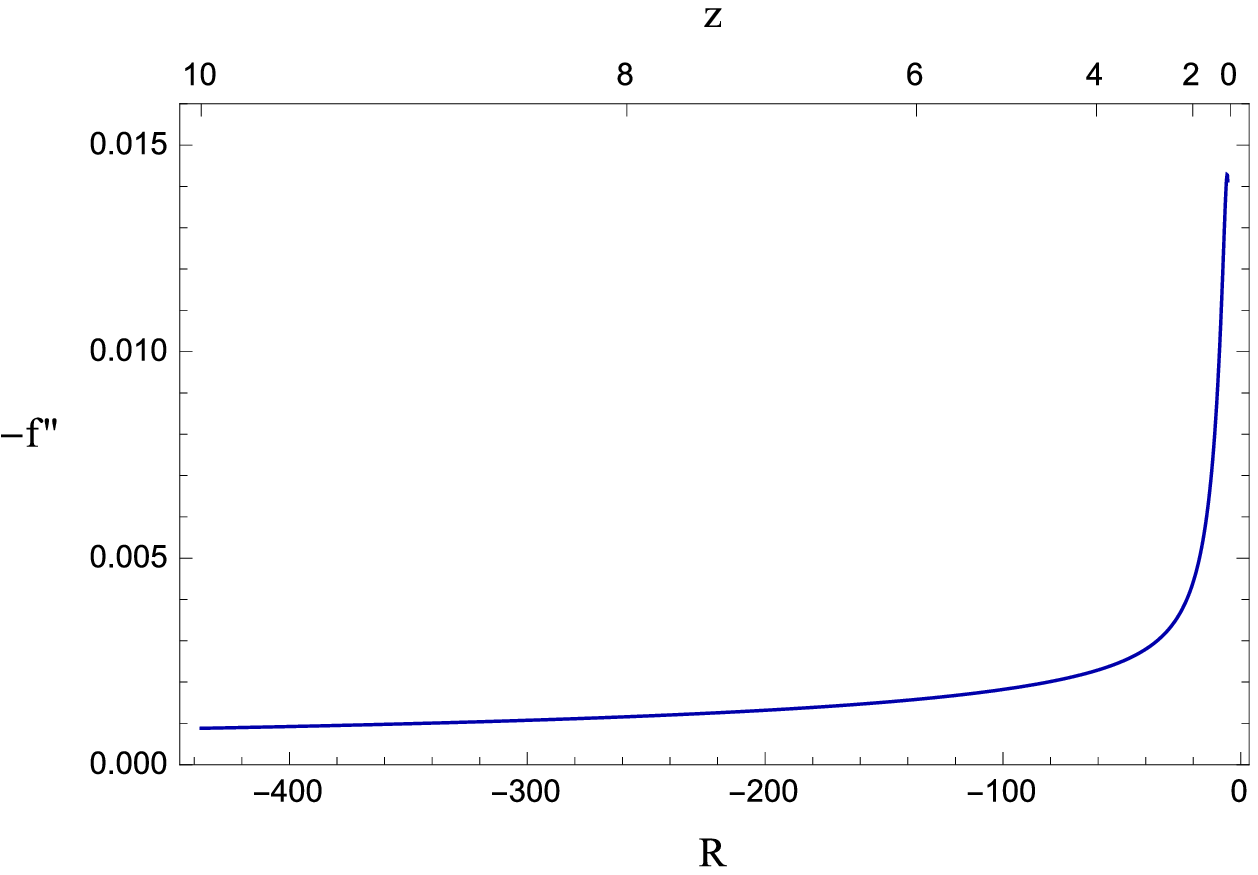}
\caption{Behaviour of $|d^2f/dR^2|$ as byproduct of (2,1) Pad{\'e} polynomial inside $z\in[0,10]$.}
\label{fig:f''(R)}
\end{center}
\end{figure}

\begin{figure}[h!]
\begin{center}
\includegraphics[width=0.7\textwidth]{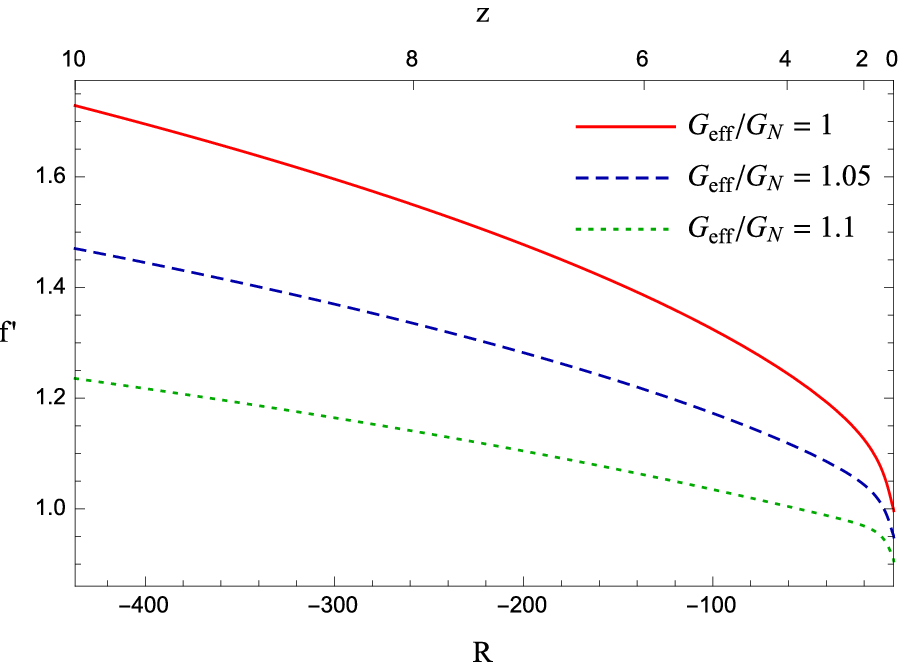}
\caption{Behaviour of $f'(R)$ for different values of $G_{eff}$.}
\label{fig:Geff}
\end{center}
\end{figure}

From \Cref{fig:f'(R)} and \Cref{fig:f''(R)} we can see that our model fulfils the viability conditions discussed in \Cref{fRmetrico}.
However, \Cref{fig:f'(R)} indicates that $f'(R)$ becomes higher than one for large curvatures, due to the condition imposed on $f'(R_0)$.
A correct asymptotic behaviour can be found by relaxing the assumption $f'(R_0)=1$,  i.e. requiring that $G_{eff}$ is slightly different from $G$, within the limits imposed by the most recent measurements \cite{Martins17}.
In light of this, one can modify \Cref{f(R):f0,f(R):fp0} as follows:
\begin{align}
&f_0=f'(R_0)(6 H_0^2 + R_0) - 6 H_0^2 \Omega_{m0}\ , \\
&f_z\big|_{z=0}=f'(R_0)\ R_z\big|_{z=0}\ .
\end{align}
In \Cref{fig:Geff} we show the results we obtain by using the above relations to find the auxiliary function $f(z)$.
Finally, it is important to stress that the asymptotic value of $f'(R)$ depends on the accuracy of the cosmographic series at high redshifts.
The predictive power and the convergence radius of the Pad\'e polynomials could be further improved by considering high-order terms, so that the difference $f'(R)_{numerical}-f'(R)_{exact}$ at large curvatures can be make smaller up to the desired level.

\begin{figure}[h!]
\begin{center}
\includegraphics[width=0.7\textwidth]{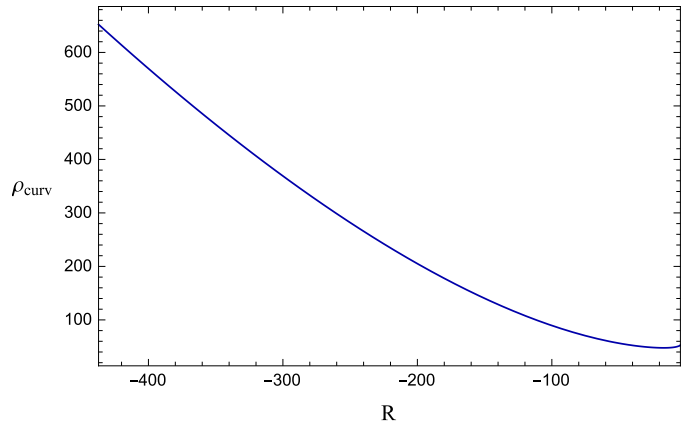}
\caption{Curvature density with Pad{\'e} approximation. Here we use $\lambda=100$.}
\label{fig:rho(R)}
\end{center}
\end{figure}

\begin{figure}[h!]
\begin{center}
\includegraphics[width=0.7\textwidth]{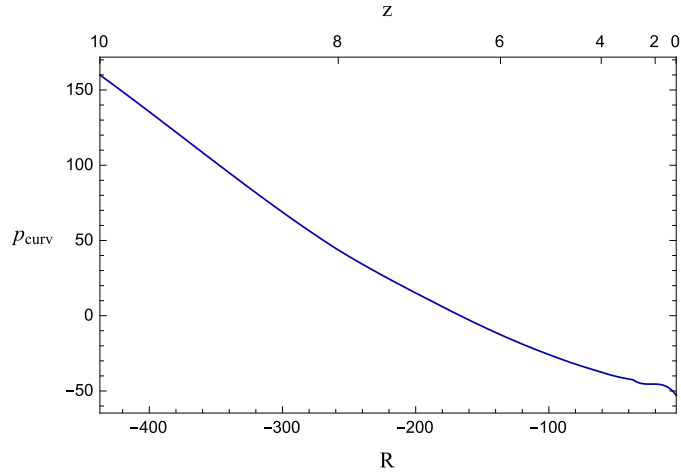}
\caption{Curvature pressure with Pad{\'e} approximation. Here we use $\lambda=100$.}
\label{fig:p(R)}
\end{center}
\end{figure}

\begin{figure}[h!]
\begin{center}
\includegraphics[width=0.7\textwidth]{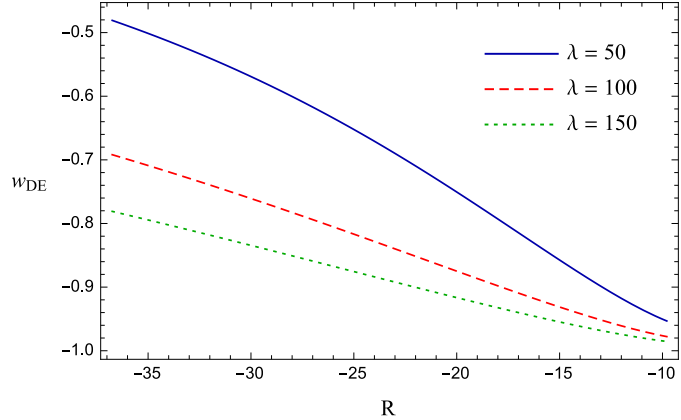}
\caption{Dark energy equation of state with different values of $\lambda$ in Pad{\'e} approximation}
\label{fig:w(R)}
\end{center}
\end{figure}

We can now study the behaviour of the dark energy equation of state $w_{DE}$ inferred from the reconstructed $f(R)$ function.
For this purpose, we rescaled \Cref{f(R):approx f} to take into account the error propagation in the numerical procedure:
\begin{equation}
f(z)\longrightarrow \lambda+f(z)\,.
\label{eq:rescale}
\end{equation}
Here, $\lambda$ does not come as vacuum energy contribution but it plays the role of a scaling constant which guaranties the matching between the numerical value of $f(z)$ at $z=0$ and the physical condition $f'(R_0)=1$.
The value of $\lambda$ can be found by imposing the condition of present acceleration:
\begin{equation}
-1\leq w_{DE}\Big|_{z=0}<-\dfrac{1}{3}\,,
\label{eq:constraint on w}
\end{equation}
which yields
\begin{equation}
\lambda \gtrsim 19.3\,.
\label{eq:constraint on lambda}
\end{equation}
The behaviours of curvature density and curvature pressure are displayed inn \Cref{fig:rho(R)} and \Cref{fig:p(R)} for an indicative value of $\lambda=100$.
\Cref{fig:w(R)} shows the effective dark energy equation of state parameter for various values of $\lambda$ according to (\ref{eq:constraint on lambda}).
\\

\begin{figure}[h!]
\begin{center}
\includegraphics[width=0.7\textwidth]{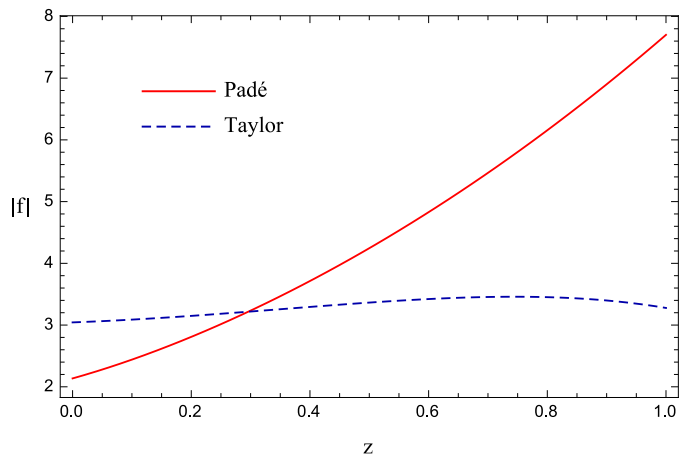}
\caption{Comparing $|f(z)|$ with (2,1) Pad{\'e} (solid red) and third-order Taylor (dashed blue) approximations.}
\label{fig:f(z) Pade vs Taylor}
\end{center}
\end{figure}

\begin{figure}[h!]
\begin{center}
\includegraphics[width=0.7\textwidth]{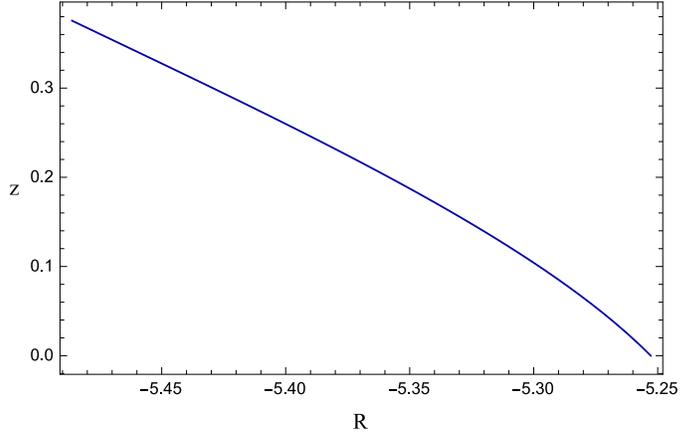}
\caption{Numerical shape of $z(R)$ third-order Taylor approximation.}
\label{fig:z(R) Taylor}
\end{center}
\end{figure}

\begin{figure}[h!]
\begin{center}
\includegraphics[width=0.7\textwidth]{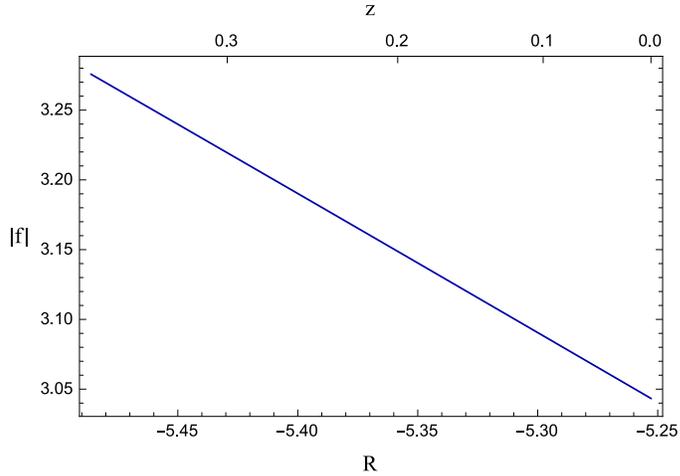}
\caption{Reconstructing $f(R)$ models with third-order Taylor approximation.}
\label{fig:f(R) Taylor}
\end{center}
\end{figure}

Finally, to better check the benefits of the analysis based on Pad{\'e} approximations with respect to the standard approach based on the Taylor series, we used \Cref{eq:H_Taylor} to solve \Cref{f(R):f(z)} with the best-fit results of (\ref{eq:cosmogr param Taylor}).
The comparison between the two methods are shown in \Cref{fig:f(z) Pade vs Taylor}, from which we see that the Taylor approach is no longer predictive at $z\gtrsim 0.3$.
Inverting numerically \Cref{eq:H_Taylor} with the use of \Cref{eq:R-H} gives $z(R)$ (see \Cref{fig:z(R) Taylor}), which we inserted into $f(z)$ to find the $f(R)$ function in the case of the Taylor approximation (see \Cref{fig:f(R) Taylor}).


\begin{figure}[h!]
\begin{center}
\includegraphics[width=0.8\textwidth]{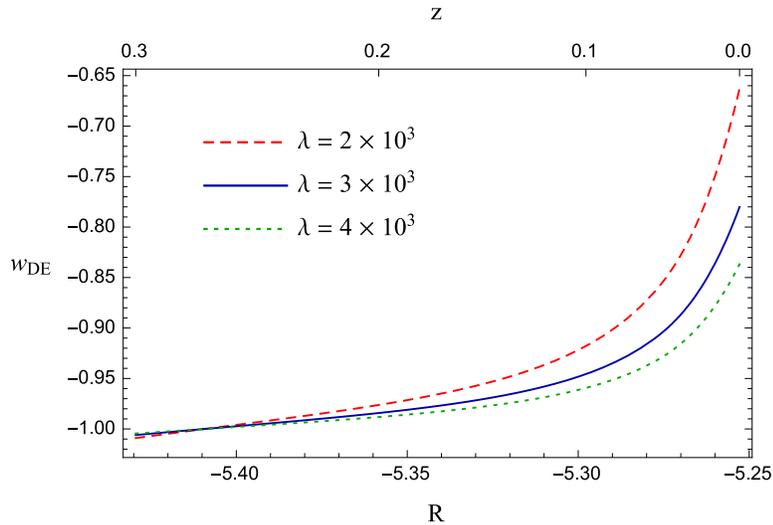}
\caption{Dark energy barotropic factor for third-order Taylor approximation with several values of  $\lambda$.}
\label{fig:w_Taylor}
\end{center}
\end{figure}

We shall now study the dark energy equation of state parameter for the Taylor approach and compare it with the results of the Pad{\'e} approximation.
Applying to the Taylor approximation the rescaling (\ref{eq:rescale}) and the condition (\ref{eq:constraint on w}), one gets
\begin{equation}
\lambda \gtrsim 1196\,.
\label{eq:constraint on lambda 2}
\end{equation}
The dark energy equation of state parameter shown in \Cref{fig:w_Taylor} experiences a the phantom-line crossing at $z\sim 0.3$ . This confirms the problems of the Taylor approach to account for high-redshift observations.

 At this point, some important remarks are in order. As discussed in \refcite{Diego}, a cosmological reconstruction scheme for $f(R)$ gravity can be  developed in terms of e-folding (or, redshift). In such an approach  FLRW cosmology emerges from specific $f(R)$ models. The application of this scheme allows a viable  unification of  inflation with dark energy bypassing the shortcoming related to the Taylor series adopted for the luminosity distance. The  reconstruction scheme may be generalized in  presence of scalar fields \cite{Diego}. By reconstruction techniques applied to $f(R)$ gravity,    the transition from matter dominated epoch to dark energy universe can be also achieved \cite{Troisi,transition}. This fact is extremely relevant in order to obtain viable cosmological models.

\subsection{The Palatini formalism case}

This section is dedicated to the reconstruction of $f(\cal{R})$ cosmology within the Palatini formalism.
 Rational polynomials \cite{rocco5}, such as Pad\'e and ratios of Chebyshev polynomials, can be used to provide accurate information on the thermal properties of the effective cosmic fluid entering the energy-momentum tensor of Palatini's gravity.

In this case, to obtain the cosmological solutions we consider the metric
\begin{equation}
ds^2=-dt^2+a(t)^2\delta_{ij}dx^idx^j\ ,
\end{equation}
and the energy-momentum tensor for a perfect fluid of density $\rho$ and pressure $p$ given as
\begin{equation}
T_{\mu\nu}=\text{diag}(-\rho,p,p,p)\ .
\end{equation}
We thus write down the relation between the Ricci scalar and the Hubble rate in the metric formalism:
\begin{equation}
R=6(\dot{H}+2H^2)\ .
\label{eq:R-H metric}
\end{equation}
Converting the time derivative into derivative with respect to the redshift, we get
\begin{equation}
R(z)=-6(1+z)H(z)H'(z) +12H(z)^2\ ,
\label{eq:new}
\end{equation}
where the `prime' indicates derivative with respect to $z$.  Inserting \Cref{eq:new} into \Cref{f(R):Ricci scalar Palatini}, we get
\begin{equation}
\mathcal{R}(z)=12H^2-6(1+z)H\left(\dfrac{HF'+FH'}{F}\right) +3(1+z)^2\left[\dfrac{2HF(HF''+H'F')-H^2{F'}^2}{2F^2}\right].
\label{eq:R-H Palatini}
\end{equation}
Combining \Cref{f(R):gen Friedmann Palatini 2,eq:R-H Palatini} one then obtains
\begin{equation}
F''-\dfrac{3}{2}\dfrac{{F'}^2}{F}+\left(\dfrac{H'}{H}+\dfrac{2}{1+z}\right)F'-\dfrac{2H'}{H(1+z)}F +3\Omega_{m0}(1+z)\left(\dfrac{H_0}{H}\right)^2=0\ .
\label{eq:F-z}
\end{equation}
Hence, after extracting $H(z)$ from data, we can substitute $z(\mathcal{R})$ found from \Cref{eq:R-H Palatini} into the solution of \Cref{eq:F-z} to get $F(\mathcal{R})$. Then, the $f(\mathcal{R})$ function can be finally obtained by integrating numerically $F(\mathcal R)$.

\begin{figure}
\begin{center}
\includegraphics[width=0.7\textwidth]{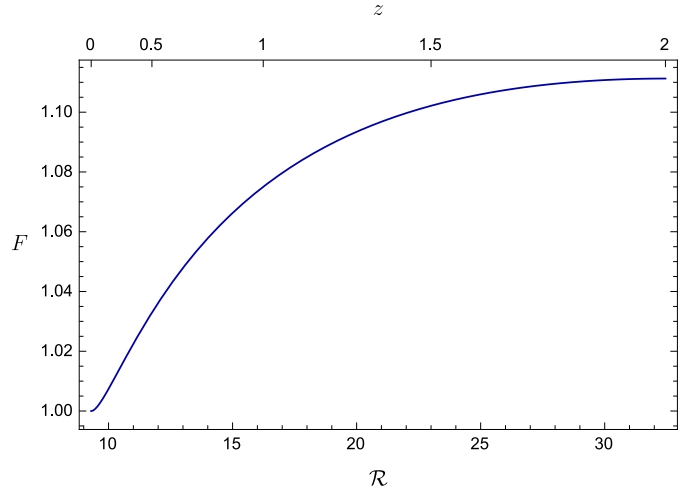}
\caption{Pad\'e approximation of $F(\mathcal R)$ inside $[0,2]$.}
\label{fig:F(R) Pade}
\end{center}
\end{figure}

\begin{figure}
\begin{center}
\includegraphics[width=0.7\textwidth]{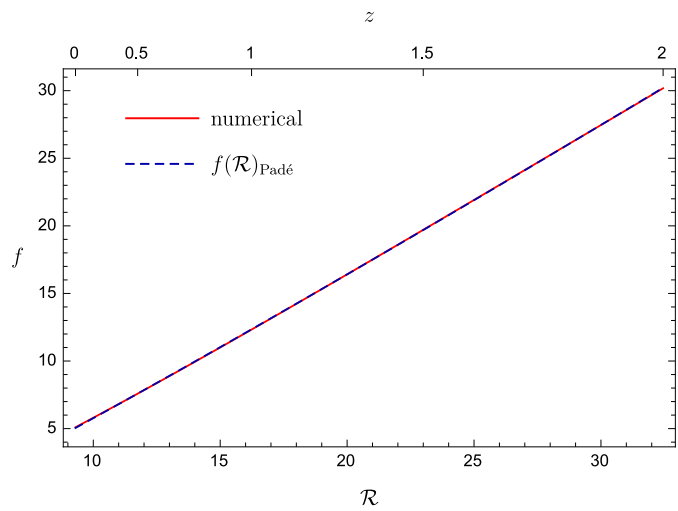}
\caption{Pad\'e reconstruction of $f(\mathcal{R})$ with the corresponding most suitable  analytical approximation (cf. \Cref{eq:best f(R) Pade}).}
\label{fig:f(R) Pade}
\end{center}
\end{figure}

In view of the treatment we proposed in \refcite{rocco3}, we considered the $(2,2)$ Pad\'e and the $(2,1)$ rational Chebyshev approximations of $d_L(z)$ (see \ref{sec:Pade approx} and \ref{sec:Cheb approx}), from which one can infer the corresponding $H(z)$ by means of \Cref{eq:Hubble rate}. In the following, we fix $\Omega_{m0}=0.3$ and adopt the best-fit results obtained in \refcite{rocco3} for the cosmographic parameters.
Thus, in the case of the Pad\'e approximation we used
\begin{equation}
\left\{
\begin{aligned}
&h_0=0.6494^{+0.0211}_{-0.0202}\ , \\
&q_0=-0.285^{+0.040}_{-0.046}\ , \\
&j_0=0.545^{+0.463}_{-0.652}\ , \\
&s_0=0.118^{+0.451}_{-1.600}\ .
\end{aligned}
\right.
\label{cosm param Pade}
\end{equation}
The initial conditions to solve \Cref{eq:F-z} are found by requiring that $G_\text{eff}=G$\footnote{Relaxing this condition and allowing for slight departures from $G$ \cite{Martins17} would ensure to recover the asymptotic $\Lambda$CDM behaviour \cite{Hu07,Appleby07}.} \cite{Dick04,Dominguez04}. This implies
\begin{equation}
F\big|_{z=0}=1\ , \hspace{0.5cm} F'\big|_{z=0}=0\ .
\end{equation}
\Cref{fig:F(R) Pade} shows the behaviour of $F(R)$ using the central values of (\ref{cosm param Pade}) for the numerical integration of \Cref{eq:F-z}
Then, we integrate this solution by means of the initial condition
\begin{equation}
f_0=6(\Omega_{m0}-1)+\mathcal{R}_0\  ,
\end{equation}
obtained from evaluating \Cref{f(R):gen Friedmann Palatini 2} at the present time.
The analytical match to the numerical solution is provided by
\begin{equation}
f(\mathcal R)_\text{Pad\'e}=a+b\mathcal R^n\ ,
\label{eq:best f(R) Pade}
\end{equation}
where
\begin{equation}
(a,\ b,\ n)=(-1.627,\ 0.866,\ 1.074)\ .
\end{equation}
We show the Pad\'e reconstruction of $f(\mathcal R)$ in \Cref{fig:f(R) Pade}.
Taking into account the $1\sigma$ values of \ref{cosm param Pade}, one finds the following bounds:
\begin{equation}
\left\{
\begin{aligned}
&a\in[-1.627,\ -1.326]\ ,\\
&b\in[0.733,\ 0.951]\ , \\
&n\in[1.025,\ 1.123]\ .
\end{aligned}
\right.
\end{equation}

\begin{figure}[h!]
\begin{center}
\includegraphics[width=0.7\textwidth]{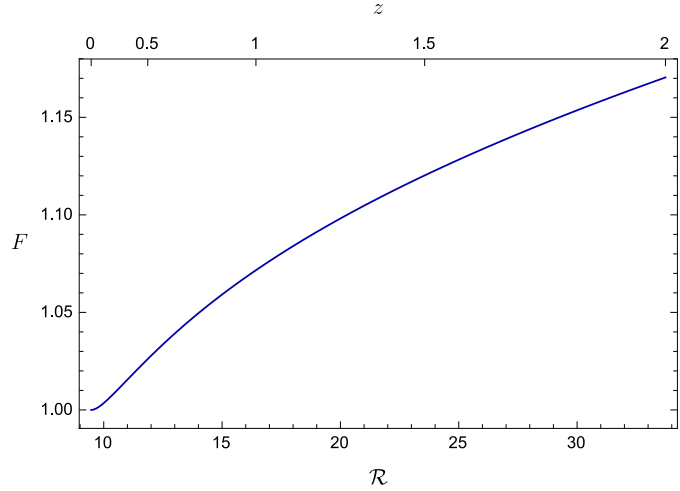}
\caption{Chebyshev reconstruction of $F(\mathcal R)$ inside $z\in[0,2]$}
\label{fig:F(R) Chebyshev}
\end{center}
\end{figure}

\begin{figure}[h!]
\begin{center}
\includegraphics[width=0.7\textwidth]{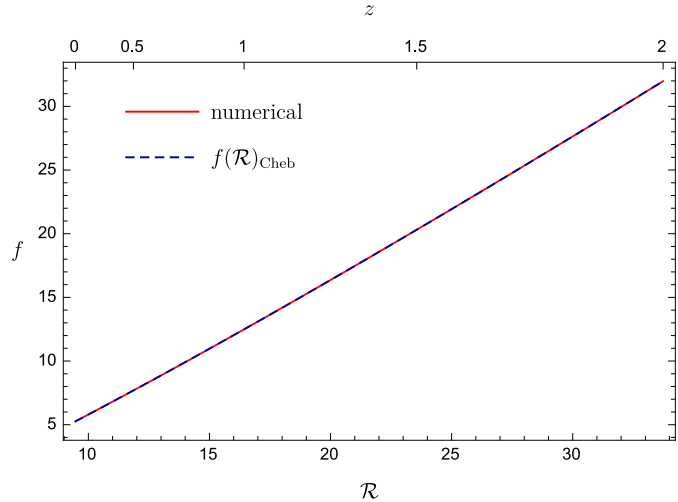}
\caption{Comparison between the rational Chebyshev approximation on $f(\mathcal R)$ with its most suitable analytical approximation (cf. \Cref{eq:best f(R) Cheb}).}
\label{fig:f(R) Chebyshev}
\end{center}
\end{figure}

In the case of rational Chebyshev approximation we used \cite{rocco3}
\begin{equation}
\left\{
\begin{aligned}
&h=0.6495^{+0.0189}_{-0.0194}\ , \\
&q_0=-0.278^{+0.021}_{-0.021}\ , \\
&j_0=1.585^{+0.497}_{-0.914}\ , \\
&s_0=1.041^{+1.183}_{-1.784}\ .
\end{aligned}
\right.
\label{cosm param Cheb}
\end{equation}
Following a similar procedure as the one seen above, we show in \Cref{fig:F(R) Chebyshev} $F(\mathcal R)$ reconstructed using the central values of (\ref{cosm param Cheb}).
In this case, the analytical function matching the numerical integration of $F(\mathcal{R})$ is
\begin{equation}
f(\mathcal R)_\text{Cheb}=\alpha +\beta \mathcal{R}^m\ ,
\label{eq:best f(R) Cheb}
\end{equation}
where
\begin{equation}
(\alpha,\ \beta,\ m)=(-1.332, \ 0.749, \ 1.124)\ .
\end{equation}
The rational Chebyshev reconstruction of $f(\mathcal R)$ is finally shown in \Cref{fig:f(R) Chebyshev}.
If also the lower and upper $1\sigma$ bounds are considered, we find
\begin{equation}
\left\{
\begin{aligned}
&\alpha\in[-1.481,\ -1.332]\ ,\\
&\beta\in[0.749,\ 0.818]\ , \\
&m\in[1.096,\ 1.124]\ .
\end{aligned}
\right.
\end{equation}

\begin{figure}[h!]
\begin{center}
\includegraphics[width=0.7\textwidth]{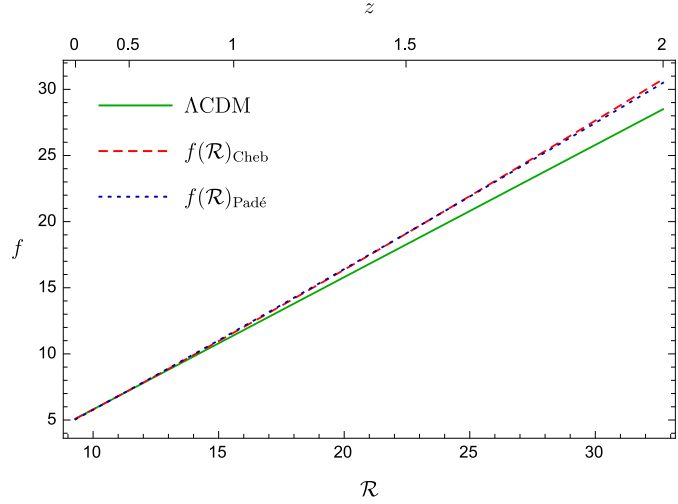}
\caption{Confront among $f(\mathcal R)$ actions of $\Lambda$CDM, Pad\'e and rational Chebyshev approximations.}
\label{fig:f(R) comparison}
\end{center}
\end{figure}

\begin{figure}[h]
\begin{center}
\includegraphics[width=0.75\textwidth]{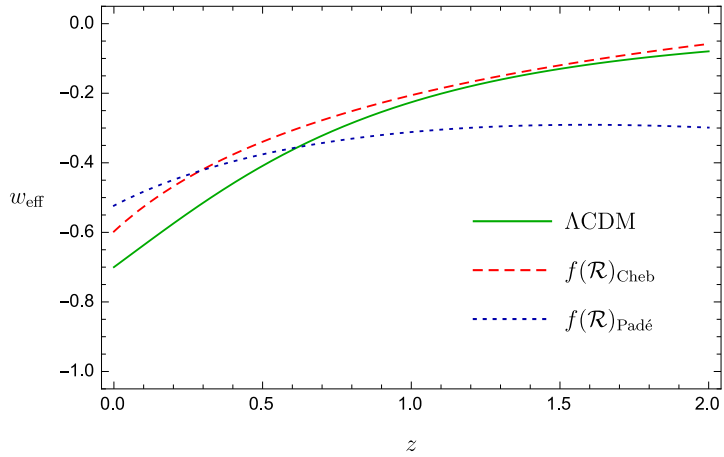}
\caption{Comparison between the $\Lambda$CDM barotropic factor and the Pad\'e and the rational Chebyshev approximations.}
\label{fig:w_eff comparison}
\end{center}
\end{figure}

An interesting exercise is to compare the obtained results with the cosmological predictions of the standard $\Lambda$CDM model.
\Cref{fig:f(R) comparison} shows the best $f(\mathcal{R})$ reconstructed through the rational approximations compared to the action of $\Lambda$CDM assuming $\{h,\Omega_{m0}\}=\{0.7,0.3\}$.
Moreover, one can calculate the effective equation of state parameter as
\begin{equation}
w_\text{eff}(z)=-1+\dfrac{2}{3}(1+z)\dfrac{H'(z)}{H(z)}\ .
\end{equation}
In \Cref{fig:w_eff comparison} we show the results for the different models.

\section{Model-independent reconstruction of $f(T)$ gravity}
\label{sec:f(T) cosmography}

\begin{figure}[h]
\begin{center}
\includegraphics[width=0.7\textwidth]{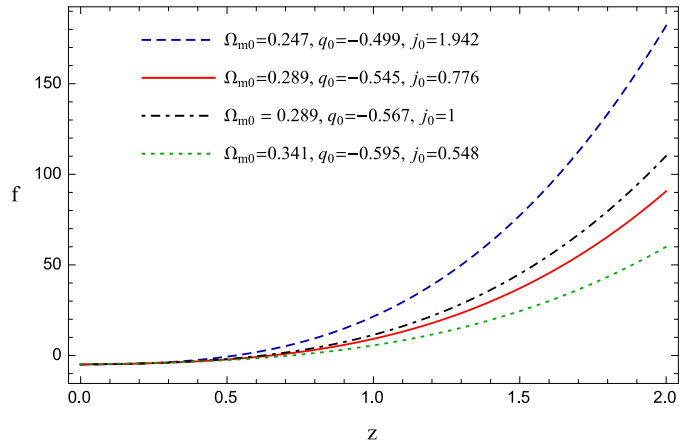}
\caption{Reconstructing $f(z)$ for different $(\Omega_{m0},q_0,j_0)$ based on Table 6 outcomes in \protect\refcite{delaCruz16}. The solid red, the dashed blue and the dotted green lines correspond,  respectively,  to the the best-fit values, the upper $2\sigma$ bounds and the lower $2\sigma$ constraints imposed over $(\tilde{\Omega}_{m0},\tilde{F}_2,\tilde{F}_3)$. The $\Lambda$CDM paradigm is due to dot-dashed black line and $h=0.692$.}
\label{fig:f(T) f_z}
\end{center}
\end{figure}

\begin{figure}[h!]
\begin{center}
\includegraphics[width=0.7\textwidth]{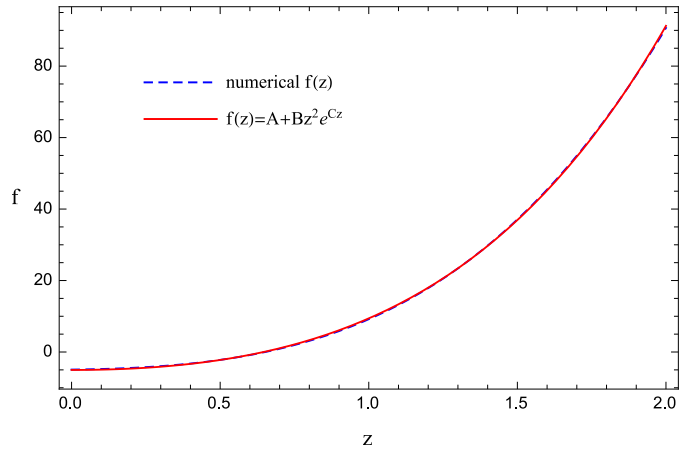}
\caption{Confront between numerical and analytical solutions on $f(z)$, with best fit parameters $(\Omega_{m0},h,q_0,j_0)$ as in \protect\refcite{delaCruz16}, while the coefficients $( A,  B, C)$ are given in \Cref{eq:coeff 1}. }
\label{fig:f(T) num vs analy}
\end{center}
\end{figure}

In this section, we apply the cosmographic method to reconstruct the $f(T)$ function of teleparallel gravity in a model-independent way.  As above, we reconstruct \cite{rocco} the $f(T)$ function from cosmological observations, without \textit{a priori} assumptions over the model. To do that, we assume the flat FLRW metric and consider the Taylor series expansion of the luminosity distance up to the fourth order (cf. \Cref{eq:luminosity distance}).  Then, making use of \Cref{eq:Hubble rate} we can write down the Taylor expansion of the Hubble rate as a function of the cosmographic parameters:
\begin{equation}
H(z)\simeq H_0 \left[1 + z (1 + q_0) + \dfrac{z^2}{2} (j_0 - q_0^2) + \dfrac{z^3}{6} \left(-3 q_0^2 - 3 q_0^3 + j_0 (3 + 4 q_0) + s_0\right)\right].
\label{eq:H_Taylor}
\end{equation}
Hence, once the parameters $(H_0, q_0, j_0,s_0)$ are known, one can infer $f(T(z))=f(z)$ numerically by combining the modified Friedmann equations (\ref{f(T):Fried1}) and (\ref{f(T):Fried2}).
To do that, one need to convert the derivatives with respect to time and the derivatives with to respect to the torsion scalar into derivatives with respect to the redshift. For any function $\mathfrak{F}(z)$, we have
\begin{align}
\dfrac{d}{dt}\mathfrak{F}(z)&=-(1+z)H(z)\dfrac{d}{dz}\mathfrak{F}(z) \ ,	\label{eq:dt-dz}	\\
\,\nonumber\\
\dfrac{\partial}{\partial T}\mathfrak{F}(z)&=-12H(z)H'(z)\dfrac{d}{dz}\mathfrak{F}(z)\ , \label{eq:dT-dz}
\end{align}
where we have used \Cref{eq:T-H} in the latter equation.
Thus, if we combine \Cref{f(T):Fried1,eq:dt-dz} we obtain a differential equation for $f(z)$:
\begin{equation}
\left(\frac{df}{dz}\right)^{-1}\left[H (1+z) \frac{d^2f}{dz^2}+3 f \frac{dH}{dz}\right]=\frac{1}{H}\left(\frac{dH}{dz}\right)^{-1}\left[3 \frac{dH}{dz}+(1+z) \frac{d^2H}{dz^2}\right].
\label{f(T):f(z)}
\end{equation}
One possibility to solve this equation is to assume a particular cosmological model and impose the form of $H(z)$. This, however, would introduce a bias in the analysis and would drive the resulting dynamics towards solutions only slightly deviating from the postulated model.
Our idea is, instead, to reconstruct the Hubble expansion as model-independent as possible, and this can be done through cosmography.
It is, in fact, possible to perform a model-independent procedure by using the kinematic expansion given in \Cref{eq:H_Taylor} to solve \Cref{f(T):f(z)} as a function of $z$ only.
Specifically, two initial conditions are needed to find $f(z)$. The first one is obtained by imposing the equivalence between the effective gravitation constant and the Newton constant:
\begin{equation}
\frac{df}{dz}\Big|_{z=0}=1 \ .
\label{eq:fp0}
\end{equation}
The second initial condition comes from \Cref{eq:T-H,eq:fp0}:
\begin{equation}
f(T(z=0))=f(z=0)=6{H_0}^2(\Omega_{m0}-2)\,.
\label{eq:f0}
\end{equation}
Therefore, we follow the strategy presented in \refcite{delaCruz16} and recast the cosmographic parameters as
\begin{align}
q_0=&-1+\dfrac{3\tilde{\Omega}_{m0}}{2(1+2\tilde{F}_2)} \ ,	\\
\,\nonumber\\
j_0=\ &1-\dfrac{9\tilde{\Omega}_{m0}^2(3\tilde{F}_2+2\tilde{F}_3)}{2(1+2\tilde{F}_2)^3} \ ,\\
\,\nonumber\\
s_0=&\ 1-\dfrac{9\tilde{\Omega}_{m0}}{2(1+2\tilde{F}_2)}+\dfrac{45\tilde{\Omega}_{m0}^2(3\tilde{F}_2+2\tilde{F}_3)}{2(1+2\tilde{F}_2)^3} +\dfrac{27\tilde{\Omega}_{m0}^3(3\tilde{F}_2+12\tilde{F}_3+4\tilde{F}_4)}{4(1+2\tilde{F}_2)^4} \nonumber \\
&-\dfrac{81\tilde{\Omega}_{m0}^3(3\tilde{F}_2+2\tilde{F}_3)^2}{2(1+2\tilde{F}_2)^5} \ ,	
\end{align}
where
\begin{equation}
\begin{aligned}
&\tilde{\Omega}_{m0}  =\dfrac{\Omega_{m0}}{F_1}\ , 	\hspace{0.5cm} \tilde{F}_i  =\dfrac{F_i}{F_1} \hspace{0.3cm} (i=2,3,4) \\
&F_i  =T_0^{i-1} f^{(i)}(T_0) \hspace{0.3cm}  (i=1,2,3,4) \ .
\end{aligned}
\end{equation}
In order to numerically solve \Cref{f(T):f(z)}, we adopt the following results \cite{delaCruz16}:
\begin{equation}
\left\{
\begin{aligned}
&\Omega_{m0}=0.289 \\
&h=0.692 \\
&q_0=-0.545\\
&j_0= 0.776\\
&s_0=-0.192
\end{aligned}
\right .
\label{eq:best-fit}
\end{equation}
where $h\equiv H_0/(100\ \text{km/s/Mpc})$.
These values are compatible with the observational bounds obtained through a comparison with different data sets \cite{Muthukrishna16}.
Then, the first step is to consider a second-order expansion of the Hubble rate, \emph{i.e.} up to the jerk coefficient of the cosmographic series. In  \Cref{fig:f(T) f_z} we show the results for different sets of cosmographic parameters.
To match the numerical behaviours, we used the following test-functions\footnote{The forms of the test-functions have been chosen \emph{a posteriori} to match the shapes of the curves gotten from the numerical analysis.}:
\begin{subequations}
\begin{align}
f_1(z)&=\mathcal A+z(\mathcal B+\mathcal C z)\ln(1+z^2)   \label{eq:ln}\\
f_2(z)&=\mathcal A+\mathcal B z^2e^{\mathcal C z}	           \label{eq:exp}\\
f_3(z)&=\mathcal A z^2+\mathcal B z\sin(1+\mathcal C z^2)	    \label{eq:sin}\\
f_4(z)&=\mathcal A+\mathcal B z^2\cos(1+\mathcal C z) 		       \label{eq:cos}\\
f_5(z)&=\mathcal A+\mathcal B\sinh(1+\mathcal C z)				\label{eq:sinh}\\
f_6(z)&=\mathcal A+\mathcal B z^3\tanh(\mathcal C z^2)		\label{eq:tanh}
\end{align}
\end{subequations}
where the values of the free coefficients $\mathcal A$, $\mathcal B$ and $\mathcal C$ are found through a comparison with the numerical curves.
To obtain information on how well the test-functions approximate the numerical $f(z)$, we performed the $\mathcal{R}^2$-test \cite{Draper98}. If $f_i^{obs}$ are the numerical values of $f(z_i)$, and $f_i$ are the correspondent analytical values, one can define
\begin{equation}
\mathcal{R}^2\equiv1-\dfrac{\sum_{i=1}^n( f_i^{obs}-f_i)^2}{\sum_{i=1}^n (f_i^{obs}-\bar{f})^2}\ ,
\label{eq:R test}
\end{equation}
where
\begin{equation}
\bar{f}=\dfrac{1}{n}\sum_{i=1}^nf_i^{obs}\ ,
\end{equation}
being $n$ the number of points. The ideal case $\mathcal{R}^2=1$ occurs when the test-function and $f(z)$ agree exactly.
In the case of the best-fit red curve of \Cref{fig:f(T) f_z}, the $\mathcal{R}^2$-test indicates the function (\ref{eq:exp}) as the most suitable choice (cf. \Cref{tab:test R^2}):
\begin{equation}
f(z)=\mathcal A+ \mathcal Bz^2e^{ \mathcal C z}\ ,
\label{eq:approx f}
\end{equation}
where
\begin{equation}
(\mathcal A, \mathcal B, \mathcal C)=(-5.024,\ 8.651,\ 0.512) \ .
\label{eq:coeff 1}
\end{equation}
We show in \Cref{fig:f(T) num vs analy} the comparison between the numerical solution of $f(z)$ and its best analytical approximation given by \ref{eq:approx f} .
From \Cref{tab:test R^2} we note that also the functions \ref{eq:ln} and \ref{eq:tanh} represents very good approximations of $f(z)$: their $\mathcal{R}^2$ values are  only $0.024\%$ and $0.038\%$ far from the best one, respectively.
\begin{table}[h]
\small
\setlength{\tabcolsep}{1.5em}
\renewcommand{\arraystretch}{1.5}
\tbl{Outcomes of the $\mathcal{R}^2$-test on the test-functions \Crefrange{eq:ln}{eq:tanh} for the best-fit curve of \Cref{fig:f(T) f_z} .}
{\begin{tabular}{c| c| c}
\hline
\hline
Test-function & $(\mathcal A, \mathcal B, \mathcal C)$ & $\mathcal{R}^2$  \\
\hline
$f_1(z)$ & $(-3.897,\ 10.88,\ 7.185)$ & 0.99974   \\
$f_2(z)$ & $(-5.024,\ 8.651,\ 0.512)$ & 0.99997   \\
$f_3(z)$ & $(15.73,\ -9.286,\ 1.112)$ & 0.99102   \\
$f_4(z)$ & $(-3.152,\ -21.52,\ 1.114)$ & 0.99630   \\
$f_5(z)$ & $(-10.93,\ 3.173,\ 1.593)$ & 0.99909   \\
$f_6(z)$ & $(-3.463,\ 11.89,\ 4.143)$ & 0.99959   \\
\hline
\hline
\end{tabular}
\label{tab:test R^2}
}
\end{table}

\begin{figure}[h!]
\begin{center}
\includegraphics[width=0.7\textwidth]{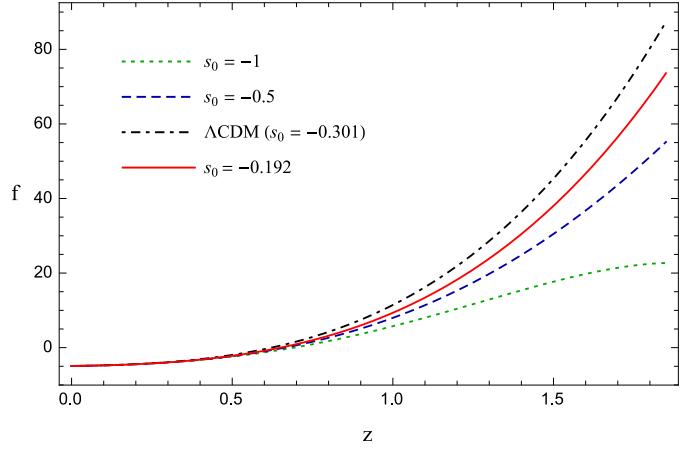}
\caption{Different $f(z)$ shapes with distinct  snap parameters and $(\Omega_{m0},h,q_0,j_0)$ fixed by best-fit results as in \protect\refcite{delaCruz16}. The solid red line is due to  the best-fit value of $s_0$, while the dot-dashed black line to the $\Lambda$CDM model, assuming that  $\Omega_{m0}=0.289$. }
\label{fig:f(T) var s0}
\end{center}
\end{figure}

The second step of the analysis considers the expansion of $H(z)$ up to the third order. We thus show in \Cref{fig:f(T) var s0} the behaviour of $f(z)$ for different values of $s_0$ within the interval $[-1,0]$, while we fix the other cosmographic parameters to their best-fit results.
As in the previous case, the best approximation of $f(z)$ is given by the function \ref{eq:approx f} with the following values of the free parameters:
\begin{equation}
(\mathcal A, \mathcal B, \mathcal C)=(-5.022,\ 8.577,\ 0.532)\ .
\label{eq:coeff 2}
\end{equation}

We finally describe the strategy to reconstruct the function $f(T)$. We plug the expansion of $H(z)$ up to the snap parameter into \Cref{eq:T-H}, which can be inverted to find $z(T)$. Therefore, inserting the obtained result back into \Cref{eq:approx f} provides us with the function $f(T)$. In this procedure one takes into account the error propagation due to the uncertainties in cosmographic parameters through redefining $f(z)$ by a rescaling factor $\alpha$:
\begin{equation}
\alpha f(z)\longrightarrow f(z)\ .
\end{equation}
The value of the constant $\alpha$ will be determined from cosmological constraints.
One thus gets
\begin{align}
z(T)=&\ \dfrac{1}{2\mathcal{Q}}\bigg[2(q_0^2-j_0)+\left(\dfrac{4\mathcal{M}(T)}{H_0^3}\right)^{1/3}+\left(\dfrac{16H_0^3}{\mathcal{M}(T)}\right)^{1/3}\Big(j_0^2+q_0^2(6+12q_0+7q_0^2)\nonumber \\
&-2j_0(3+7q_0+5q_0^2)-2s_0(1+q_0)\Big)\bigg] ,
\label{eq:z(T)}
\end{align}
and
\begin{align}
f(T)=&\ \alpha \mathcal{A}+\dfrac{\alpha \mathcal{B}}{4\mathcal{Q}^2}\bigg[2(q_0^2-j_0)+\left(\dfrac{4\mathcal{M}(T)}{H_0^3}\right)^{1/3}+\left(\dfrac{16H_0^3}{\mathcal{M}(T)}\right)^{1/3}\Big(j_0^2+q_0^2(6+12q_0+7q_0^2) \nonumber \\
&-2j_0(3+7q_0+5q_0^2) -2s_0(1+q_0)\Big)\bigg]^2 \exp\Bigg\{\frac{\mathcal{C}}{2\mathcal{Q}}\bigg[2(q_0^2-j_0)+\left(\frac{4\mathcal{M}(T)}{H_0^3}\right)^{1/3}\nonumber \\
&+\left(\frac{16H_0^3}{\mathcal{M}(T)}\right)^{1/3}\Big(j_0^2+q_0^2(6+12q_0+7q_0^2)-2j_0(3+7q_0+5q_0^2)-2s_0(1+q_0)\Big)\bigg]\Bigg\} ,
\label{eq:f(T)}
\end{align}
where
\begin{equation}
\mathcal{M}(T)\equiv H_0^2\sqrt{2\mathcal{P}(T)}\mathcal{Q}-2H_0^3\mathcal{N}+\sqrt{-6T}H_0^2 \mathcal{Q}^2 \ ,
\label{eq:M}
\end{equation}

\begin{align}
\mathcal{P}(T)\equiv&\ 2H_0^2\bigg[6 j_0^3 - 6 q_0^2 \big(2 + q_0 (4 + q_0)\big)^2+3 j_0^2 \big(8 + q_0 (28 + 17 q_0)\big)+4  \Big(2 + q_0 \big(6 - q_0 (3	\nonumber \\
&+ 7 q_0)\big)\Big)s_0+9 s_0^2+2 j_0 \Big(6 (2 + 3 s_0) + q_0 \big(52 + q_0 (60 - q_0 (6 + 17 q_0)\big) + 27 s_0\Big)\bigg] \nonumber \\
&-2 \sqrt{-6T} H_0 \Big(j_0^3+3 j_0^2 (1 + q_0) (6 + 11 q_0) -3 j_0 q_0^2 \big(12 + q_0 (29 + 16 q_0)\big)+q_0^4 \big(18 \nonumber \\
&+ q_0 (36 + 17 q_0)\big)-15 q_0^2 (1 + q_0) s_0+3 j_0 (5 + 7 q_0) s_0 + 3 s_0^2\Big)-3 T \big(-3 q_0^2 (1 + q_0) \nonumber \\
& + j_0 (3 + 4 q_0) + s_0\big)^2\ ,
\label{eq:P}
\end{align}

\begin{equation}
\mathcal{Q}\equiv -3 q_0^2 (1 + q_0) + j_0 (3 + 4 q_0) + s_0 \ ,
\label{eq:Q}
\end{equation}

\begin{align}
\mathcal{N}\equiv &\ j_0^3 + 3 j_0^2 (1 + q_0) (6 + 11 q_0) - 3 j_0 q_0^2 \big(12 + q_0 (29 + 16 q_0)\big) + q_0^4 \big(18 + q_0 (36 + 17 q_0)\big) \nonumber \\
& - 15 q_0^2 (1 + q_0) s_0  +3 j_0 (5 + 7 q_0) s_0 + 3 s_0^2\ .
\label{eq:N}
\end{align}
\\

\begin{figure}[h!]
\begin{center}
\includegraphics[width=0.7\textwidth]{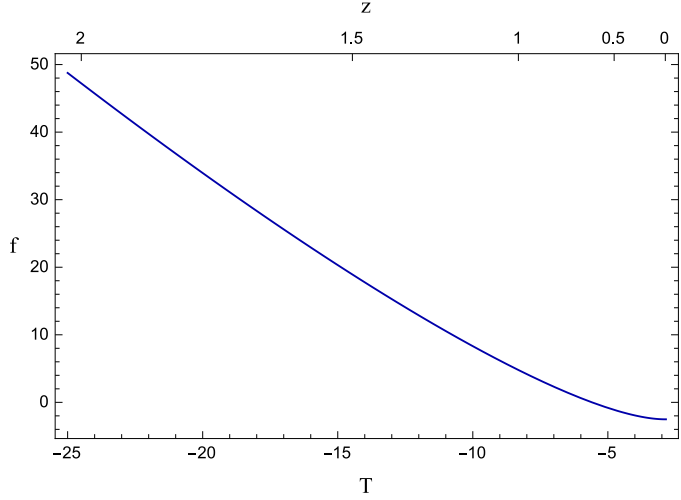}
\caption{Numerical shape of $f(T)$ imposed from best-fit values and $\alpha=0.5$ inside $0\leq z \leq 2$. }
\label{fig:f(T)}
\end{center}
\end{figure}

\begin{figure}[h!]
\begin{center}
\includegraphics[width=0.7\textwidth]{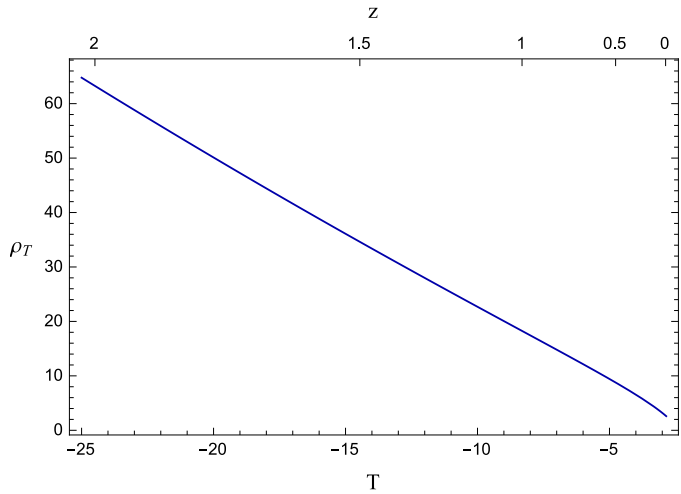}
\caption{Torsion density with mean values of cosmographic parameters got from the experimental analysis with $\alpha=0.5$ inside $0\leq z\leq 2$.}
\label{fig:rho-T}
\end{center}
\end{figure}

\begin{figure}[h!]
\begin{center}
\includegraphics[width=0.7\textwidth]{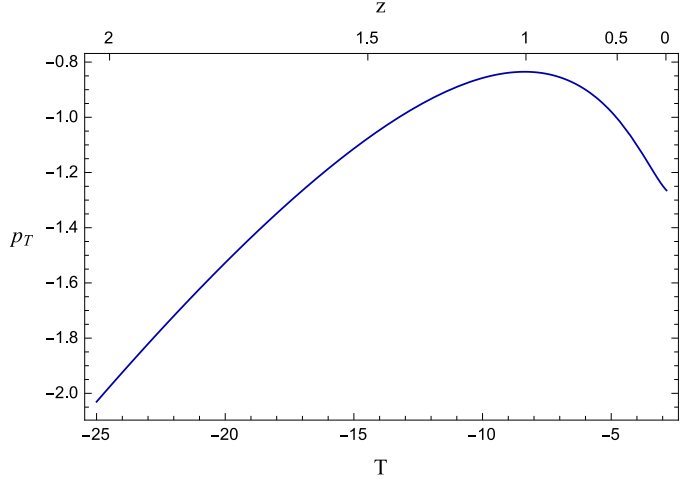}
\caption{$P_T$ with the best-fit values of the cosmographic parameters inside  $0\leq z\leq 2$.}
\label{fig:p-T}
\end{center}
\end{figure}

\begin{figure}[h]
\begin{center}
\includegraphics[width=0.7\textwidth]{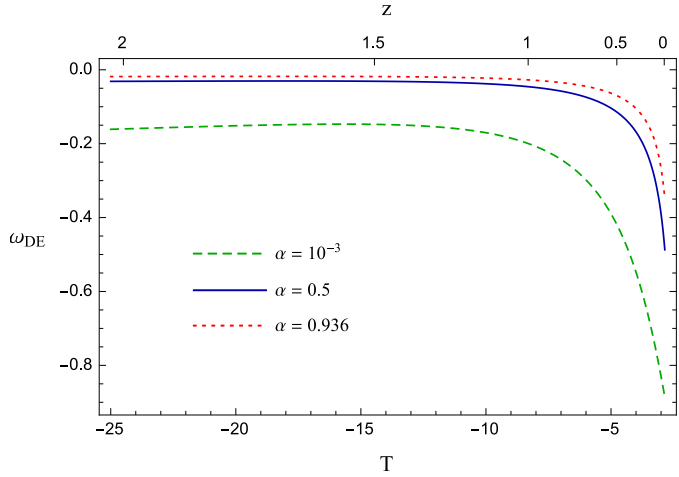}
\caption{Dark energy equation of state inside  $0\leq z\leq 2$, with different values of $\alpha$ (cf. \Cref{eq:rescale factor}).}
\label{fig:w_DE-T}
\end{center}
\end{figure}

The obtained $f(T)$ can be used to study the torsional density and pressure, and compare the  above results with the cosmological findings in the literature \cite{Nesseris13}.
From \Cref{eq:rho_T,eq:p_T} we get
\begin{align}
\rho_T=&-\dfrac{1}{2}\left[T+\alpha\left(\mathcal{A}+\dfrac{\mathcal{B}\ \xi(T)}{4\mathcal{Q}^2}e^{\frac{\mathcal{C} \xi(T)}{2\mathcal{Q}}}\right)\right]+\dfrac{2^{1/3} \alpha \mathcal{B} T G(T) }{24H_0^3\mathcal{Q}^3\mathcal{M}(T)^2} X(T) Y(T) \mathcal{M}'(T)\ e^{\frac{\mathcal{C} \xi(T)}{2\mathcal{Q}}} \ ,
\label{eq:rho(T)}
\end{align}

\begin{align}
p_T=&\Bigg[e^{-\frac{\mathcal{C} \xi(T)}{2\mathcal{Q}}} H_0^4 Q^4 \mathcal{M}(T)^{10/3}\Bigg(72\mathcal{A}+\dfrac{18\mathcal{B} \xi(T)^2}{\mathcal{Q}^2}e^{\frac{\mathcal{C} \xi(T)}{2\mathcal{Q}}}-\dfrac{3\times 2^{1/3}\mathcal{B} T G(T) X(T) Y(T) \mathcal{M}'(T)}{H_0^3\mathcal{Q}^3\mathcal{M}(T)^2} \nonumber \\
&\times e^{\frac{\mathcal{C} \xi(T)}{2\mathcal{Q}}}+\dfrac{2^{1/3}\mathcal{B} T^2 }{H_0^3\mathcal{Q}^4\mathcal{M}(T)^{10/3}}\times e^{\frac{\mathcal{C} \xi(T)}{2\mathcal{Q}}} \bigg(2^{1/3}\mathcal{C} G(T)^2  X(T) Y(T)\mathcal{M}'(T)^2+2^{7/3}H_0\mathcal{Q} \nonumber \\
&\times X(T) Y(T) \mathcal{M}(T) \mathcal{M}'(T)^2 +2H_0\mathcal{Q} G(T) \mathcal{M}(T)^{1/3} \times \Big(-6 X(T) Y(T) \mathcal{M}'(T)^2 \nonumber \\
&+2^{5/3} \big(\mathcal{C} X(T)+Y(T)\big)\mathcal{M}(T)^{2/3} \mathcal{M}'(T)^2 +2H_0\big((2\mathcal{Q}+\mathcal{C}(q_0^2-j_0))X(T)+(q_0^2-j_0)\nonumber \\
&\times Y(T)\big) \mathcal{M}(T)^{1/3}\mathcal{M}'(T)^2 + 3X(T) Y(T) \mathcal{M}(T) \mathcal{M}''(T)\Big)\bigg)\Bigg)\Bigg]\times \Bigg[2^{4/3}\mathcal{B}\bigg(2^{1/3}\mathcal{C} G(T)^2 \nonumber \\
&\times  X(T) Y(T)\mathcal{M}'(T)^2+2^{7/3}H_0\mathcal{Q} T  X(T) Y(T)	 \mathcal{M}(T) \mathcal{M}'(T)^2+H_0\mathcal{Q} G(T) \mathcal{M}(T)^{1/3} \nonumber \\
&\times \Big(-12 T X(T) Y(T) \mathcal{M}'(T)^2+2^{7/3} \big(\mathcal{C} X(T) +Y(T)\big) T  \mathcal{M}(T)^{2/3} \mathcal{M}'(T)^2+4H_0 \nonumber \\
&\times \big((2\mathcal{Q}+\mathcal{C}(q_0^2-j_0))X(T)+(q_0^2-j_0)Y(T)\big)T \mathcal{M}(T)^{1/3}\mathcal{M}'(T)^{2} + 3X(T) Y(T) \mathcal{M}(T) \nonumber \\
&\times \big(\mathcal{M}'(T)+2 T \mathcal{M}''(T)\big)\Big)\bigg)\Bigg] ,
\label{eq:p(T)}
\end{align}
where
\begin{equation}
\xi(T)\equiv 2(q_0^2-j_0)+\left(\frac{4\mathcal{M}(T)}{H_0^3}\right)^{1/3}+\left(\frac{16H_0^3}{\mathcal{M}(T)}\right)^{1/3}\ ,
\label{eq:xi}
\end{equation}

\begin{equation}
X(T)\equiv 2^{4/3}H_0^2\mu+2H_0(q_0^2-j_0)\mathcal{M}(T)^{1/3}+2^{2/3}\mathcal{M}(T)^{2/3}\ ,
\label{eq:X}
\end{equation}

\begin{equation}
Y(T)\equiv 2^{4/3}\mathcal{C} H_0^2\mu+2H_0\left(2\mathcal{Q}+\mathcal{C}(q_0^2-j_0)\right)\mathcal{M}(T)^{1/3}+2^{2/3}\mathcal{C} \mathcal{M}(T)^{2/3}\ ,
\label{eq:Y}
\end{equation}

\begin{equation}
G(T)\equiv -2H_0^2\mu+2^{1/3}\mathcal{M}(T)^{2/3}\ ,
\label{eq:G}
\end{equation}

\begin{equation}
\mu\equiv  j_0^2+q_0^2(6+12q_0+7q_0^2)-2j_0(3+7q_0+5q_0^2) -2s_0(1+q_0)\ ,
\label{eq:mu}
\end{equation}
and
\begin{equation}
\mathcal{M}'(T)\equiv\dfrac{\partial \mathcal{M}}{\partial T}=\dfrac{1}{\sqrt{2}}\left[\dfrac{\sqrt{3}H_0^2\mathcal{Q}^2 T}{(-T)^{3/2}}+\dfrac{H_0^2\mathcal{Q}}{\sqrt{\mathcal{P}(T)}}\mathcal{P}'(T)\right] ,
\label{eq:Mp}
\end{equation}

\begin{equation}
\mathcal{M}''(T)\equiv\dfrac{\partial^2 \mathcal{M}}{\partial T^2}= -\dfrac{\sqrt{-3T}H_0^2\mathcal{Q}^2\mathcal{P}(T)^2+H_0^2\mathcal{Q}T^2 \mathcal{P}(T)^{1/2}\mathcal{P}'(T)^2-2H_0^2\mathcal{Q}T^2\mathcal{P}(T)^{3/2}\mathcal{P}''(T)}{2\sqrt{2}T^2\mathcal{P}(T)^2}
\label{eq:Mpp}
\end{equation}

\begin{align}
\mathcal{P}'(T) \equiv &\ \dfrac{\partial \mathcal{P}}{\partial T}= -3 \left(-3 q_0^2 (1 + q_0) + j_0 (3 + 4 q_0) + s_0\right)^2+\dfrac{\sqrt{6}H_0}{\sqrt{-T}}\Big[j_0^3 + 3 j_0^2 (1 + q_0) (6 + 11 q_0) \nonumber \\
&- 3 j_0 q_0^2 \big(12+ q_0 (29 + 16 q_0)+q_0^4 \big(18 + q_0 (36 + 17 q_0)-15 q_0^2 (1 + q_0) s_0\big) \nonumber \\
&+3 j_0 (5 + 7 q_0) s_0 + 3 s_0^2\Big] ,
\label{eq:Pp}
\end{align}

\begin{align}
\mathcal{P}''(T)\equiv &\ \dfrac{\partial^2 \mathcal{P}}{\partial T^2}=\sqrt{\dfrac{3}{2}}\dfrac{H_0}{(-T)^{3/2}}\Big[j_0^3 + 3 j_0^2 (1 + q_0) (6 + 11 q_0)-3 j_0 q_0^2 \big(12 + q_0 (29 + 16 q_0)\big) \nonumber \\
&+q_0^4 \big(18 + q_0 (36 + 17 q_0)\big) -15 q_0^2 (1 + q_0) s_0 + 3 j_0 (5 + 7 q_0) s_0 + 3 s_0^2\Big] .
\label{eq:Ppp}
\end{align}
\\
We note that $p_T$ does not actually depend on the rescaling factor $\alpha$.
To constrain the value of this coefficient, we impose the condition $w_{DE}<-1/3$ implying present accelerated expansion. We thus find
\begin{equation}
0<\alpha \lesssim 0.936\ .
\label{eq:rescale factor}
\end{equation}
In \Cref{fig:f(T)} we show the reconstructed $f(T)$ assuming an indicative $\alpha=0.5$.
\Cref{fig:rho-T,fig:p-T} show the behaviours of the $\rho_T$ and $p_T$, while the effective dark energy equation of state parameter for different values of $\alpha$ is displayed in  \Cref{fig:w_DE-T}.


\subsection{Comparison with previous $f(T)$ models}

\begin{figure}[h!]
\begin{center}
\includegraphics[width=0.7\textwidth]{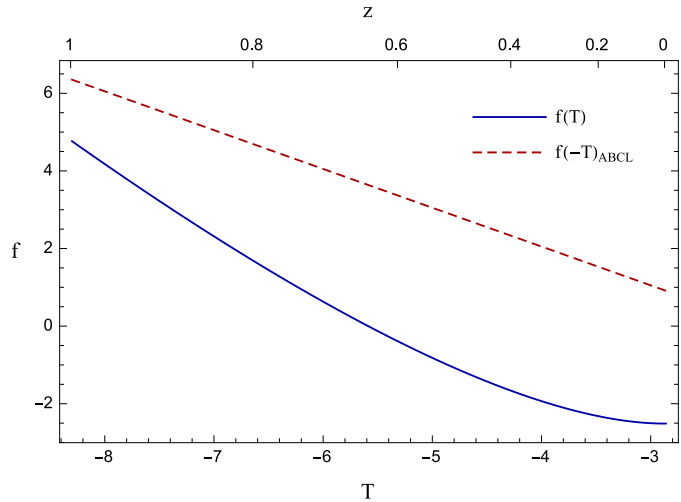}
\caption{Comparison between $f(T)$ functions. The first got for $\alpha=0.5$ (solid blue line), whereas the second by \protect\refcite{Aviles13} (dashed red line) inside $0\leq z \leq 1$.}
\label{fig:comp f(T)}
\end{center}
\end{figure}

We here compare the $f(T)$ model we have obtained with the cosmographic reconstruction found in \refcite{Aviles13}.
In fact, testing the consistency of the two models in the valid redshift interval provides a measure of the goodness of the present approach.
To this end, we report the model proposed in \refcite{Aviles13}:
\begin{align}
f(T)_\text{{\tiny ABCL}}&=	c_0T+(T-T_0)\Big[c_1+c_3\cosh(T-T_0)+(T-T_0)\Big(c_2+c_4(T-T_0)\sinh(T-T_0)\Big)\Big]
\label{eq:ABCL f(T)}
\end{align}
where $T_0=-6H_0^2$. The above model is made of different functions which dominate over each other in different redshift domains.
The free coefficients $c_i\  (i=0,\hdots,4)$ have been determined through imposing the conditions $f(T_0)=6H_0^2(\Omega_{m0}-2)$ and $f'(T_0)=1$ as for the initial settings. In particular, an experimental analysis performed on different data sets provided
\begin{subequations}
\begin{align}
&c_0=2-\Omega_{m0}\ ,\\
&c_1=\Omega_{m0}-1\ ,\\
&c_2=-3\times 10^{-6}\ ,\\
&c_3=\dfrac{1}{15}\times 10^{-9}\ ,\\
&c_4=\dfrac{3}{4}\times 10^{-14}\ .
\end{align}
\end{subequations}
along with
\begin{equation}
\Omega_{m0}=0.364\ , \hspace{0.5cm}  H_0=71.47\ \text{km/s/Mpc} \ .
\end{equation}
Let us compare \Cref{eq:ABCL f(T)} to the above model without caring about the sign of $T$. \Cref{fig:comp f(T)} clearly shows the compatibility between the two models for  $z\leq 1$.
The $10\% - 15\%$ level discrepancies are due to bigger uncertainties in the estimate of the cosmographic series present in the previous model. Both curves indicate slight departures from the standard model, which become more evident as the redshift increases.
We can then conclude that the limits of the model \eqref{eq:ABCL f(T)} are overcome by adopting the numerical analysis performed in here.

\section{Final outlook}

In this review paper, we discussed the cosmographic method and its applications to cosmological models derived from extended or modified theories of gravity. The philosophy is  going beyond Einstein's gravity to cure the shortcomings of the standard cosmological model \cite{Capozziello-Faraoni10}. Specifically, the observational evidence that almost the entire energy density of the cosmic fluid is in the form of dark components might be the sign that standard  GR  breaks down at IR cosmological scales. One way to address this issue is to modify the Einstein-Hilbert action and introduce higher-order curvature invariants and minimally or non-minimally coupled terms between scalar fields and geometry.  The ETG scenarios represent a semi-classical approach where GR is recovered in the low-energy limit.

With this recipe in mind, we focused on $f(R)$ gravity in the metric and Palatini formalisms. We showed the equivalence between these paradigms and the scalar-tensor theories, analyzing the role of conformal transformations in the Einstein and Jordan frames.  We thus described the dynamics of the $f(R)$ cosmological models and discussed the observational viability of such theories.

Then, we discuss an alternative description of the gravitational interaction in terms of torsion.
Assuming torsion instead of curvature as dynamical field  leads to the  equivalent teleparallel formulation of GR (TEGR).
We show that the features of a late-time accelerating universe can be reproduced by modifying the gravity action to include a generic function of the torsion scalar. Moreover, scalar field non-minimally coupled to torsion can be considered.

Furthermore, it is possible to  develope model-independent techniques to describe the expansion of the universe without postulating the dark terms a priori. We showed how cosmography can be used to break the degeneracy among cosmological models. We thus proposed the use of rational polynomials to overcome the convergence issues and reduce the error propagations typical of the standard cosmographic approach. The latest cosmic data can be used to place bounds on the cosmographic series through different  Monte Carlo integrations. This new cosmographic method can be adopted to accurately describe the late-time history of the universe.

These cosmographic techniques can be adopted to derive cosmological consistent models of  $f(R)$ and $f(T)$ gravity. The approach  allows to reconstruct the extended/modified gravity actions in different formalisms by  assuming only the validity of the cosmological principle.
Hence, we discussed the dynamical features of the reconstructed models and their consequences at the level of background cosmology. The results indicated slight departures from GR (that is the $\Lambda$CDM model) and  dark energy terms evolving in time.

To conclude, the current state of the art suggests that it is not yet possible to falsify the $\Lambda$CDM model by robust statements  adopting only  cosmography. The degeneracy among the dark energy models is still unavoidable and the need of more precise measurements (eventually at very  high redshift)  from forthcoming observations remains crucial. As reported in \refcite{j36}, a redshift of the order $z\sim 1.5$ is crucial to discriminate among concurring models while the issue of determining the physical frame, i.e. the Einstein or the Jordan frame, may be addressed in cosmography \cite{Lor}. Finally, the models have also to be confronted with perturbations to be consistent with observations of large scale structure.  These topics will be the arguments of future works.

\section*{Acknowledgments}

This work is based upon the  COST action CA15117 (CANTATA), supported by COST (European Cooperation in Science and Technology). S.C. is supported by Istituto Nazionale di Fisica Nucleare (INFN), {\it iniziative specifiche} QGSKY and MOONLIGHT2.
R.D. thanks Anna Silvia Baldi for useful discussions. O.L. acknowledges the financial support by MES of the RK, Program ‘Center of Excellence for Fundamental and Applied Physics’ IRN: BR05236454, and by the MES Program IRN: BR05236494.

\newpage

%\section{Appendices}
%
%Appendices should be used only when absolutely necessary. They
%should come\break
%before the References. If there is more than one
%appendix, number them alphabetically. Number displayed equations
%occurring in the Appendix in this way, e.g.~(\ref{app1}),\break
%(\ref{app2}), etc.
%\begin{eqnarray}	%A.1
%\begin{array}{rcl}
%g_{\mu_1\mu_2} &=& g_{axby}=-\displaystyle{\epsilon_{abc}}{4\pi}\,
%\frac{(x-y)^c}{|x-y|^3}\,, \\[8pt]
%h_{\mu_1\mu_2\mu_3} &=& \epsilon^{\alpha_1 \alpha_2 \alpha_3}
%g_{\mu_1\alpha_1}g_{\mu_2\alpha_2}g_{\mu_3\alpha_3}
%\end{array}
%\label{app1}
%\end{eqnarray}
%with
%\begin{eqnarray}	%A.2
%\epsilon^{\alpha_1 \alpha_2 \alpha_3} = \epsilon^{b_1y_1b_2y_2cx} =
%\epsilon^{b_1b_2c}\delta(x-y_1)\delta(x-y_2)\,.
%\label{app2}
%\end{eqnarray}
%
%If the atom remains in the cavity for a time $1/2\nu_R$, then it will
%contribute a photon by stimulated emission.
%Furthermore, the atom and photon field will exhibit a phase shift due
%to the zero point energy of the quantum gravity vacuum.

\appendix

\section{Experimental data compilations}
\label{sec:data}
\begin{table}[h!]
\setlength{\tabcolsep}{1.5em}
\small
\tbl{$H(z)$ measurements got from differential age treatment, in which $H(z)$ is given by km/s/Mpc.}
{\begin{tabular}{c c c }
\hline
\hline
 $z$ &$H \pm \sigma_H$ &  Ref. \\
\hline
0.0708	& $69.00 \pm 19.68$ & \cite{Zhang14} \\
0.09	& $69.0 \pm 12.0$ & \cite{Jimenez02} \\
0.12	& $68.6 \pm 26.2$ & \cite{Zhang14} \\
0.17	& $83.0 \pm 8.0$ & \cite{Simon05} \\
0.179 & $75.0 \pm	4.0$ & \cite{Moresco12} \\
0.199 & $75.0	\pm 5.0$ & \cite{Moresco12} \\
0.20 &$72.9 \pm 29.6$ & \cite{Zhang14} \\
0.27	& $77.0 \pm 14.0$ & \cite{Simon05} \\
0.28	& $88.8 \pm 36.6$ & \cite{Zhang14} \\
0.35	& $82.1 \pm 4.85$ & \cite{Chuang12}\\
0.352 & $83.0	\pm 14.0$ & \cite{Moresco16} \\
0.3802	& $83.0 \pm 13.5$ & \cite{Moresco16}\\
0.4 & $95.0	\pm 17.0$ & \cite{Simon05} \\
0.4004	& $77.0 \pm 10.2$ & \cite{Moresco16} \\
0.4247	& $87.1 \pm 11.2$  & \cite{Moresco16} \\
0.4497 &	$92.8 \pm 12.9$ & \cite{Moresco16}\\
0.4783	 & $80.9 \pm 9.0$ & \cite{Moresco16} \\
0.48	& $97.0 \pm 62.0$ & \cite{Stern10} \\
0.593 & $104.0 \pm 13.0$ & \cite{Moresco12} \\
0.68	& $92.0 \pm 8.0$ & \cite{Moresco12} \\
0.781 & $105.0 \pm 12.0$ & \cite{Moresco12} \\
0.875 & $125.0 \pm 17.0 $ & \cite{Moresco12} \\
0.88	& $90.0 \pm 40.0$ & \cite{Stern10} \\
0.9 & $117.0 \pm 23.0$ & \cite{Simon05} \\
1.037 & $154.0 \pm 20.0$ & \cite{Moresco12} \\
1.3 & $168.0 \pm 17.0$ & \cite{Simon05} \\
1.363 & $160.0 \pm 33.6$ & \cite{Moresco15} \\
1.43	& $177.0 \pm18.0$ & \cite{Simon05} \\
1.53	& $140.0	\pm 14.0$ & \cite{Simon05} \\
1.75	 & $202.0 \pm 40.0$ & \cite{Simon05} \\
1.965& $186.5 \pm 50.4$ & \cite{Moresco15} \\
\hline
\hline
\end{tabular}
 \label{tab:OHD}
 }
\end{table}

\begin{table}
\setlength{\tabcolsep}{1.5em}
\small
\tbl{Baryon acoustic oscillations measurements.}
{\begin{tabular}{ c c c }
\hline
\hline
 $z$ &$d_V \pm \sigma_{d_V}$  &  Ref. \\
\hline
0.106 & 0.336 $\pm$  0.015    &  \cite{Beutler11}\\
0.15 & 0.2239 $\pm$ 0.0084 & \cite{Ross15} \\
0.32 &  0.1181 $\pm$ 0.0023  & \cite{Anderson14} \\
0.57 & 0.0726 $\pm$ 0.0007 & \cite{Anderson14} \\
2.34 & 0.0320 $\pm$ 0.0016  & \cite{Delubac15} \\
2.36 &  0.0329 $\pm$ 0.0012 & \cite{Font-Ribera14}\\
\hline
\hline
\end{tabular}
 \label{tab:BAO}
 }
\end{table}

\section{Pad\'e approximations of the luminosity distance}
\label{sec:Pade approx}

\small
\begin{equation}
P_{1,1}(z)=\dfrac{1}{H_0}\left[\dfrac{2z}{2+(-1+q_0)z}\right] .
\label{eq:P_11}
\end{equation}

\begin{equation}
P_{1,2}(z)=\dfrac{1}{H_0}\left[\dfrac{12z}{12 + 6 (-1 + q_0) z + (5 + 2 j_0 - q_0 (8 + 3 q_0)) z^2}\right] .
\label{eq:P_12}
\end{equation}

\begin{equation}
P_{2,1}(z)=\dfrac{1}{H_0}\left[\dfrac{z (6 (-1 + q_0) + (-5 - 2 j_0 + q_0 (8 + 3 q_0)) z)}{-2 (3 + z + j_0 z) + 2 q_0 (3 + z + 3 q_0 z)}\right] .
\label{eq:P_21}
\end{equation}

\begin{align}
P_{1,3}(z)&=\dfrac{24z}{H_0}\Big/(24 + 12 (-1 + q_0) z + 2 (5 + 2 j_0 - q_0 (8 + 3 q_0)) z^2 - (9 + j_0 (9 + 6 q_0)- q_0 (19  \nonumber \\
&+2 q_0 (7 + 3 q_0)) + s_0) z^3)\ .
\label{eq:P_13}
\end{align}

\begin{align}
P_{2,2}(z)&=\dfrac{1}{H_0}(6 z (10 + 9 z - 6 q_0^3 z + s_0 z - 2 q_0^2 (3 + 7 z) - q_0 (16 + 19 z) +
j_0 (4 + (9 + 6 q_0) z))\Big/ \nonumber \\	
&(60 + 24 z + 6 s_0 z - 2 z^2	+ 4 j_0^2 z^2 - 9 q_0^4 z^2 - 3 s_0 z^2 + 6 q_0^3 z (-9 + 4 z) + q_0^2 (-36 - 114 z + 19 z^2) \nonumber \\
& +j_0 (24 + 6 (7 + 8 q_0) z + (-7 - 23 q_0 + 6 q_0^2) z^2) +  q_0 (-96 - 36 z + (4 + 3 s_0) z^2))\ .
\label{eq:P_22}
\end{align}

\begin{align}
P_{3,1}(z)&= \dfrac{1}{H_0}(z (24 (1 + j_0 - q_0 (1 + 3 q_0)) + 6 (4 + j_0 (7 + 8 q_0) - q_0 (6 + q_0 (19 + 9 q_0)) + s_0) z \nonumber \\
&+(2 - 4 j_0^2 + j_0 (7 + (23 - 6 q_0) q_0) +  q_0 (-4 + q_0 (-19 + 3 q_0 (-8 + 3 q_0)) - 3 s_0) + 3 s_0) z^2))\Big/ \nonumber \\
&(24 (1 + j_0 - q_0 (1 + 3 q_0)) + 6 (2 + 5 j_0 (1 + 2 q_0) - q_0 (2 + 15 q_0 (1 + q_0)) + s_0) z)\ .
\label{eq:P_31}
\end{align}

\section{Rational Chebyshev approximations of the luminosity distance}
\label{sec:Cheb approx}

 \begin{align}
 R_{1,1}(z)&=-\dfrac{1}{H_0}\Big[(3 (2 - 2 q_0 - 15 q_0^2 - 15 q_0^3 + 5 j_0 (1 + 2 q_0) + s_0) +(1/(7 - j_0 + q_0 + 3 q_0^2))(-72 (7 \nonumber \\
& - j_0 + q_0 + 3 q_0^2)^2+ 3 (18 + 5 j_0 (1 + 2 q_0) - 3 q_0 (6 + 5 q_0 (1 + q_0)) + s_0) (14 + 5 j_0 (1 + 2 q_0)\nonumber \\
& - q_0 (14 + 15 q_0 (1 + q_0)) + s_0)+2 (14 + 5 j_0 (1 + 2 q_0) - q_0 (14 + 15 q_0 (1 + q_0)) + s_0)^2) z)\Big/ \nonumber \\
&(576 (1 - ((14 + 5 j_0 (1 + 2 q_0) - q_0 (14 + 15 q_0 (1 + q_0)) + s_0) z)/( 3 (7 - j_0 + q_0 + 3 q_0^2))))\Big].
\label{eq:R11}
 \end{align}

  \begin{align}
 R_{1,2}(z)&=\dfrac{1}{H_0}(-((44184 + 5 j_0^3 (1 + 2 q_0) (1 + 10 q_0) (9 + 10 q_0) - 3 q_0 (32024 + q_0 (26948 + q_0 (4780 \nonumber \\
 &+q_0 (-15938 +q_0 (2134+ 15 q_0 (565 + 3 q_0 (249 + 25 q_0 (3 + q_0))))))))+7148 s_0 + q_0 (-5272 \nonumber \\
 &+ q_0 (-16 + 3 q_0 (-32 + 3 q_0 (439 + 75 q_0 (2 + q_0))))) s_0- 3 (-38 + q_0 (38 + 15 q_0 (1 + q_0))) s_0^2 \nonumber \\
 &+ s_0^3+j_0 (49884 +  q_0 (34880 + q_0 (-41632 + q_0 (-9472 + 135 q_0 (125 + q_0 (332 + 25 q_0 (5  \nonumber \\
  &+ 2 q_0)))))) +916 s_0 - 2 q_0 (-586 + 3 q_0 (439 + 75 q_0 (3 + 2 q_0))) s_0+15 (1 + 2 q_0) s_0^2)+j_0^2 (-222 \nonumber \\
&+ 59 s_0+ q_0 (8422 - 5 q_0 (17 + 15 q_0 (211 + 60 q_0 (2 + q_0)) - 60 s_0) + 300 s_0)))/(8 (-1 - j_0 + q_0 \nonumber \\
   &+ 3 q_0^2) (7 - j_0 + q_0 + 3 q_0^2) +24 (7 - j_0 + q_0 + 3 q_0^2)^2 - (18 + 5 j_0 (1 + 2 q_0) - 3 q_0 (6 + 5 q_0 (1\nonumber \\
   & + q_0))+ s_0)(14 + 5 j_0 (1 + 2 q_0) - q_0 (14 + 15 q_0 (1 + q_0)) s_0)))+4 (271 - 17 j_0 + 17 q_0 + 51 q_0^2 \nonumber \\
 &  +(4 (39106 - 56 j_0^3 + j_0^2 (1665 + q_0 (193 + 454 q_0)) + 5 s_0 -  j_0 (14469 - 5 s_0+ q_0 (3492 + q_0 (10127 \nonumber \\
 &+ q_0 (1033 + 1287 q_0)) + 5 s_0))+q_0 (14282 - 10 s_0 + q_0 (45365 - 10 s_0 +  q_0 (10501 + 3 q_0 (5082 \nonumber \\
 &+q_0 (479 + 429 q_0)) + 15 s_0)))))/(-868 + j_0^2 (-7 + 100 q_0 (1 + q_0)) +  q_0 (-888+ q_0 (-1412 \nonumber \\
& + 3 q_0 (-64 + q_0 (139 + 75 q_0 (2 + q_0))))) +  32 s_0 - 2 q_0 (16 + 15 q_0 (1 + q_0)) s_0+s_0^2 + j_0 (544  \nonumber \\
   & + 10 s_0- 2 q_0 (q_0 (139 + 75 q_0 (3 + 2 q_0)) - 2 (56+ 5 s_0))))) z)\Big/(576 (1 - (4 (214 - 5 j_0^2 (1 + 2 q_0) \nonumber \\
    &+ j_0 (65 + 5 q_0 (33 + q_0 (8 + 9 q_0)) - s_0) + 15 s_0 + q_0 (-204 - 5 q_0 (41+ 3 q_0 (18 + q_0 (4 + 3 q_0))) \nonumber \\
    & + s_0 + 3 q_0 s_0)) z)/(8 (-1 - j_0 + q_0 + 3 q_0^2) (7 -j_0 + q_0 + 3 q_0^2) + 24 (7 - j_0 + q_0 + 3 q_0^2)^2-(18 \nonumber \\
    & + 5 j_0 (1 + 2 q_0) - 3 q_0 (6 + 5 q_0 (1 + q_0)) + s_0) (14 + 5 j_0 (1 + 2 q_0)  - q_0 (14 + 15 q_0 (1 + q_0)) + s_0))\nonumber \\
    & -(4 (12 (-1-j_0 + q_0 + 3 q_0^2) (7 - j_0 + q_0 + 3 q_0^2) +  4 (1 + j_0 - q_0 (1 + 3 q_0))^2 - (14 + 5 j_0 (1\nonumber \\
   &+ 2 q_0) - q_0 (14 + 15 q_0 (1 + q_0)) + s_0)^2) (-1 + 2 z^2))/(3 (8 (-1 - j_0 + q_0 + 3 q_0^2) (7 - j_0 + q_0 \nonumber \\
   & + 3 q_0^2)+ 24 (7 - j_0 + q_0 + 3 q_0^2)^2 - (18 + 5 j_0 (1 + 2 q_0) - 3 q_0 (6 + 5 q_0 (1 + q_0)) + s_0) (14 \nonumber \\
   & + 5 j_0 (1 + 2 q_0) - q_0 (14 + 15 q_0 (1 + q_0)) + s_0)))))\ .
   \label{eq:R12}
 \end{align}

 \begin{align}
 R_{2,1}(z)&=\dfrac{1}{H_0}(-((3 (16 (-1 - j_0 + q_0 + 3 q_0^2) (7 - j_0 + q_0 + 3 q_0^2) - (18 + 5 j_0 (1 + 2 q_0) - 3 q_0 (6 \nonumber \\
  &+ 5 q_0 (1 + q_0)) + s_0) (14 + 5 j_0 (1 + 2 q_0) - q_0 (14 + 15 q_0 (1 + q_0)) + s_0)))/(14 + 5 j_0 (1 + 2 q_0) \nonumber \\
  &- q_0 (14 + 15 q_0 (1 + q_0)) + s_0))+4 (47 - j_0 + q_0 + 3 q_0^2 - (12 (-1 + q_0) (1 + j_0 - q_0 (1 + 3 q_0)))/\nonumber \\
  &(14 + 5 j_0 (1 + 2 q_0) - q_0 (14 + 15 q_0 (1 + q_0)) + s_0)) z -(4 (12 (-1- j_0 + q_0 + 3 q_0^2) (7 - j_0 + q_0 \nonumber \\
  &+ 3 q_0^2) + 4 (1 + j_0 - q_0 (1 + 3 q_0))^2 - (14 + 5 j_0 (1 + 2 q_0) - q_0 (14  + 15 q_0 (1 + q_0))+ s_0)^2) (-1 \nonumber \\
  &+ 2 z^2))/(14 + 5 j_0 (1 + 2 q_0) - q_0 (14 + 15 q_0 (1 + q_0)) + s_0))\Big/(192 (1 + (4 (1 + j_0 - q_0 (1 \nonumber \\
  & + 3 q_0)) z)/(14+5 j_0 (1 + 2 q_0) - q_0 (14 + 15 q_0 (1 + q_0)) + s_0))) \ .
  \label{eq:R21}
 \end{align}

 \begin{align}
 R_{2,2}(z)&=-\dfrac{1}{H_0}\Big[(2808 + 3828 s_0 + 474 s_0^2 + 15 s_0^3 + 891712 z + 115776 s_0 z + 4560 s_0^2 z + 753744 z^2  \nonumber \\
  &+ 142392 s_0 z^2 + 4284 s_0^2 z^2 +90 s_0^3 z^2 - 50625 q_0^9 (1 + 6 z^2) -10125 q_0^8 (15 - 16 z + 90 z^2) \nonumber \\
  &- 135 q_0^7 (1723 - 2800 z + 12738 z^2) +135 q_0^6 (-845 + 10688 z - 7470 z^2 + 75 s_0 (1 + 6 z^2))\nonumber \\
  &-q_0^3 (-13140 + 440320 z - 79608 z^2 + 675 s_0^2 (1 + 6 z^2) +576 s_0 (2 + 275 z + 12 z^2)) + 18 q_0^5 (4463 \nonumber \\
  & + 117160 z + 23082 z^2 + 75 s_0 (15 - 16 z + 90 z^2))+5 j_0^3 (183 - 688 z + 1098 z^2 + 3000 q_0^3 (1 \nonumber \\
  &+ 6 z^2) +  300 q_0^2 (15 - 16 z + 90 z^2) + 6 q_0 (311 - 800 z + 1866 z^2))-3 q_0^2 (9420 + 555072 z \nonumber \\
 & + 541384 z^2 +15 s_0^2 (15 -16 z + 90 z^2) + 8 s_0 (385 + 5300 z + 2718 z^2)) + 9 q_0^4 (15158 + 229520 z \nonumber \\
 &+ 214372 z^2 + s_0 (2513 - 3200 z + 19878 z^2)) -6 q_0 (s_0^2 (79 - 40 z + 714 z^2) + 4 s_0 (319 + 4624 z \nonumber \\
 &+ 5594 z^2) + 4 (351 + 61768 z + 69386 z^2))-3 j_0^2 (s_0 (-311 + 800 z  - 1866 z^2) + 22500 q_0^5 (1 \nonumber \\
 &+ 6 z^2) + 3000 q_0^4 (15 - 16 z + 90 z^2) + 125 q_0^3 (313 - 544 z + 2262 z^2) - 2 (1079 + 6040 z \nonumber \\
  &+ 12986 z^2)-5 q_0^2 (-293 + 37264 z - 1758 z^2 + 300 s_0 (1 + 6 z^2)) - 2 q_0 (6181 + 71720 z  \nonumber \\
  &+62254 z^2 + 50 s_0 (15 - 16 z  + 90 z^2)))+3 j_0 (33750 q_0^7 (1 + 6 z^2) + 5625 q_0^6 (15 - 16 z + 90 z^2) \nonumber \\
  &+ 5 s_0^2 (15 - 16 z + 90 z^2) + 4 s_0 (363 + 3400 z + 3634 z^2) + 60 q_0^5 (1723 - 2800 z + 12738 z^2) \nonumber \\
  &+4 (1019 + 62496 z + 78914 z^2) - 15 q_0^4 (-1745 + 39664 z - 15270 z^2 + 300 s_0 (1 + 6 z^2)) \nonumber \\
  &+2 q_0 (75 s_0^2 (1 + 6 z^2) + 64 (36 + 1895 z + 1520 z^2) + 2 s_0 (427 + 8200 z + 3762 z^2))-2 q_0^3 (225 s_0 (15 \nonumber \\
  &- 16 z + 90 z^2) + 8 (2477 + 42040 z + 19398 z^2))-2 q_0^2 (s_0 (2513 - 3200 z + 19878 z^2) + 12 (1433 \nonumber \\
  &+ 18780 z + 21502 z^2))))\Big/(384 (-2348 - 324 s0 - 15 s_0^2 - 784 z - 200 s_0 z + 104 z^2 + 136 s_0 z^2  \nonumber \\
  &+ 10 s_0^2 z^2 + 1125 q0^6 (-3 + 2 z^2) + 90 q_0^5 (-75 - 4 z + 50 z^2) + q_0^4 (-6795 - 480 z + 3138 z^2) \nonumber \\
  &-6 q_0^3 (25 s_0 (-3 + 2 z^2) + 8 (-20 - 35 z + 16 z^2))  - 4 q_0 (-982 - 296 z + 52 z^2 + s_0 (-81 - 2 z \nonumber \\
  &+ 34 z^2))+j_0^2 (-215 - 40 z + 122 z^2 + 500 q_0^2 (-3 + 2 z^2) + 20 q_0 (-75 - 4 z + 50 z^2)) + q_0^2 (8 (578 \nonumber \\
  &+ 475 z - 146 z^2) - 6 s_0 (-75- 4 z + 50 z^2))-2 j_0 (1034 + 700 z - 212 z^2 + s_0 (75 + 4 z -50 z^2) \nonumber \\
  &+ 750 q_0^4 (-3 + 2 z^2) + 45 q_0^3 (-75 - 4 z + 50 z^2) + q_0^2 (-2265 - 160 z + 1046 z^2)+q_0 (970 + 780 z \nonumber \\
  &- 468 z^2  - 50 s_0 (-3 + 2   z^2)))))\Big] .
  \label{eq:R22}
 \end{align}

\break

%\section{References}
%
%References are to be listed in the order cited in the text in Arabic
%numerals.  They should be listed according to the style shown in the
%References. Typeset references in 9 pt roman.
%
%References in the text can be typed in superscripts,
%e.g.: ``$\ldots$ have proven\cite{edbk,rvo,seri} that
%this equation $\ldots$'' or after punctuation marks:
%``$\ldots$ in the statement.\cite{seri}'' This is
%done using LaTeX command: ``\verb|\cite{name}|''.
%
%When the reference forms part of the sentence, it should not
%be typed in superscripts, e.g.: ``One can show from
%Ref.~\refcite{edbk} that $\ldots$'', ``See
%Refs.~\citen{jpap,colla,edbk}, \citen{seri}
%and \citen{publ} for more details.''
%This is done using the LaTeX
%command: ``\verb|Ref.~\refcite{name}|''.


\begin{thebibliography}{0}    %for 1 digit

%%%journal paper
%\bibitem{jpap} R. Loren and D. B. Benson, {\it J. Comput.
%System Sci.} {\bf 27} (1983) 400.
%
%%%collaboration
%\bibitem{colla} OPAL Collab. (G. Abbiendi {\it et al}.),
%{\it Eur. J. Phys. C} {\bf 11} (1999) 217.
%
%%%normal book (editors)
%\bibitem{edbk} R. Loren and D. B. Benson (eds.), {\it Introduction to
%String Field Theory}, 2nd edn. (Springer-Verlag, New York, 1999).
%
%%%review volume
%\bibitem{rvo} C. M. Wang, J. N. Reddy and K. H. Lee, New set of
%buckling parameters, in {\it Shear Deformable Beams}, ed.~T. Rex
%(Elsevier, Oxford, 2000), p.~201.
%
%%%book in a series
%\bibitem{seri} R. Loren, J. Li and D. B. Benson, Deterministic flow-chart
%interpretations, in {\it Introduction to String Field Theory},
%Advanced Series in Mathematical Physics, Vol.~3 (Springer-Verlag, New York, 1999),
%p.~401.
%
%%%proceedings
%\bibitem{pro} R. Loren, J. Li and D. B. Benson, Deterministic
%flow-chart interpretations, in {\it Proc. 3rd Int. Conf.
%Entity-Relationship Approach}, eds. C. G. Davis and R. T. Yeh
%(North-Holland, Amsterdam, 1983), p.~421.
%
%%%to be published
%\bibitem{publ} R. Loren, J. Li and D. B. Benson, Deterministic
%flow-chart interpretations, to appear in {\it J. Comput. System Sci.}

\bibitem{Weinberg72}
S. Weinberg, {\it Gravitation and Cosmology} (Wiley, New York, 1972).

\bibitem{Guth81}
A. H. Guth, {Phys. Rev. D} \textbf{23} (1981)
347.

\bibitem{Capozziello-Faraoni10}
S. Capozziello and V. Faraoni, \textit{Beyond Einstein Gravity: A Survey of Gravitational Theories for Cosmology and Astrophysics}, {Fundam. Theor. Phys.} \textbf{170} (2010).

\bibitem{Capozziello08}
S. Capozziello and M. Francaviglia, {Gen. Rel. Grav.} \textbf{40} (2008) 35.

\bibitem{Buchbinder92}
I. L. Buchbinder, S. D. Odintsov and I. Shapiro, \textit{Effective Action in Quantum Gravity} (IOP Publishing, Bristol, 1992).

\bibitem{Birrell82}
N. D. Birrell and P. C. W. Davies, \textit{Quantum Fields in Curved Space} (Cambridge University Press, Cambridge, 1982).

\bibitem{Perlmutter99}
S. Perlmutter \textit{et al.}, {Astrophys. J.} \textbf{517} (1999) 565.

\bibitem{Riess-Schmidt98}
A. G. Riess \textit{et al}., {Astron. J.} \textbf{116} (1998) 1009;
B. Schmidt \textit{et al.}, {Astrophys. J.} \textbf{507} (1998) 46.

\bibitem{roccoacc}
B. S. Haridasu, V. V. Lukovi\'c, R. D'Agostino and N. Vittorio, {Astron. Astrophys.} \textbf{600} (2017) L1.


\bibitem{WMAP9}
 WMAP Collab. (G. Hinshaw {\it et al}.), {Astrophys. J. Supp. Ser.} \textbf{208} (2013) 2.

\bibitem{Planck15}
Planck Collab. (P. A. R. Ade \textit{et al.}), {Astron. Astrophys.} \textbf{594} (2016) A132015.

\bibitem{Sahni00}
V. Sahni and A. Starobinsky, {Int. J. Mod. Phys. D} \textbf{9} (2000) 373.

\bibitem{Weinberg89}
S. Weinberg, {Rev. Mod. Phys.} \textbf{61} (1989) 1.

\bibitem{Peebles03}
P. J. E. Peebles and B. Ratra, {Rev. Mod. Phys.} \textbf{75} (2003) 559.

\bibitem{Zlatev99}
I. Zlatev, L. Wang and P. J. Steinhardt, {Phys. Rev. Lett.} \textbf{82} (1999) 896.

\bibitem{Sahni02}
V. Sahni, {Class. Quant. Grav.} \textbf{19} (2002) 3435.

\bibitem{Carroll00}
S. M. Carroll, {Living Rev. Rel.} \textbf{4} (2001) 1.

\bibitem{Huterer01}
D. Huterer and M. S. Turner, {Phys. Rev. D} \textbf{64} (2001) 123527.

\bibitem{Padmanabhan03}
T. Padmanabhan, {Phys. Rept.} \textbf{380} (2003) 235.

\bibitem{Caldwell03}
R. R. Caldwell, M. Kamionkowski and N. N. Weinberg, {Phys. Rev. Lett.} \textbf{91} (2003) 071301.

\bibitem{Copeland06}
E. J. Copeland, M. Sami and S. Tsujikawa, {Int. J. Mod. Phys. D} \textbf{15} (2006) 1753.

\bibitem{Li11}
M. Li, X. D. Li, S. Wang and Y. Wang, {Commun. Theor. Phys.} \textbf{56} (2011) 525.

\bibitem{Kamenshchik01}
A. Y. Kamenshchik, U. Moschella and V. Pasquier,  {Phys. Lett. B} \textbf{511} (2001) 265.

\bibitem{Padmanabhan02}
T. Padmanabhan, {Phys. Rev. D} \textbf{66} (2002) 021301.

\bibitem{Capozziello02}
S. Capozziello, {Int. J. Mod. Phys. D} \textbf{11} (2002) 483.

\bibitem{Cai15}
Y. F. Cai, S. Capozziello, M. De Laurentis and E. N. Saridakis, {Rept. Prog. Phys.} \textbf{79}  (2016) 106901.

\bibitem{Friedmann}
A. Friedmann, {Zeitschrift f\"ur Physik} \textbf{10} (1922) 377.

\bibitem{Lemaitre31}
G. Lema\^itre, {Mon. Not. Roy. Astron. Soc.}, \textbf{91} (1931) 483.

\bibitem{Robertson35}
H. P. Robertson, {Astrophys. J.} \textbf{82} (1935) 284.

\bibitem{Walker37}
A. G. Walker, {Proc. London Math. Soc.} \textbf{s2-42} (1937) 90.

\bibitem{roccoH0}
V. V. Lukovi\'c, R. D'Agostino and N. Vittorio, {Astron. Astrophys.} \textbf{595} (2016) A109.

\bibitem{Hubble29}
E. Hubble, {Proc. Nat. Acad. Sci.}, \textbf{15} (1929) 168.

\bibitem{Freedman01}
W. L. Freedman \textit{et al.}, {Astrophys. J.} \textbf{553} (2001) 47.

\bibitem{Riess16}
A. G. Riess \textit{et al.}, {Astrophys. J.} \textbf{826}  (2016) 56.

\bibitem{Bonvin16}
V. Bonvin \textit{et al.}, {Mon. Not. Roy. Astron. Soc.} \textbf{465} (2017) 4914.

\bibitem{LMmodel}
O. Luongo, M. Muccino, {Phys. Rev. D} \textbf{98} (2018) 103520.

\bibitem{dopolm1}
O. Luongo, H. Quevedo, {Int. J. Mod. Phys. D}, {\bf 23}, (2014) 1450012.


\bibitem{dopolm2}
O. Luongo, H. Quevedo, {Astroph. sp. sci.}, {\bf 338}, (2012) 345.



\bibitem{dopolm3}
P. K. S. Dunsby, O. Luongo, L. Reverberi, {Phys. Rev. D}, {\bf 94}, (2016) 083525.



\bibitem{dopolm4}
A. Aviles, N. Cruz, J. Klapp, O. Luongo, {Gen. Rel. Grav.}, {\bf 47}, (2015), 63.



\bibitem{dopolm5}
S. Capozziello, R. D'Agostino, O. Luongo, {Phys. Dark Univ.}, {\bf 20}, (2018), 12.



\bibitem{dopolm6}
S. Capozziello, R. D'Agostino, R. Giambo', O. Luongo, {Phys. Rev. D}, {\bf 99}, (2019), 023532.


\bibitem{dopolm7}
K. Boshkayev, R. D'Agostino, O. Luongo, ArXiv[gr-qc]:1901.01031, (2019).


\bibitem{Linde82}
A. D. Linde, {Phys. Lett. B}  \textbf{108} (1982) 389.

\bibitem{Albrecht82}
A. Albrecht and P. J. Steinhardt, {Phys. Rev. Lett.} \textbf{48} (1982) 1220.

\bibitem{Peebles-Ratra88}
P. J. Peebles and R. Ratra, {Astrophys. J.} \textbf{325} (1988) 1220;
B. Ratra and P. J. E. Peebles, {Phys. Rev. D} \textbf{37} (1988) 3406.

\bibitem{Caldwell98}
R. R. Caldwell, R. Dave and P. J. Steinhardt,  {Phys. Rev. Lett.} \textbf{80} (1998) 1582.

\bibitem{Armendariz00}
C. Armendariz-Picon, V. Mukhanov and  P. J. Steinhardt, {Phys. Rev. Lett.} \textbf{85} (2000) 4438.

\bibitem{Ferreira98}
P. G. Ferreira and M. Joyce, {Phys. Rev. D} \textbf{58} (1998) 023503.

\bibitem{Copeland98}
E. J. Copeland, A. R. Liddle and  D. Wands, {Phys. Rev. D} \textit{57} (1998) 4686.

\bibitem{Steinhardt99}
P. J. Steinhardt, L. Wang and I. Zlatev, {Phys. Rev. D} \textbf{59} (1999) 123504.

\bibitem{Lemaitre33}
G. Lema\^{i}tre, {Ann. Soc. Sci. Bruxelles} \textbf{53} (1933) 51.

\bibitem{Tolman34}
R. C. Tolman, {Proc. Nat. Acad. Sci.} \textbf{20} (1934) 169.

\bibitem{Bondi47}
H. Bondi, {Mon. Not. Roy. Astron. Soc.} \textbf{107} (1947) 410.

\bibitem{Nadathur11}
S. Nadathur and S. Sarkar, {Phys. Rev. D} \textbf{83} (2011) 063506.

\bibitem{review1}
S. Capozziello and M. De Laurentis, {Phys. Rept.} \textbf{509} (2011) 167.

\bibitem{review2}
S. Nojiri, S. D. Odintsov and V. K. Oikonomou, {Phys. Rept.} \textbf{692} (2017) 1.

\bibitem{Nojiri07}
S. Nojiri and S. D. Odintsov, {Int. J. Geom. Meth. Mod. Phys.} \textbf{4} (2007) 115.

\bibitem{GW}
LIGO Scientific and Virgo Collab. (B. P. Abbott \textit{et al.}), {Phys. Rev. Lett.} \textbf{116} (2016) 061102.

\bibitem{Starobinsky80}
A. A. Starobinsky, {Phys. Lett. B} \textbf{91} (1980) 99.

\bibitem{Duruisseau83}
J. P. Duruisseau, R. Kerner and P. Eysseric, {Gen. Rel. Grav.} \textbf{15} (1983) 797.
	
\bibitem{La89}
D. La and P. J. Steinhardt, {Phys. Rev. Lett.} \textbf{62} (1989) 1066.

\bibitem{Maeda89}
K. Maeda, {Phys. Rev. D} \textbf{39} (1989) 3159.

\bibitem{Wands94}
D. Wands, {Class. Quant. Grav.} \textbf{11} (1994) 269.

\bibitem{Capozziello98}
S. Capozziello, R. de Ritis and A. A. Marino, {Gen. Rel. Grav.} \textit{30} (1998) 1247.

\bibitem{Utiyama62}
R. Utiyama and B. S. De Witt,  {J. Math. Phys.} \textbf{3} (1962) 608.

\bibitem{Capozziello}
S. Capozziello, {Int. J. Mod. Phys. D} \textbf{11} (2002) 483;
S. Capozziello, V. F. Cardone, S. Carloni and A. Troisi, {Int. J. Mod. Phys. D} \textbf{12} (2003) 1969;
S. Capozziello, V. F. Cardone and A. Trosi, {Phys. Rev. D} \textbf{71} (2005) 043503.

\bibitem{Carroll04}
S. M. Carroll, V. Duvvuri, M. Trodden and M. S. Turner, {Phys. Rev. D} \textbf{70} (2004) 043528.

\bibitem{Sotiriou10}
T. P. Sotiriou and V. Faraoni, {Rev. Mod. Phys.} \textbf{82} (2010) 451.

\bibitem{DeFelice10}
A. De Felice and S. Tsujikawa, {Living Rev. Rel.} \textbf{13} (2010) 3.

\bibitem{Nojiri11}
S. Nojiri and S. D. Odintsov, {Phys. Rept.} \textbf{505} (2011) 59.

\bibitem{Palatini19}
A. Palatini, {Rend. Circ. Mat. Palermo} \textbf{43} (1919) 203.

\bibitem{Ferraris94}
M. Ferraris, M. Francaviglia and I. Volovich, {Class. Quant. Grav.} \textbf{11} (1994) 1505.

\bibitem{Magnano94}
G. Magnano and L. M. Sokolowski, {Phys. Rev. D} \textbf{50} (1994) 5039.

\bibitem{Borunda08}
M. Borunda, B. Janssen and M. Bastero-Gil,  {J. Cosm. Astrop. Phys.} \textbf{0811} (2008) 008.

\bibitem{Hehl78}
F. W. Hehl and G. D, Kerlick, {Gen. Rel. Gravit.} \textbf{9} (1978) 691.

\bibitem{Capozziello07}
S. Capozziello, R. Cianci, C. Stornaiolo and S. Vignolo, {Class. Quant. Grav.} \textbf{24} (2007) 6417.

\bibitem{Sotiriou07}
T. P. Sotiriou and S. Liberati, {Ann. Phys.} \textbf{322} (2007) 935.

\bibitem{Wang90}
Y. Wang,  {Phys. Rev. D} \textbf{42} (1990) 2541.

\bibitem{Nojiri03}
S. Nojiri and S. D. Odintsov,  {Phys. Rev. D} \textbf{68} (2003) 123512.

\bibitem{Dolgov03}
A. D. Dolgov and M. Kawasaki, {Phys. Lett. B}, \textbf{1573} (2003) 1.

\bibitem{Soussa04}
M. E. Soussa and R. P. Woodard,  {Gen. Rel. Grav.} \textbf{36} (2004) 855.

\bibitem{Amendola07}
L. Amendola, D. Polarski and S. Tsujikawa, {Phys. Rev. Lett.} \textbf{98} (2007) 131302;
L. Amendola, D. Polarski, and S. Tsujikawa,  {Int. J. Mod. Phys. D} \textbf{16} (2007) 1555.

\bibitem{Troisi}
%\cite{Capozziello:2006dj}
%\bibitem{Capozziello:2006dj}
  S.~Capozziello, S.~Nojiri, S.~D.~Odintsov and A.~Troisi,
  %``Cosmological viability of f(R)-gravity as an ideal fluid and its compatibility with a matter dominated phase,''
  Phys.\ Lett.\ B {\bf 639} (2006) 135.

\bibitem{Olmo05}
G. J. Olmo, {Phys. Rev. D} \textbf{72} (2005) 083505.

\bibitem{Faraoni06}
V. Faraoni, {Phys. Rev. D} \textbf{74} (2006) 023529.

\bibitem{Amendola07c}
L. Amendola, R. Gannouji, D. Polarski and S. Tsujikawa, {Phys. Rev. D}, \textbf{75} (2007) 083504.

\bibitem{Carroll06}
S. M. Carroll, I. Sawicki, A. Silvestri, M. Trodden, New J. Phys., \textbf{8}, 323 (2006).

\bibitem{Song07}
Y. S. Song, W. Hu, I. Sawicki, Phys. Rev. D, \textbf{75}, 044004 (2007).

\bibitem{Bean07}
R. Bean, D. Bernat, L. Pogosian, A. Silvestri, M. Trodden, Phys. Rev. D, \textbf{75}, 064020 (2007).

\bibitem{Chiba07}
T. Chiba, T. L. Smith and A. L. Erickcek, {Phys. Rev. D}, \textbf{75}, 124014 (2007).

\bibitem{Hu07}
W. Hu and I. Sawicki, {Phys. Rev. D} \textbf{7676} (2007) 064004.

\bibitem{Appleby07}
S. A. Appleby and R. A. Battye, {Phys. Lett. B} \textbf{654} (2007) 7.

\bibitem{Cognola}
G. Cognola, E. Elizalde, S. Nojiri, S.D. Odintsov, L. Sebastiani, S. Zerbini,  Phys.Rev. D \textbf{77} (2008) 046009.
%DOI: 10.1103/PhysRevD.77.046009
%e-Print: arXiv:0712.4017
\bibitem{Noj1}
S. Nojiri, S. D. Odintsov,  Phys.Rev. D \textbf{77} (2008) 026007.

%DOI: 10.1103/PhysRevD.77.026007
%e-Print: arXiv:0710.1738

\bibitem{Emilio}
E. Elizalde, S. Nojiri, S.D. Odintsov, L. Sebastiani, S. Zerbini, Phys. Rev. D {\bf 83} (2011) 086006.

\bibitem{Vollick03}
D. N. Vollick, {Phys. Rev. D} \textbf{68} (2003) 063510.

\bibitem{Meng03}
X. Meng and P. Wang, {Gen. Rel. Grav.} \textbf{36} (2004) 2673.

\bibitem{Brans61}
C. Brans and R. H. Dicke, {Phys. Rev.} \textbf{124} (1961) 925.

\bibitem{Chiba03}
T. Chiba, {Phys. Lett. B} \textbf{575} (2003) 1.

\bibitem{Faraoni07}
V. Faraoni and S. Nadeau, Phys. Rev. D \textbf{75} (2007) 023501.

\bibitem{vignolo-cauchy}
%%\cite{Capozziello:2009pi}
%\item%{Capozziello:2009pi}
%{\bf ``The Cauchy problem for metric-affine f(R)-gravity in presence of perfect-fluid matter''}
  S.~Capozziello and S.~Vignolo,
Class.\ Quant.\ Grav.\  {\bf 26}, 175013 (2009).


\bibitem{Briscese} 
F. Briscese, E. Elizalde, S. Nojiri, S.D. Odintsov,  Phys.Lett. B {\bf 646} (2007) 105.
%DOI: 10.1016/j.physletb.2007.01.013
%e-Print: hep-th/0612220


\bibitem{Baha1}
S. Bahamonde, S.D. Odintsov, V.K. Oikonomou,  Annals Phys. {\bf 373} (2016) 96.
%DOI: 10.1016/j.aop.2016.06.020
%e-Print: arXiv:1603.05113

\bibitem{Baha2}
S. Bahamonde, S. D. Odintsov, V.K. Oikonomou, P. V. Tretyakov, Phys.Lett. B {\bf 766} (2017) 225.
%DOI: 10.1016/j.physletb.2017.01.012
%e-Print: arXiv:1701.02381

%\cite{Harko:2011nh}
\bibitem{hybrid1}
  T.~Harko, T.~S.~Koivisto, F.~S.~N.~Lobo and G.~J.~Olmo,
  %``Metric-Palatini gravity unifying local constraints and late-time cosmic acceleration,''
  Phys.\ Rev.\ D {\bf 85} (2012) 084016
  %doi:10.1103/PhysRevD.85.084016
  %[arXiv:1110.1049 [gr-qc]].
  %%CITATION = doi:10.1103/PhysRevD.85.084016;%%
  %86 citations counted in INSPIRE as of 25 Mar 2019

\bibitem{hybrid2}
%{\bf ``Hybrid metric-Palatini gravity''}
  S.~Capozziello, T.~Harko, T.~S.~Koivisto, F.~S.~N.~Lobo and G.~J.~Olmo,
 % \\{}arXiv:1508.04641 [gr-qc]
  %\\{}DOI:10.3390/universe1020199
  Universe {\bf 1},  199 (2015).
 % \\{}NORDITA-2015-92
%(Aug 19, 2015)
%\inspireurl{http://inspirehep.net/record/1388489}
%\citations{45 citations counted in INSPIRE as of 25 Mar 2019}

\bibitem{hybrid3}
%{\bf ``Cosmology of hybrid metric-Palatini f(X)-gravity''}
  S.~Capozziello, T.~Harko, T.~S.~Koivisto, F.~S.~N.~Lobo and G.~J.~Olmo.
 % \\{}arXiv:1209.2895 [gr-qc]
  %\\{}DOI:10.1088/1475-7516/2013/04/011
  JCAP {\bf 1304}, 011 (2013).
%(Sep 2012)
%\inspireurl{http://inspirehep.net/record/1185371}
%\citations{54 citations counted in INSPIRE as of 25 Mar 2019}


%\cite{Capozziello:2013uya}
\bibitem{hybrid4}
  S.~Capozziello, T.~Harko, F.~S.~N.~Lobo and G.~J.~Olmo,
  %``Hybrid modified gravity unifying local tests, galactic dynamics and late-time cosmic acceleration,''
  Int.\ J.\ Mod.\ Phys.\ D {\bf 22} (2013) 1342006.
  %doi:10.1142/S0218271813420066
  %[arXiv:1305.3756 [gr-qc]].
  %%CITATION = doi:10.1142/S0218271813420066;%%
  %29 citations counted in INSPIRE as of 25 Mar 2019

\bibitem{Meng1}
X. H. Meng and P. Wang, {Class. Quant. Grav.} \textbf{20} (2003) 4949.

\bibitem{Meng2}
X. H. Meng and P. Wang, {Class. Quant. Grav.} \textbf{21} (2004) 951.

\bibitem{Fay07}
S. Fay, R. Tavakol and S. Tsujikawa, {Phys. Rev. D} \textbf{75} (2007) 063509.

\bibitem{AmarzguiouiPalatini}
M. Amarzguioui, O. Elgaroy, D. F. Mota and T. Multamaki, {Astron. Astrophys.} \textbf{454} (2006) 707.

\bibitem{Koivisto06}
T. Koivisto, {Phys. Rev. D} \textbf{73} (2006) 083517.

\bibitem{LiPalatini}
B. Li and M. C. Chu, {Phys. Rev. D} \textbf{74} (2006) 104010.

\bibitem{TsujikawaPalatini}
S. Tsujikawa, K. Uddin and R. Tavakol, {Phys. Rev. D} \textbf{77} (2008) 043007.

\bibitem{Flanagan04}
E. E. Flanagan, {Phys. Rev. Lett.} \textbf{92} (2004) 071101.

\bibitem{Hehl76}
F. W. Hehl,  P. Von Der Heyde,  G. D. Kerlick and J. M. Nester, {Rev.  Mod. Phys.}, \textbf{48} (1976) 393.

\bibitem{Green87}
M. B. Green, J. H. Schwarz and E. Witten, \textit{Superstring Theory} (Cambridge University Press, Cambridge, 1987).

\bibitem{Hammond94}
R. Hammond, {Nuovo Cim. B} \textbf{109} (1994) 319.

\bibitem{deSabbata91}
V. de Sabbata and C. Sivaram, {Ann. Phys.} \textbf{48} (1991) 419.

\bibitem{deSabbata96}
V. de Sabbata, C. Sivaram and Twistors, {Nuovo Cim. A}, \textbf{109} (1996) 377.

\bibitem{Howe98}
P. S. Howe, A. Opfermann and G. Papadopoulos,  {Commun. Math. Phys.} \textbf{197} (1998) 713.

\bibitem{Hull93}
C. M. Hull, G. Papadopoulos and P. K. Townsend, {Phys. Lett. B} \textbf{316} (1993) 291.

\bibitem{Papadopoulos95}
G. Papadopoulos and P. K. Townsend, {Nucl. Phys. B}, \textbf{444} (1995) 245.

\bibitem{Gonner84}
H. Gonner and F. Mueller-Hoissen, {Class. Quant. Grav.} \textbf{1} (1984) 651.

\bibitem{Chatterjee93}
P. Chatterjee and B. Bhattacharya, {Mod. Phys. Lett. A} \textbf{8} (1993) 2249.

\bibitem{Wolf95}
C. Wolf, {Gen. Rel. Grav.} \textbf{27} (1995) 1031.


\bibitem{Ross89}
D. K. Ross, {Int. J. Theor. Phys.} \textbf{28} (1989) 1333.

\bibitem{deAndrade97}
L. C. Garcia de Andrade, {Mod. Phys. Lett. A} \textbf{12} (1997) 2005.

\bibitem{Vignolo15}
S. Vignolo, S. Carloni and L. Fabbri, {Phys. Rev. D} \textbf{91} (2015) 043528.

\bibitem{Capozziello98b}
S. Capozziello and C. Stornaiolo, {Nuovo Cim. B} \textbf{113} (1998) 879.

\bibitem{Capozziello01}
S. Capozziello, G. Lambiase and C. Stornaiolo, {Ann. Phys.} \textbf{10} (2001) 713.

\bibitem{Trautman73}
A. Trautman, {Nature} \textbf{242} (1973) 7.


\bibitem{Hawking73}
S. W. Hawking and G. F. R. Ellis, \textit{The large scale structure of space-time}  (Cambridge University Press, Cambridge, 1973).

\bibitem{Moller61}
C. M{\o}ller, {Nordita Publ.} \textbf{64} (1961) 50.

\bibitem{Maluf94}
J. W. Maluf, {J. Math. Phys.} \textbf{35} (1994) 335.

\bibitem{Unzicker05}
A. Unzicker and T. Case, physics/0503046 (2005).

\bibitem{Weitzenbock23}
R. Weitzenb\"ock, \textit{Invariantentheorie} (Noordhoff, Groningen, 1923).

\bibitem{Kobayashi63}
S. Kobayashi and K. Nomizu, \textit{Foundations of Differential Geometry} (Intersciense, New York, 1963).

\bibitem{Andrade00}
V. C. de Andrade, L. C. T. Guillen and  J. G. Pereira, {Phys. Rev. Lett.} \textbf{84} (2000) 4533.

\bibitem{Virbhadra90}
K. S. Virbhadra, {Phys. Rev. D} \textbf{42} (1990) 2919.

\bibitem{Shirafuji96}
T. Shirafuji, G. G. L. Nashed and Y. Kobayashi, {Prog. Theor. Phys.} \textbf{96} (1996) 933.

\bibitem{rocco}
S. Capozziello, R. D'Agostino and O. Luongo, {Gen. Rel. Grav.} \textbf{49} (2017) 141.

\bibitem{Capozziello15}
S. Capozziello, O. Luongo and E. N. Saridakis, {Phys. Rev. D} {\bf 91} (2015) 124037.

\bibitem{Aviles13}
A. Aviles, A. Bravetti, S. Capozziello and O. Luongo,  {Phys. Rev. D} {\bf 87} (2013) 064025.

\bibitem{Abedi17}
H. Abedi, M. Wright and  A. M. Abbassi, {Phys. Rev. D} {\bf 95}  (2017) 064020.

\bibitem{Ferraro07}
R. Ferraro and F. Fiorini,  {Phys. Rev. D} \textbf{75} (2007) 084031.

\bibitem{Bengochea09}
G. R. Bengochea and R. Ferraro,  {Phys. Rev. D} \textbf{79} (2009) 124019.

\bibitem{Linder10}
E. V. Linder,  {Phys. Rev. D} \textbf{81} (2010) 127301.

\bibitem{Geng12}
C. Q. Geng, C. C. Lee, E. N. Saridakis and Y. P. Wu, {Phys. Lett. B} \textbf{704} (2011) 384.

\bibitem{Xu12}
C. Xu, E. N. Saridakis and G. Leon, {J. Cosm. Astrop. Phys.} \textbf{07} (2012) 005.

\bibitem{D'Agostino18}
R. D'Agostino and O. Luongo, {Phys. Rev. D} \textbf{98} (2018) 124013.


\bibitem{Uzan99}
J. P. Uzan,  {Phys. Rev. D} \textbf{59} (1999) 123510.

\bibitem{Bartolo99}
N. Bartolo and M. Pietroni, {Phys. Rev. D} \textbf{61} (1999) 023518.

\bibitem{Faraoni00}
V. Faraoni, {Phys. Rev. D} \textbf{62} (2000) 023504.

\bibitem{Vardanyan15}
V. Vardanyan and L. Amendola, {Phys. Rev. D} \textbf{92} (2015) 024009.

\bibitem{Kaiser10}
D. I. Kaiser, {Phys. Rev. D} \textbf{81} (2010) 084044.

\bibitem{Abedi15}
H. Abedi and A. M. Abbassi, {J. Cosm. Astrop. Phys.} \textbf{05} (2015) 026.

\bibitem{Yang11}
R. J. Yang, {Europhys. Lett.} \textbf{93} (2011) 60001.

\bibitem{Wright16}
M. Wright, {Phys. Rev. D} \textbf{93} (2016) 103002.

\bibitem{Abedi18}
H. Abedi, S. Capozziello, R. D'Agostino and O. Luongo, {Phys. Rev. D} \textbf{97} (2018) 084008.

\bibitem{Visser}
M. Visser, {Gen. Rel. Grav.} \textbf{37} (2005) 1541;
M. Visser, {Class. Quant. Grav.} \textbf{32} (2015) 135007.

\bibitem{Dunsby16}
P. K. S. Dunsby and O. Luongo, {Int. J. Geom. Meth. Mod. Phys.} \textbf{13} (2016) 1630002.

\bibitem{Harrison76}
E. R. Harrison, {Nature} \textbf{260} (1976) 591.

\bibitem{j1}
O. Luongo, {Phys. Lett. A}, {\bf 28} (2013) 1350080.

\bibitem{j2}
S. Capozziello, M. De Laurentis, O. Luongo and A. C. Ruggeri, {Galaxies} {\bf 1} (2013) 216.

\bibitem{j3}
A. Mukherjee, N. Paul and H. K. Jassal, {J. Cosm. Astrop. Phys.} \textbf{1901} (2019) 005.

\bibitem{j4}
 S. Capozziello, O. Farooq, O. Luongo and B. Ratra, {Phys. Rev D}  \textbf{90} (2014) 044016.

\bibitem{j5}
A. Aviles, A. Bravetti, S. Capozziello and O. Luongo, {Phys. Rev. D} \textbf{87} (2013) 044012.

\bibitem{j6}
S. Capozziello, M. De Laurentis and O. Luongo, {Int. J. Mod. Phys. D} \textbf{24} (2014) 1541002.

\bibitem{j7}
 O. Luongo, {Entropy} \textbf{19} (2017) 55.

\bibitem{j8}
O. Luongo and H. Quevedo, {Gen. Rel. Grav.} \textbf{46}  (2014) 1649.


\bibitem{j9}
H. Velten, S. Gomes and V C. Busti, {Phys. Rev. D} \textbf{97} (2018) 083516.


\bibitem{j10}
U. Andrade, C. A. P. Bengaly, J. S. Alcaniz and B. Santos, {Phys. Rev. D} \textbf{97} (2018)  083518.

\bibitem{j11}
C. Rodrigues Filho, Edesio M. Barboza, {J. Cosm. Astrop. Phys.} \textbf{1807} (2018) 037.

\bibitem{j12}
O. Luongo, G. B. Pisani and A. Troisi, {Int. J. Mod. Phys. D} \textbf{26} (2016) 1750015.

\bibitem{j13}
Z. Y. Yin and H. Wei, e-Print: arXiv:1902.00289, (2019).

\bibitem{j14}
A. Al Mamon and K. Bamba,  {Eur. Phys. J. C} \textbf{78} (2018) 862.

\bibitem{j15}
A. Piloyan, S. Pavluchenko and L. Amendola, {Particles} \textbf{1} (2018)  23.


\bibitem{j17}
F. Montanari and S. Rasanen, {J. Cosm. Astrop. Phys.} 1711 (2017)  032.

\bibitem{j18}
X. B. Zou, H. K. Deng, Z. Y Yin and H. Wei,  {Phys. Lett. B} \textbf{776} (2018) 284.

\bibitem{j19}
W. Yang, L. Xu, H. Li, Y. Wu and J. Lu, {Entropy} \textbf{19} (2017) 327.

\bibitem{j20}
C. S. Carvalho, S. Basilakos,  {Astron. Astrophys.} \textbf{592} (2016) A152.

\bibitem{j21}
O. Luongo and H. Quevedo, {Found. of Phys.} \textbf{48} (2017) 1.

\bibitem{j22}
I. Semiz and A. K. Camlibel, {J. Cosm. Astrop. Phys.} \textbf{1512} (2015) 038.

\bibitem{j23}
Y. L. Bolotin, V. A. Cherkaski and O. A. Lemets, {Int. J. Mod. Phys. D} \textbf{25} (2016) 1650056.

\bibitem{j24}
B. Bochner, D. Pappas, M. Dong, {Astrophys. J.} \textbf{814} (2015) 7.

\bibitem{j25}
S. Nesseris and J. Garcia-Bellido,  {Phys. Rev. D} \textbf{88} (2013) 063521.

\bibitem{j26}
O. Farooq, S. Crandall and B. Ratra, {Phys. Lett. B} \textbf{726} (2013) 72.

\bibitem{j27}
F. A. Teppa Pannia and S. E. Perez Bergliaffa, {J. Cosm. Astrop. Phys.} \textbf{1308} (2013) 030.

\bibitem{j28}
C. J. A. P. Martins, M. Martinelli, E. Calabrese and M. P. L. P. Ramos, {Phys. Rev. D} \textbf{94} (2016) 043001.

\bibitem{j29}
F. Piazza and T. Schucker, {Gen. Rel. Grav.} \textbf{48} (2016) 41.

\bibitem{j30}
K. Bamba, S. Capozziello, S. Nojiri and S. D. Odintsov, {Astrop. Sp. Sci.} \textbf{342} (2012) 155.

\bibitem{j31}
S. Capozziello, V. F. Cardone, H. Farajollahi and A. Ravanpak, {Phys. Rev. D} 84 (2011) 043527.

\bibitem{j32}
J. C. Carvalho and J. S. Alcaniz, {Mon. Not. Roy. Astron. Soc.} \textbf{418} (2011) 1873.

\bibitem{j33}
M. Bouhmadi-Lopez, S. Capozziello and V. F. Cardone, {Phys. Rev. D} \textbf{82} (2010) 103526.

\bibitem{j34}
S. Capozziello, V. F. Cardone and V. Salzano, {Phys. Rev. D} \textbf{78} (2008) 063504.

\bibitem{j35}
M. V. John, {Astrophys. J.} \textbf{630} (2005) 667.

\bibitem{j36}
S. Capozziello, Ruchika and A. A. Sen, {Mon. Not. Roy. Astron. Soc.} \textbf{484} (2019) 4484.




\bibitem{Cattoen08}
C. Cattoen and M. Visser, {Phys. Rev. D} \textbf{78} (2008) 063501.

\bibitem{Aviles12}
A. Aviles, C. Gruber, O. Luongo and H. Quevedo, {Phys. Rev. D} \textbf{86} (2012) 123516.

\bibitem{Clarkson07}
C. Clarkson, M. Cortes and B. A. Bassett,  {J. Cosm. Astrop. Phys.} \textbf{0708} (2007) 011.

\bibitem{Pavlov13}
A. Pavlov, S. Westmoreland, K. Saaidi and B. Ratra, {Phys. Rev. D} \textbf{88} (2013) 123513.

\bibitem{Baker66}
G. A. Baker Jr. and P. Graves-Morris, \textit{Pad\'e Approximants} (Cambridge University Press, 1996).

\bibitem{Litvinov93}
G. L. Litvinov, {Appl. Russ. J. Math. Phys.} \textbf{1} (1993) 313.

\bibitem{Gruber14}
C. Gruber and O. Luongo, {Phys. Rev. D} \textbf{89} (2014) 103506.

\bibitem{Wei14}
H. Wei, X. P. Yan and Y. N. Zhou, {J. Cosm. Astrop. Phys.} \textbf{1401} (2014) 045.

\bibitem{Dutta18}
K. Dutta, Ruchika, A. Roy, A. A. Sen and M. M. Sheikh-Jabbari, e-Print: arXiv:1808.06623 (2018).

\bibitem{Aviles14}
A. Aviles, A. Bravetti, S. Capozziello and O. Luongo, {Phys. Rev. D} \textbf{90} (2014) 043531.

\bibitem{rocco3}
S. Capozziello, R. D'Agostino and O. Luongo, {Mon. Not. Roy. Astron. Soc.} \textbf{476} (2018) 3924.

\bibitem{Chebyshev}
P. L. Chebyshev,  {M{\'e}moires des Savants {\'e}trangers pr{\'e}sent{\'e}s \'a l'Acad{\'e}mie de Saint-P{\'e}tersbourg} \textbf{7} (1854) 539.

\bibitem{Obsieger13}
B. Obsieger, \textit{Numerical Methods III - Approximations of Functions} (University of Rijeka, 2013).

\bibitem{Betoule14}
M. Betoule \textit{et al.}, {Astron. Astrophys.} \textbf{568} (2014) A22.

\bibitem{Jimenez02}
R. Jimenez and A. Loeb, {Astrophys. J.} \textbf{573} (2002) 37.

\bibitem{Eisenstein05}
SDSS Collab. (D. J. Eisenstein \textit{et al.}), {Astrophys. J.} \textbf{633} (2005) 560.

\bibitem{MontePython}
B. Audren, J. Lesgourgues, K. Benabed and S. Prunet, {J. Cosm. Astrop. Phys.} \textbf{02} (2013) 001.

\bibitem{orlH}
O. Luongo, {Mod. Phys. Lett. A},  {\bf 26} (2011) 1459.

\bibitem{EIS}
A. Aviles, J. Klapp, O. Luongo, {Phys. Dark Univ.}, {\bf 17} (2017) 25.

\bibitem{Lewis:2002ah}
A.~Lewis and S.~Bridle, {Phys. Rev. D}, {\bf 66} (2002) 103511.


\bibitem{cruzless}
V. C. Busti, P. K. S. Dunsby, A. de la Cruz-Dombriz, D. Saez-Gomez, {Phys. Rev. D}, {\bf 92} (2015)  123512.



\bibitem{rocco4}
S. Capozziello, R. D'Agostino and O. Luongo,  {J. Cosm. Astrop. Phys.} \textbf{1805} (2018) 008.

\bibitem{Starobinski07}
A. A. Starobinsky, {JETP Lett.} \textbf{86} (2007) 157.

\bibitem{Cognola08}
G. Cognola, E. Elizalde, S. Nojiri, S. D. Odintsov, L. Sebastiani and S. Zerbini, {Phys. Rev. D} \textbf{77} (2008) 046009.

\bibitem{Tsujikawa08}
S. Tsujikawa, {Phys. Rev. D} \textbf{77} (2008) 023507.

\bibitem{James13}
G. James, D. Witten, T. Hastie and R. Tibshirani, \textit{An Introduction to Statistical Learning} (Springer-Verlag, New York, 2013).

\bibitem{Martins17}
C. J. A. P. Martins, {Rep. Prog. Phys.} \textbf{80} (2017) 12.

\bibitem{Diego}
S. Nojiri, S.D. Odintsov, D. Saez-Gomez,  Phys. Lett. B {\bf 681} (2009) 74.

\bibitem{transition}
S. Nojiri and S.D. Odintsov, Phys. Rev. D {\bf 74} (2006) 086005.

\bibitem{rocco5}
S. Capozziello, R. D'Agostino and O. Luongo, {Gen. Rel. Grav.} \textbf{51} (2019) 2.

\bibitem{Dick04}
R. Dick, {Gen. Rel. Grav.} \textbf{36} (2004) 217.

\bibitem{Dominguez04}
A. E. Dominguez and D. E. Barraco, {Phys. Rev. D} \textbf{70} (2004) 043505.


\bibitem{delaCruz16}
A. de la Cruz-Dombriz, P. K. S. Dunsby, O. Luongo and L. Reverberi, {J. Cosm. Astrop. Phys.} \textbf{1612} (2016) 042.

\bibitem{Muthukrishna16}
D. Muthukrishna and D. Parkinson, {J. Cosm. Astrop. Phys.} \textbf{1611} (2016) 052.

\bibitem{Draper98}
N. R. Draper and H. Smith, \textit{Applied Regression Analysis} (Wiley-Interscience, 1998).

\bibitem{Nesseris13}
S. Nesseris, S. Basilakos, E. N. Saridakis and L. Perivolaropoulos, {Phys. Rev. D} \textbf{88} (2013) 103010.

%%%%%%%%%%  tabelle Appendice %%%%%%%%%%%%

\bibitem{Zhang14}
C. Zhang \textit{et al.}, {Res. Astron. Astrophys.} \textbf{14} (2014) 1221.

\bibitem{Simon05}
J. Simon, L. Verde and R. Jimenez, {Phys. Rev. D} \textbf{71} (2005) 123001.

\bibitem{Moresco12}
M. Moresco \textit{et al.}, {J. Cosm. Astrop. Phys.} \textbf{8} (2012) 006.

\bibitem{Chuang12}
C. H. Chuang, {Mon. Not. Roy. Astron. Soc.} \textbf{426} (2012) 006.

\bibitem{Moresco16}
M. Moresco \textit{et al.}, {J. Cosm. Astrop. Phys.} \textbf{05} (2016) 014.

\bibitem{Stern10}
D. Stern \textit{et al.}, {J. Cosm. Astrop. Phys.} \textbf{1002} (2010) 008.

\bibitem{Moresco15}
M. Moresco,  {Mon. Not. Roy. Astron. Soc.} \textbf{450} (2015) L16.

\bibitem{Beutler11}
F. Beutler \textit{et al.}, {Mon. Not. Roy. Astron. Soc.} \textbf{416} (2011) 3017.

\bibitem{Ross15}
A. Ross \textit{et al.}, {Mon. Not. Roy. Astron. Soc.} \textbf{449} (2015) 835.

\bibitem{Anderson14}
BOSS Collab. (L. Anderson \textit{et al.}),  {Mon. Not. Roy. Astron. Soc.} \textbf{441} (2014) 24.

\bibitem{Delubac15}
BOSS Collab. (T. Delubac \textit{et al.}) {Astron. Astrophys.} \textbf{574} (2015) A59.

\bibitem{Font-Ribera14}
BOSS Collab. (A. Font-Ribera \textit{et al.}), {J. Cosm. Astrop. Phys.} \textbf{1405} (2014) 27.

%\cite{DelVecchio:2018abv}
\bibitem{Lor}
  L.~Del Vecchio, L.~Fatibene, S.~Capozziello, M.~Ferraris, P.~Pinto and S.~Camera,
  %``Hubble drift in Palatini $f(\mathcal{R})$-theories,''
  Eur.\ Phys.\ J.\ Plus {\bf 134} (2019)   5
  %doi:10.1140/epjp/i2019-12382-y
 % [arXiv:1810.10754 [gr-qc]].
  %%CITATION = doi:10.1140/epjp/i2019-12382-y;%%


\end{thebibliography}
\end{document}